\documentclass[12pt,preprint]{aastex}
\usepackage{graphicx}
\newcommand\ha{\hbox{H$\alpha$}}
\newcommand\ic{$I_\mathrm{C}$}

\newcommand\tm{\textit{2MASS\/}}
\newcommand\wise{\textit{WISE\/}}
\newcommand\spitzer{\textit{Spitzer\/}}
\newcommand\co{$^{13}$CO}
\newcommand\sdss{\textit{SDSS\/}}
\newcommand\msun{M$_{\sun}$}
\begin{document}
 
\title{THE YOUNG STELLAR POPULATION OF LYNDS 1340. AN INFRARED VIEW}
\author{M. Kun\altaffilmark{1}, G. Wolf-Chase\altaffilmark{2,}\altaffilmark{3}, A. Mo\'or\altaffilmark{1}, D. Apai\altaffilmark{4}, Z. Balog\altaffilmark{5}, J. O'Linger-Luscusk\altaffilmark{6}, G. Moriarty-Schieven\altaffilmark{7}}
 \email{kun@konkoly.hu}
 \altaffiltext{1}{Konkoly Observatory, Research Centre for Astronomy and Earth Sciences, Hungarian Academy of Sciences, H-1121 Budapest, Konkoly Thege \'ut  15--17, Hungary}
 \altaffiltext{2}{Astronomy Department, Adler Planetarium, 1300 South Lake Shore Drive, Chicago, IL 60605 USA}
 \altaffiltext{3}{Department of Astronomy and Astrophysics, University of Chicago, 5640 South Ellis Avenue,
Chicago, IL 60637 USA}
\altaffiltext{4}{Steward Observatory, 933 N. Cherry Av. Tucson, AZ, USA}
 \altaffiltext{5}{Max-Planck-Institut f\"ur Astronomie, K\"onigstuhl 17, D-69117 Heidelberg, Germany} 
 \altaffiltext{6}{On leave from California Institute of Technology, 1200 E
California Ave, Pasadena, CA 91125}
\altaffiltext{7}{National Research Council - Herzberg Astronomy \& Astrophysics, 5071 West Saanich Rd., Victoria, BC,
V9E 2E7, Canada}

%\email{kun@konkoly.hu}

%\date{Received  / Accepted}

\begin{abstract}
We present results of an infrared study of the molecular cloud Lynds~1340, forming three groups of low and intermediate-mass stars. Our goals are to identify and characterise the young stellar population of the cloud, study the relationships between the properties of the cloud and the emergent stellar groups, and integrate L1340 into the picture of the star-forming activity of our Galactic environment. We selected candidate young stellar objects from the \textit{Spitzer\/} and \textit{WISE\/} data bases using various published color criteria, and classified them based on the slope of the spectral energy distribution. We identified 170 {\it Class~II}, 27 {\it Flat SED}, and 45 {\it Class~0/I} sources. High angular resolution near-infrared observations of the RNO~7 cluster, embedded in L1340, revealed eight new young stars of near-infrared excess. The surface density distribution of young stellar objects shows three groups, associated with the three major molecular clumps of L1340, each consisting of $\lesssim 100$ members, including both pre-main sequence stars and embedded protostars. New Herbig--Haro objects were identified in the \textit{Spitzer\/} images. Our results demonstrate that L1340 is a prolific star-forming region of our Galactic environment in which several specific properties of the intermediate-mass mode of star formation can be studied in detail.
\end{abstract}

%\titlerunning{Lynds 1340}
%\authorrunning{Kun et al.}

\maketitle
                                                                                
\section{INTRODUCTION}
\label{Sect_1}

The star-forming history of molecular clouds, as well as early evolution of stars and protoplanetary disks depend on the environment \citep[e.g][]{Zhang2015}. Since most stars form in a clustered environment, it is important to assess how this environment influences the time scales and efficiencies of star formation, and the evolution of protoplanetary disks around young stars. 
The impact of feedback from the newborn high mass (spectral types O and early B) stars on the evolution of their natal cloud and the properties of the emergent star clusters are studied in detail by e.g. \citet{Dib2013}. Important basic properties of massive star-forming regions (MSFRs) have emerged from the MYStIX project \citep{MYStIX}. The effect of intermediate-mass stars (i.e., spectral types mid-to-late B and early A) on the ambient medium in which they are forming have attracted less interest.  There are clouds with structure and star-forming properties intermediate between the two extremes of isolated star formation (e.g., Taurus, Cepheus flare) and the rich clusters found around very massive stars (e.g., Orion). In these regions, young stars are concentrated in small clusters, whose highest mass member is usually a B-type star. Well-known nearby examples of this type are IC~348, NGC~7023, and NGC~7129. The role of this intermediate-mode of star formation in shaping the present appearance of our Galaxy is not well known. \citet{AM01} suggested that most of the Galactic stellar content might have originated from clusters containing fewer than  some 100  members. A clearer observational picture of the intermediate mode is essential to our understanding of the star formation process. 

\citet{Arvidsson10} identified a sample of 50 intermediate-mass star-forming regions (IM\,SFRs), based on {\it IRAS\/} colors, \textit{Spitzer\/} images, as well as millimeter continuum and $^{13}$CO maps. They found typical luminosities of $\sim 10^4$\,L$_{\sun}$, diameters of $\sim1$\,pc, and associated molecular clumps of mass $10^3$\,M$_{\sun}$. Recently \citet{Lund14} presented an all-sky sample of 984 candidate intermediate-mass Galactic star-forming regions, and studied in detail four of the candidates, confirming that these regions contain loose clusters of low and intermediate-mass stars. The \co\ survey of \citet{Lund15} has shown that molecular linewidth and column density correlate with the infrared luminosity of the region. Several targets of the \textit{Spitzer} survey of young stellar clusters within one kiloparsec of the Sun \citep{Gutermuth09} belong to this class of star-forming regions.
Evidence for the impact of intermediate-mass stars on their interstellar environment comes from \citet{Arce2011}, who identified a great number of bubble-like structures in Perseus, most of them around intermediate-mass stars. Examination of these star-forming regions is particularly important because it helps understand the relationship between cloud structure and star-forming mode. 

The first large-scale study of Lynds~1340 \citep[][hereinafter Paper~I]{KOS94}, including an objective prism survey for \ha\ emission, low resolution $^{12}$CO, $^{13}$CO, and C$^{18}$O maps, and IRAS data analysis, suggests that this cloud is an IM~SFR, containing a few mid-B, A and early F type stars associated with reflection nebulosities \citep{DG}.  The $^{13}$CO maps revealed three clumps, L1340\,A, L1340\,B, and L1340\,C. 
Ten dense cores have been identified in L1340 through a large-scale NH$_3$ survey \citep*[][hereinafter Paper~II]{KWT03}, with masses and  and kinetic temperatures halfway between the values obtained for the ammonia cores in Taurus and Orion. Thirteen \ha\ emission objects were identified in Paper~I, and 14, which were concentrated in the small nebulous cluster RNO~7 were identified  by \citet{MMN03}. Herbig--Haro objects and their driving sources are reported in \citet{KAY03} and \citet{MMN03}. 
Our recent paper  \citep[][hereinafter Paper~III]{Kun2016}, reports on 11 candidate intermediate-mass (2--5\,\msun) members and 60 new candidate T~Tauri stars in L1340, and presents a revised distance of 825~pc.

Whereas most of the cluster-forming molecular clouds of our Galactic neighborhood, including those studied by \citet{Arvidsson10} and \citet{Lund14} are parts of giant star-forming regions, which also contain high-mass stars \citep[e.g.][]{Ridge2003}, Lynds~1340 is an isolated molecular cloud of some 3700\,M$_{\sun}$ at a Galactic latitude of $b \approx 11.5^\mathrm{o}$, corresponding to some 160~pc distance above the Galactic plane. 
To explore the nature of interstellar processes, leading to star formation in this environment, the cloud structure and the young stellar population have to be mapped. In this paper we identify the young stellar object (YSO) population of L1340 based on \textit{Spitzer} and \textit{WISE\/} mid-infrared data, as well as on high angular resolution near-infrared imaging data of the embedded RNO~7 cluster. The goals of our studies are as follows.  
(i) Determine the properties of star formation in this cloud, such as surface distribution, mass and age spread, accretion and disk properties of young stars, efficiency of star formation; (ii) explore possible feedback from intermediate-mass stars; (iii) integrate this cloud into the picture of star formation of our 1-kpc Galactic environment. 
We describe the available data and analysis in Sect.~\ref{Sect_2}. The results are presented and discussed in Sections~\ref{Sect_3}--\ref{Sect_5}. A short summary of the results is given in Sect.~\ref{Sect_sum}.

\section{DATA}
\label{Sect_2}

\subsection{\textit{Spitzer}\/ Data}
\label{Sect_spitzer}

\begin{figure*}
\centering{
\includegraphics[width=12cm]{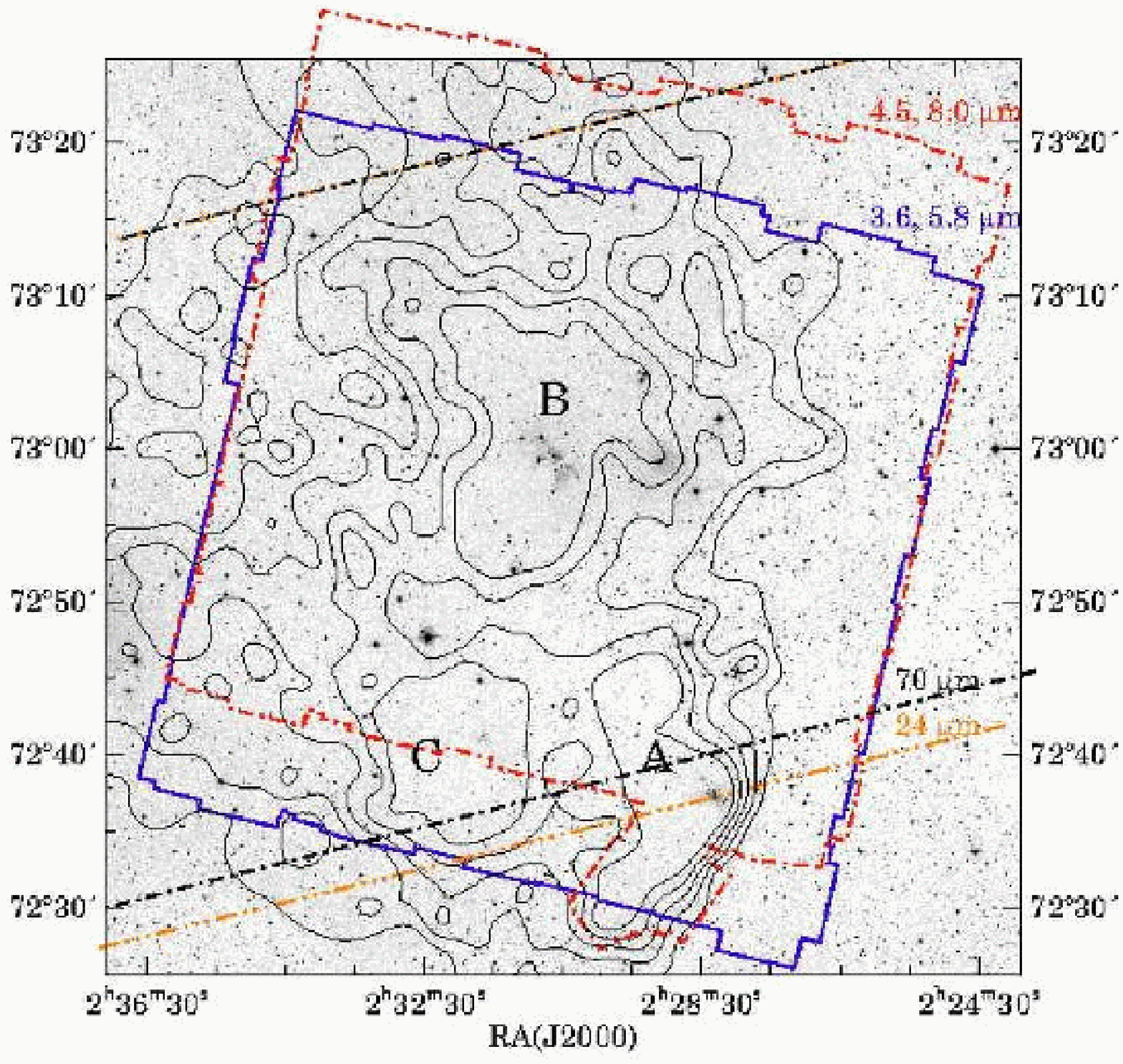}}
\caption{Outlines of the Spitzer IRAC (3.6 and 5.8~\micron: blue solid line,  4.5 and 8.0~\micron: red dash-dotted line), MIPS 24\,\micron\ (orange, dash-three dots), MIPS 70\,\micron\ (black dash-dotted line) observations, overplotted on the DSS2 red image of the region, one square degree in area. Both the lowest contour and the increment of the overlaid $^{13}$CO integrated intensity map are 0.5\,K\,km\,s$^{-1}$.}
\label{fig1}
\end{figure*}

L1340 was observed by the \textit{Spitzer Space Telescope} using \spitzer's Infrared Array Camera  \citep[IRAC;][]{Fazio2004}  on 2009 March 16 and the Multiband Imaging Photometer for Spitzer  \citep[MIPS;][]{Rieke2004} on 2008 November 26 (Prog. ID: 50691, PI: G. Fazio). The IRAC observations covered $\sim 1$~deg$^2$ in all four bands. Moreover, a small part of the cloud, centered on RNO~7, was observed in the four IRAC bands on 2006 September 24 (Prog. ID: 30734, PI: D. Figer). Figure~\ref{fig1} shows the areas of the {Spitzer\/} observations, overplotted on the {\it DSS2 red} image of the region. $^{13}$CO contours from Paper~I are drawn to indicate the boundaries of the molecular cloud, and the L1340\,A, L1340\,, and L1340\,C clumps are marked. The centers of the 3.6 and 5.8~\micron\ images are slightly displaced from those of the 4.5 and 8~\micron\ images, therefore part of the clump L1340\,C is outside of the 4.5 and 8~\micron\ maps. Moreover, the 24 and 70\,\micron\ images do not cover the southern half of L1340\,A.
The data of the four IRAC and MIPS 24~\micron\ bands were processed by the Spitzer Science Center (SSC) and the resulting Super Mosaics and Source List are available at \url{http://irsa.ipac.caltech.edu/data/SPITZER/Enhanced/SEIP/}.
We selected candidate YSOs from the {\it Spitzer Enhanced Imaging Products\/} (SEIP) Source List, containing 19745 point sources in the target field. 

We followed the methods described in \citet{Gutermuth09} for removing probable extragalactic, stellar, and interstellar sources and selecting candidate YSOs based on color indices. 
We identified 98 candidate YSOs detected in each of the four IRAC bands (Phase~1 criteria of \citealt{Gutermuth09}). Phase~2 criteria, based on \tm, 3.6 and 4.5\,\micron\ data, resulted in 44 new YSO candidates. Based on their high MIPS 24\,\micron\ fluxes and very red $[24]-[IRAC_\mathrm{i}]$ color (Phase 3 criteria), we identified 46 additional sources that were missing one or more IRAC band data. Four additional sources obeyed the criteria $[4.5] - [8.0] > 0.5$ and $[8.0] < 14 - ([4.5] - [8.0])$, set by \citet{Harvey06}. A sizeable area of the cloud was observed only at 3.6 and 5.8\,\micron. We regarded sources, located in this area and having $[3.6]-[5.8] > 0.50$, as candidate YSOs. Thirteen new objects were selected by this criterion. Most of them have associated \sdss, \tm, and/or \wise\ data, which help confirm their candidate YSO nature. 
We also subjected the SEIP Source List of L1340 to the criteria established by \citet{Kryukova} for selecting protostars. Of the 116 sources meeting the color criteria, there are 19 not selected during the previous steps and located within the lowest significant C$^{18}$O contours of the cloud clumps. These sources were also included into the candidate YSO list.

\begin{figure}
 \centering{
 \includegraphics[width=8cm]{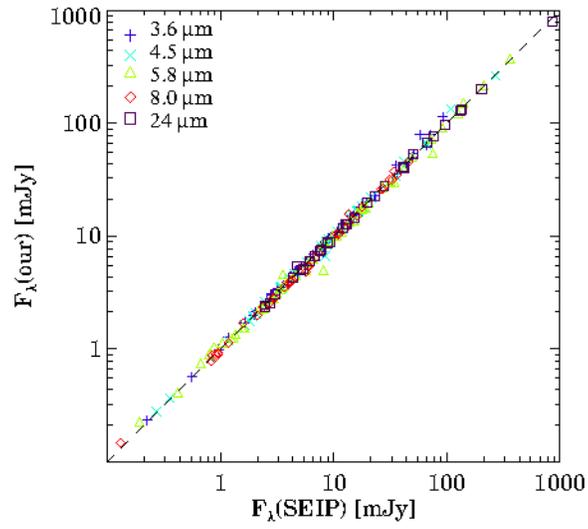}}
 \caption{SEIP Source List fluxes of a sample of candidate YSOs plotted against those measured during the present work.} 
 \label{fig2}
 \end{figure}

Due to the strict quality requirements of the SEIP Source List several sources might have been missed in one or more bands. Furthermore, the 70-\micron\ data are not included in the SEIP data base. Therefore we checked the positions of the selected sources and performed photometry by the procedures described in \citet{Kun2014} to refill the missing flux data. Then we checked the 70-\micron\ images at each source position and measured 70-\micron\ fluxes.
Figure~\ref{fig2} compares our photometry with the SEIP Source List data. 

\placefigure{fig1}

\placefigure{fig2}

\subsection{High Angular Resolution Near-infrared Imaging}
\label{Sect_oc}

High angular resolution near-infrared images of two small regions of L1340 were obtained on 2002 October 24 in the {\it JHK\/} bands, using the near-infrared camera \textit{Omega-Cass\/}, mounted on the 3.5-m telescope at the Calar Alto Observatory, Spain. Our targets were  IRAS~02224+7227, the possible driving source of HH~487, and the compact, partly embedded cluster RNO~7, centred on IRAS~02236+7224. The results for IRAS~02224+7227 have been shown in \citet{Kun2014}. Here we present the results for RNO~7. 

Omega Cass's detector was a Rockwell $1024\times1024$ pixel HAWAII array (HgCdTe detector + Si MOSFET non-destructive readout). The plate scale was 0.1\arcsec/pixel. RNO~7 was observed at four dithering positions around the nominal position of IRAS~02234+7224, and the observations consisted of two dither cycles, and each cycle with 120~s ($4\times30$\,s in {\it J\/} and {\it H\/}, $12\times10$\,s in {\it K\/}) spent at each position. Thus, the total on-source integration time of a cycle was 480\,s in each filter. Double Correlated Read (Reset-Read-Read) was applied.

The data were reduced in IRAF. Following the flat-field correction and bad pixel removal the sky frame for each cycle was obtained by taking the minimum of the images at different dithering positions. This sky frame was subtracted from each individual image of a given cycle. Then the frames from a single cycle were combined into a mosaic image, and aperture photometry was performed on the reduced images. The instrumental magnitudes were transformed into the \textit{JHK}$_\mathrm{s}$ system by using the \tm\ magnitudes of 17 stars within the field of view. Then, in order to search for possible close visual companions, the point spread function of the images were  determined and the scaled psf of the stars were subtracted from the images. 

\subsection{Supplementary Data}
\label{Sect_add}

To classify the evolutionary status of the color-selected candidate YSOs and obtain as complete picture of the star-forming region and its YSO population as possible, we supplemented the \spitzer\ data with photometric data available in public data bases. The data bases included in our study are as follows. 

\paragraph{\tm\ and \textit{AllWISE} data} The SEIP Source List contains \wise\ and \tm\ associations of the catalogued objects. \wise\ 22-\micron\ fluxes exist for 24 \spitzer-selected candidate YSOs outside of the area of the 24-\micron\ MIPS observations. 
We included these association into the analysis, taking into account that, due to the different angular resolutions, a few 2MASS/WISE sources are associated with more than one IRAC source. Furthermore we searched the {\it AllWISE\/} Source Catalog \citep{Wright2010} for young stellar objects, using the color and flux criteria established by \citet{Koenig2012,Koenig2014}.  We identified eight new candidate YSOs, seven of which are located outside of the field of view of the \textit{Spitzer\/} observations.

\paragraph{\textit{Akari FIS/IRC\/} data} \textit{Akari\/} far-infrared all-sky survey images \citep{Doi2015}, tracing out the surface and temperature structure of the cold dust in the cloud region, are accessible at \url{http://www.ir.isas.jaxa.jp/AKARI/Archive/Images/FISMAP/}. We identified counterparts of 9 candidate YSOs the \textit{Akari/FIS Bright Source Catalogue\/} \citep{Yamamura}, containing point sources detected at 65, 90, 140, and 160~\micron. 

\paragraph{Submillimeter data\/} Part of the molecular clump L1340\,B was observed at 450 and 850\,\micron\ with the Submillimetre Common User Bolometer Array (SCUBA) on the James Clerk Maxwell Telescope. The outlines of the mapped area are shown in fig.~6 of Paper~III. The 850\,\micron\ image and positions, sizes and fluxes/upper limits of nine submillimeter sources  can be found in the {\it SCUBA Legacy Catalogues} \citep{DiFran}, at  \url{http://www3.cadc-ccda.hia-iha.nrc-cnrc.gc.ca/community/scubalegacy/}.  Four of them coincide in position with  \textit{Spitzer\/} sources. 

\paragraph{\textit{Herschel\/} data for L1340\,C}
The \textit{Planck\/}  Galactic cold clump PGCC~G130.38+11.26, associated with L1340\,C, was included in the detailed \textit{Herschel\/} study of cold clumps by \citet{Juvela12}. Far-infrared images, observed by the PACS instrument at 100 and 160\,\micron, as well as 250, 350, and 500~\micron\ images observed by the SPIRE instrument are available in the \textit{Herschel\/} Science Archive (\url{http://www.cosmos.esa.int/web/herschel/science-archive}). We found far-infrared counterparts of 20 color-selected \textit{Spitzer\/} sources in the PACS 100 and 160\,\micron\ images. We measured the fluxes of the sources on the  \texttt{level2.5 JScanam} images, downloaded from the \textit{Herschel\/} Science Archive (\emph{Galactic Cold Cores: A Herschel survey of the source populations revealed by Planck}, PI: M. Juvela). The photometry was performed using the  \texttt{L3\_multiplePointSourceAperturePhotometry.py}, supplied in HIPE 14.0 RC4 \citep[\textit{Herschel Interactive Processing Environment},][]{Ott2010}. We used 6\arcsec\  and 10\arcsec\ apertures at 100\,\micron\  and 160\,\micron, respectively, with an annulus between 35\arcsec\ and 45\arcsec\ for determining the background. The aperture correction were calculated using the values given in \citet{Balog2014}. The initial positions of the sources were taken from the SEIP Source List and were refined using a two dimensional Gaussian during the photometry. 

\paragraph{\textit{SDSS\/} data} \textit{SDSS\/} \textit{ugriz\/} magnitudes are available for each star brighter than some 25~mag in each band within the whole area of L1340 (see Paper~III). We searched for counterparts of our candidate YSOs the \textit{SDSS\/} Data Release~9 \citep{Ahn2012} within 1\arcsec\  to the SEIP Source List position. We transformed the \sdss\ magnitudes of the optical counterparts into the Johnson--Cousins {\it UBVR}$_\mathrm{C}${\it I}$_\mathrm{C}$  system, using the equations given in \citet{Ivezic} (for {\it BVR}$_\mathrm{C}${\it I}$_\mathrm{C}$) and \citet{Jordi06} (for {\it U}). We found optical counterparts for 149 of the 155 Class~II \textit{Spitzer\/} sources, and for 8 of the 26 Flat~SED sources (see Sect.~\ref{Sect_spyso}).

\section{INFRARED APPEARANCE OF L1340: THE SWAN NEBULA}
\label{Sect_3}

The extended infrared emission reveals the surface distribution of various components of the cloud. Cold ($T_\mathrm{kin} \sim $10--20\,K), big ($r \ga 0.1$\,\micron) dust grains radiate in the far-infrared, whereas extended mid-infrared emission traces out very small grains and excited PAH molecules. Heating and shocks from embedded YSOs also appear in the infrared images of a molecular cloud. 
 
\begin{figure*}
\centering{\includegraphics[width=16cm]{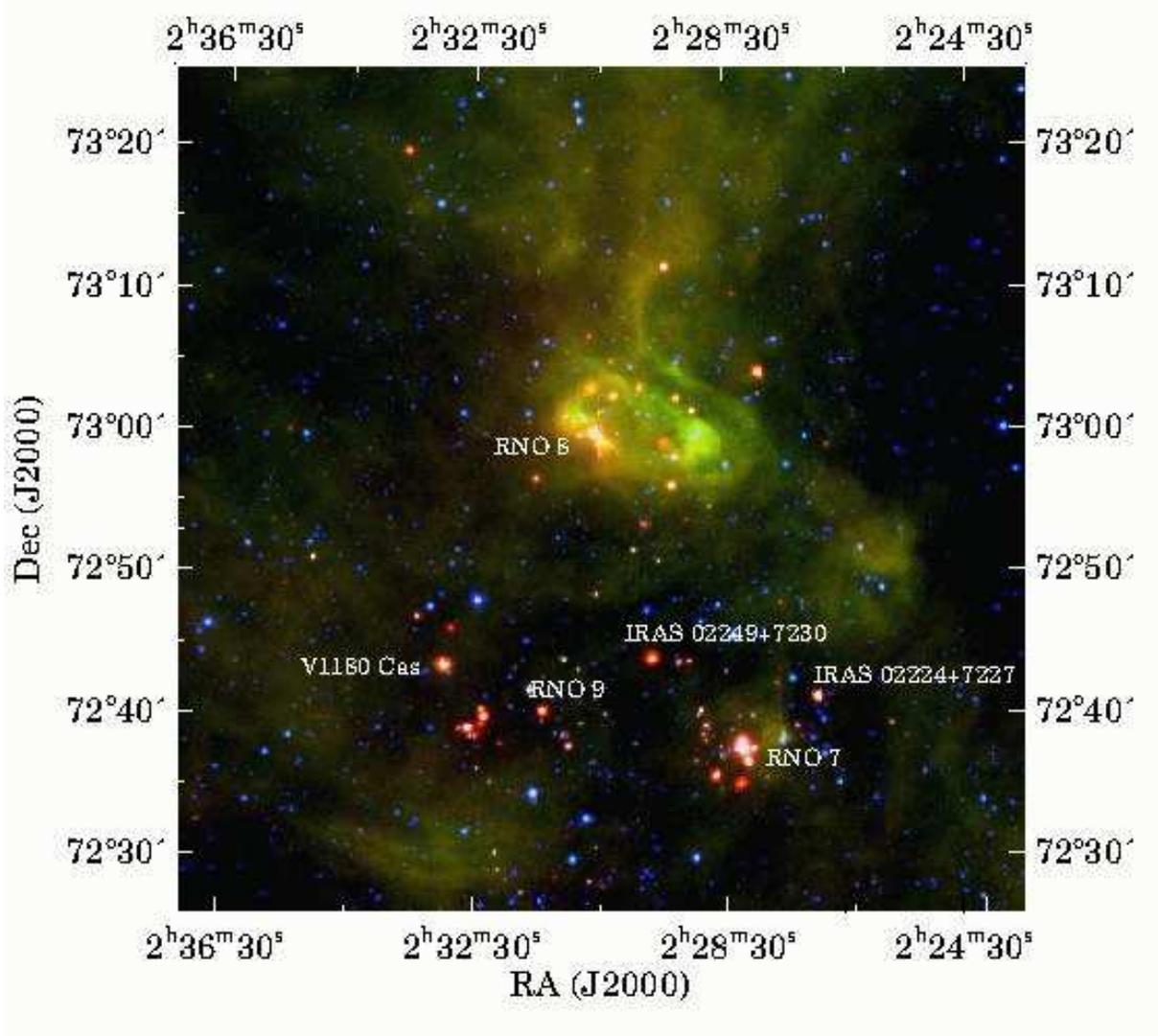}}
\caption{\textit{WISE\/} 4.6\,$\mu$m\ (blue), 12\,$\mu$m\ (green), and 22\,$\mu$m\ (red) composite image of L1340. The size of the image is 1~square degree, and is centered on  RA(J2000) = 37\fdg625, Dec(J2000) = +72\fdg933. Small groups of 22-\micron\ sources point to the three star-forming clumps of L1340. The extended 12\,\micron\ emission (the Swan nebula) is excited by the B and A type stars associated with L1340\,B. The annotated objects are the most striking signposts of low-mass star formation: IRAS~02224+7227 is a FUor-like star \citep{Kun2014}, IRAS~02249+7230 is a protostellar source exciting HH~489, RNO~7, RNO~8, and RNO~9 are nebulous, partially embedded stellar groups, and V1180~Cas is an eruptive star \citep{Kun2011}. }
\label{fig3}
\end{figure*}

Figure~\ref{fig3} shows a three-color view of L1340, composed of the \wise\ 4.6\,\micron\ (blue), 12\,\micron\ (green), and 22\,\micron\ (red) images. Striking features of this image are the bright, extended 12-\micron\ radiation, indicative of PAH emission excited by B and A type stars, and small groups of 22-\micron\ sources, associated with the three cloud clumps. The shape of the brightest part of the diffuse 12-\micron\  emission, located slightly northwest of the image centre, and associated with the clump L1340\,B, suggests the \textit{Swan nebula\/} label.

\begin{figure*}
\centering{
\includegraphics[width=14cm]{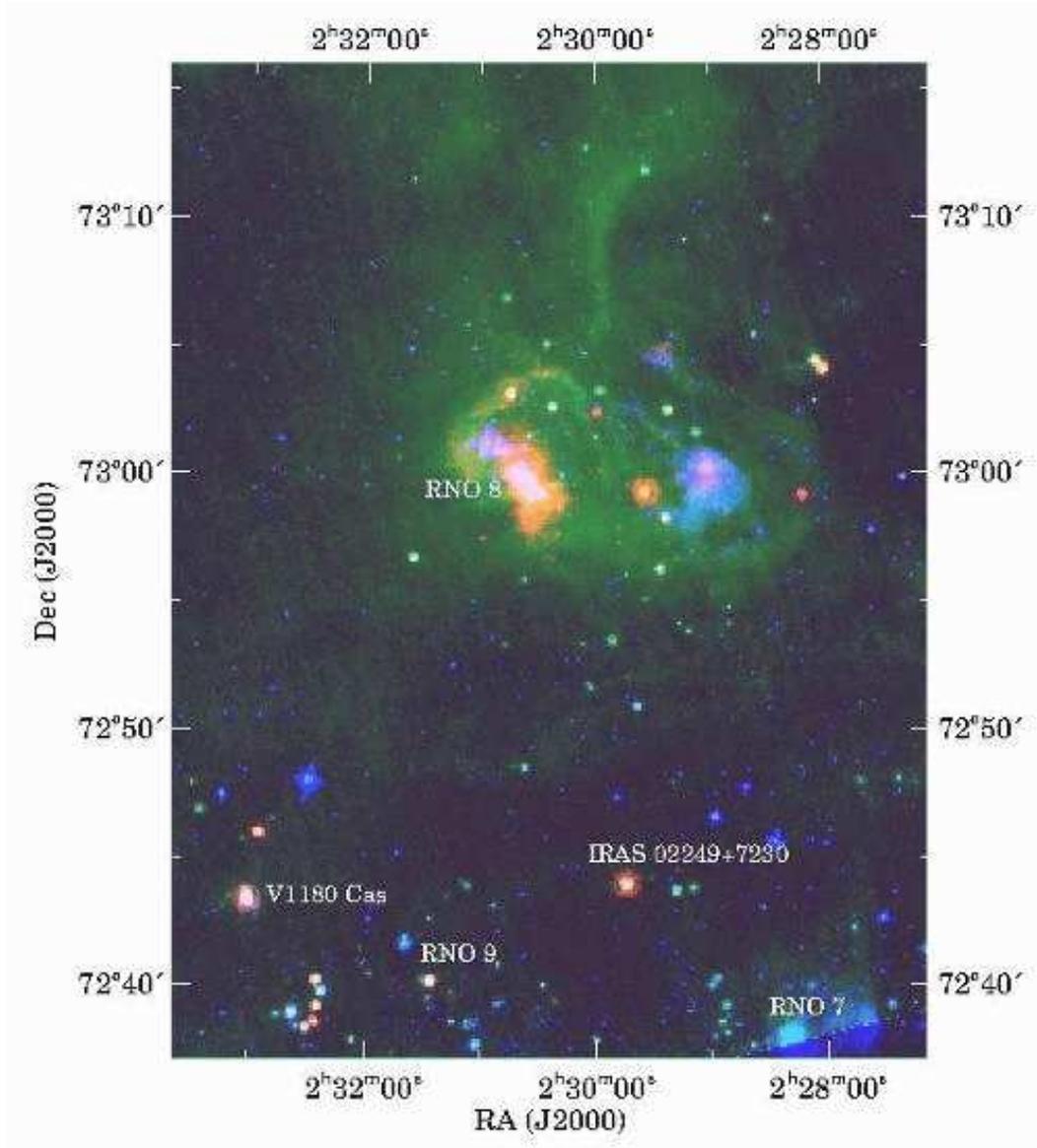}}
\caption{Three-color image of a $29\arcmin \times 39\arcmin$ size part of L1340, composed of \textit{Spitzer} IRAC 5.8\,\micron\ (blue), MIPS 24\,\micron\ (green), and MIPS 70\,\micron\ images. Part of L1340\,A is outside the field of view of the MIPS images.} 
\label{fig4}
\end{figure*}

Figure~\ref{fig4} is a composite of the 5.8\,\micron\  IRAC (blue), 24\,\micron\  MIPS (green), and 70\,\micron\  MIPS (red) images. The image reveals an extended 70-\micron\ structure associated with RNO~8, diffuse 24-\micron\ emission which delineates the Swan nebula, a bluish (5.8\,\micron) glowing around the B-type stars, and a variety of far-infrared point sources.

\begin{figure*}
\centering{
\includegraphics[width=16cm]{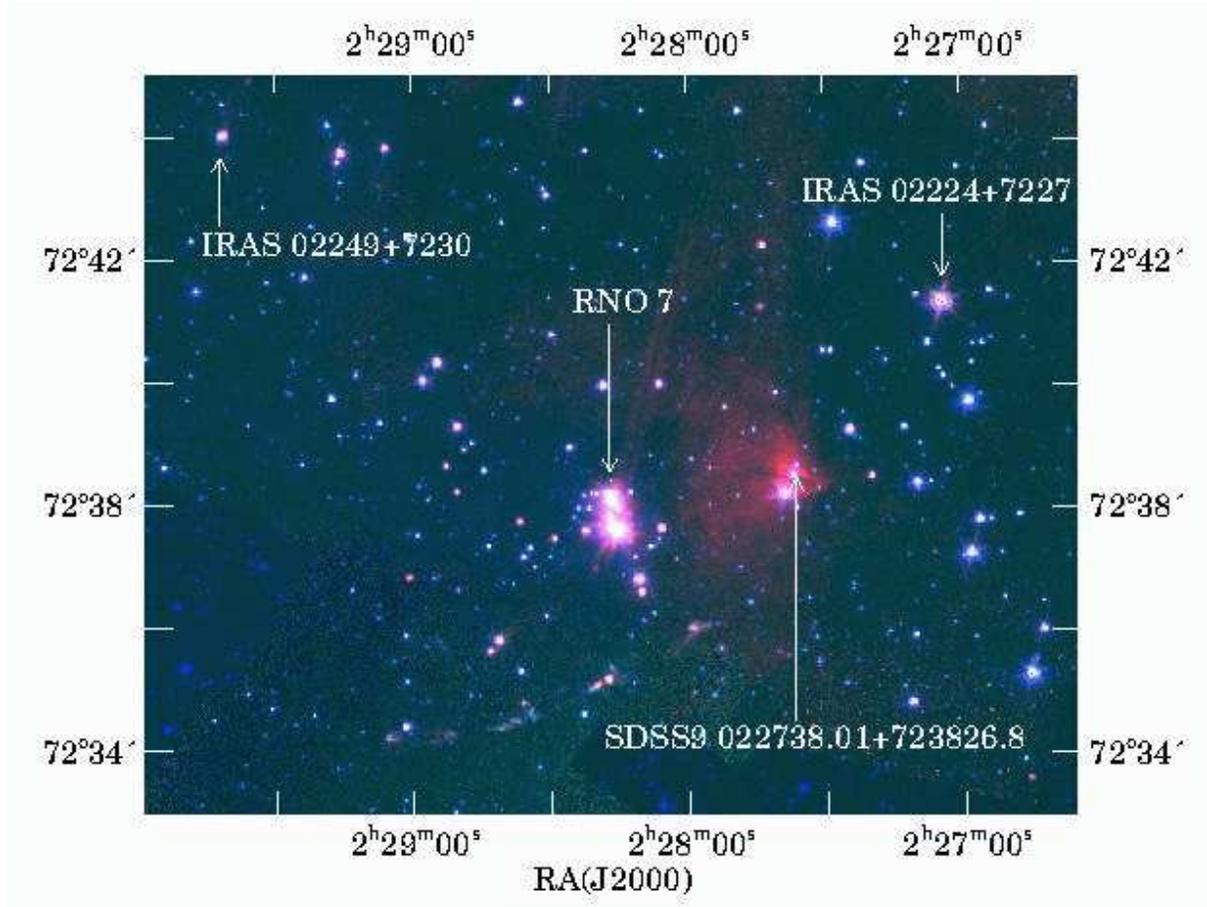}}
\caption{Three-color image of the cloud clump L1340~A, composed of {\it Spitzer} IRAC 3.6\,\micron\ (blue), 5.8\,\micron\ (green), and 8.0 \micron\ (red) images. The size of the image is about $15.6\arcmin \times 12\arcmin$. The arrows point to the positions of the FUor-like YSO IRAS 02224+7227, the A0-type star SDSS9~J022738.01+723826.8, associated with extended 8-\micron\ emission, the protostar IRAS~02249+7230, and the RNO~7 cluster.} 
\label{fig5}
\end{figure*}

\begin{figure*}
\centering{
\includegraphics[width=16cm]{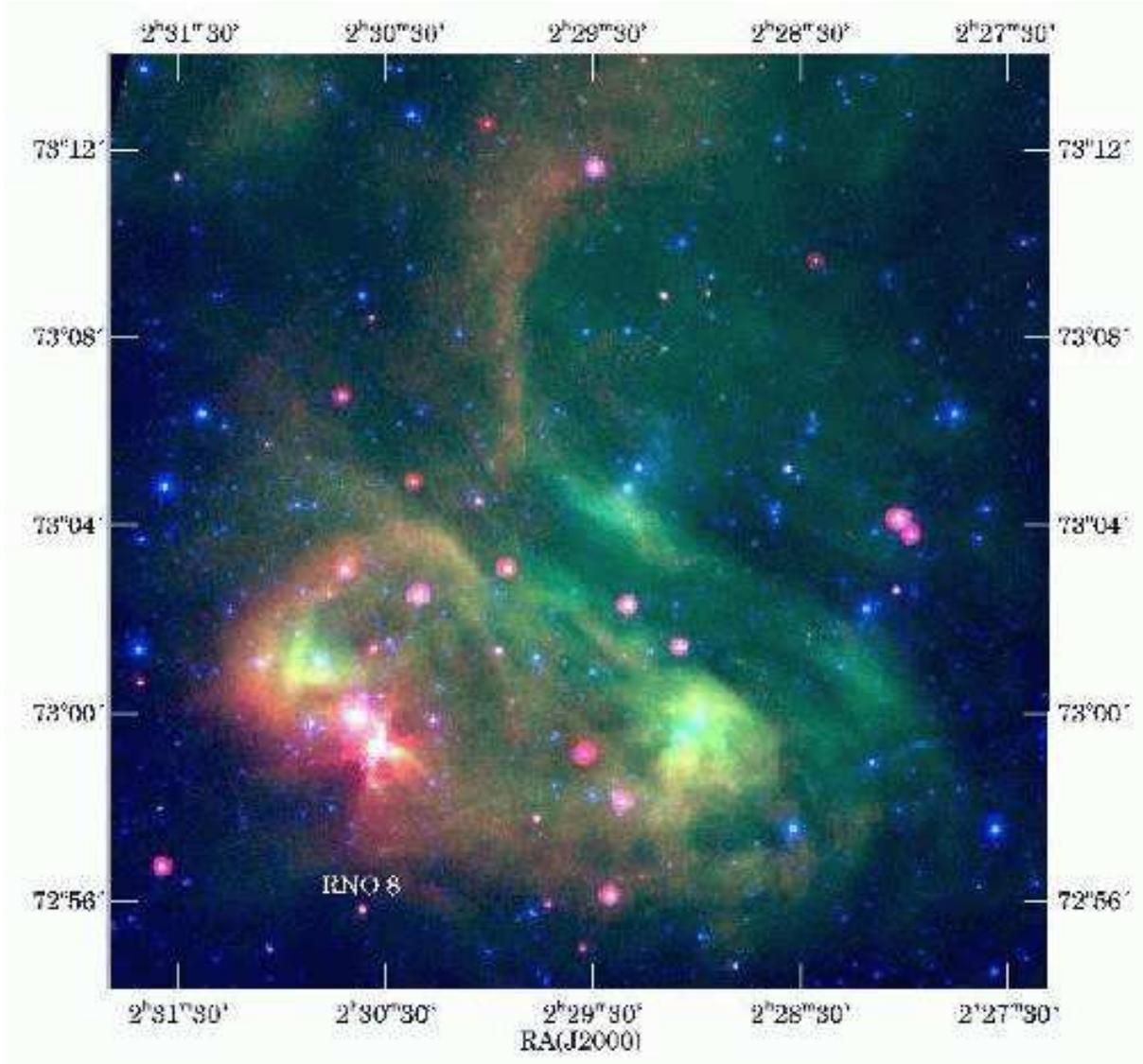}}
\caption{Three-color image of L1340\,B, the {\em Swan Nebula\/}, composed of {\it Spitzer IRAC} 4.5\,\micron\ (blue), 8.0\,\micron\ (green) and MIPS 24\,\micron\ (red) images.} 
\label{fig6}
\end{figure*}

\begin{figure*}
\centering{
\includegraphics[width=16cm]{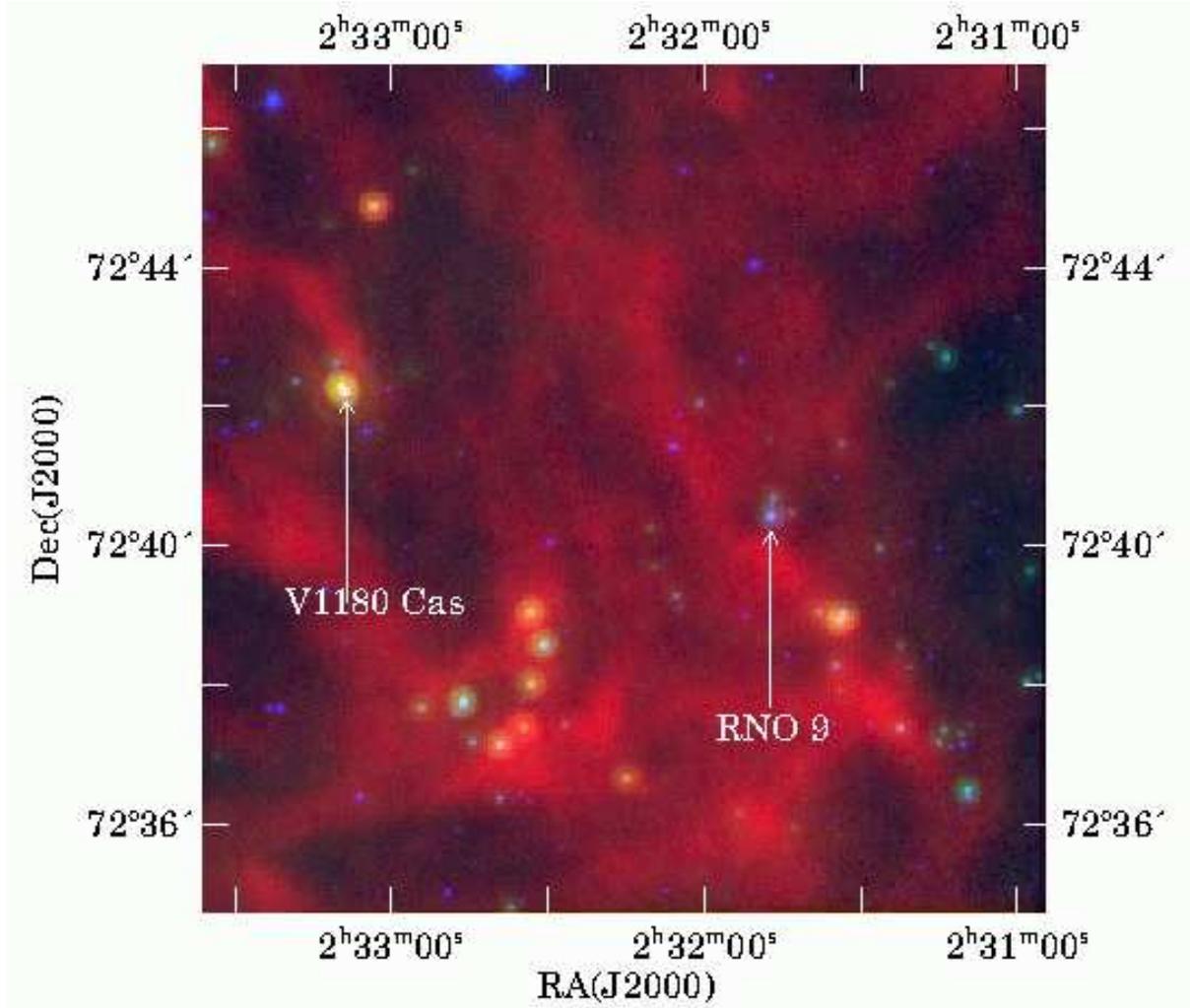}}
\caption{Three-color image of the clump L1340\,C, composed of {\it Spitzer\/} IRAC 3.6\,\micron\ (blue), MIPS 24\,\micron\ (green), and \textit{Herschel SPIRE\/} 250\,\micron\ (red) images.} 
\label{fig7}
\end{figure*}

To reveal further details of the diffuse infrared emission of L1340, we present three-color images of the clumps L1340\,A, L1340\,B, and L1340\,C in Figs.~\ref{fig5}, \ref{fig6}, and \ref{fig7}, respectively. Figure~\ref{fig5} is composed of IRAC 3.6\,\micron\ (blue), 4.5\,\micron\ (green), and 8.0\,\micron\  (red) Super Mosaic images of L1340\,A (much of this clump is outside of the MIPS images). Conspicuous features of the image are a diffuse  8-\micron\ emission around the A0 type star SDSS9 022738.01+723826.8 (Paper~III), the nebulous RNO\,7 cluster, and HH~488, stretching from NW towards SE near the southern boundary of the image. Figure~\ref{fig6} is composed of the 3.6\,\micron\  (blue), 8.0\,\micron\  (green), and 24\,\micron\ (red) images of the most massive clump L1340\,B. The wispy structure of the Swan nebula, suggesting a swirling gas cloud, becomes apparent in this image. A bow-shock like feature can be seen around the star SDSS9~J023049.80+730110.2, demonstrating supersonic motion of the gas with respect to the A2-type, young intermediate-mass star (Paper~III). The extended infrared emission from the smallest clump, L1340\,C, shows up in the \textit{Herschel\/} images, tracers of very cold dust. Figure~\ref{fig7}, composed of the 3.6\,\micron\ IRAC (blue), 24\,\micron\ MIPS (green), and 250\,\micron\ SPIRE (red) images of the central $12\arcmin\times12\arcmin$ area of L1340\,C reveals a complex network of filamentary dust formations.

\begin{figure}
\centering{
\includegraphics[width=14cm]{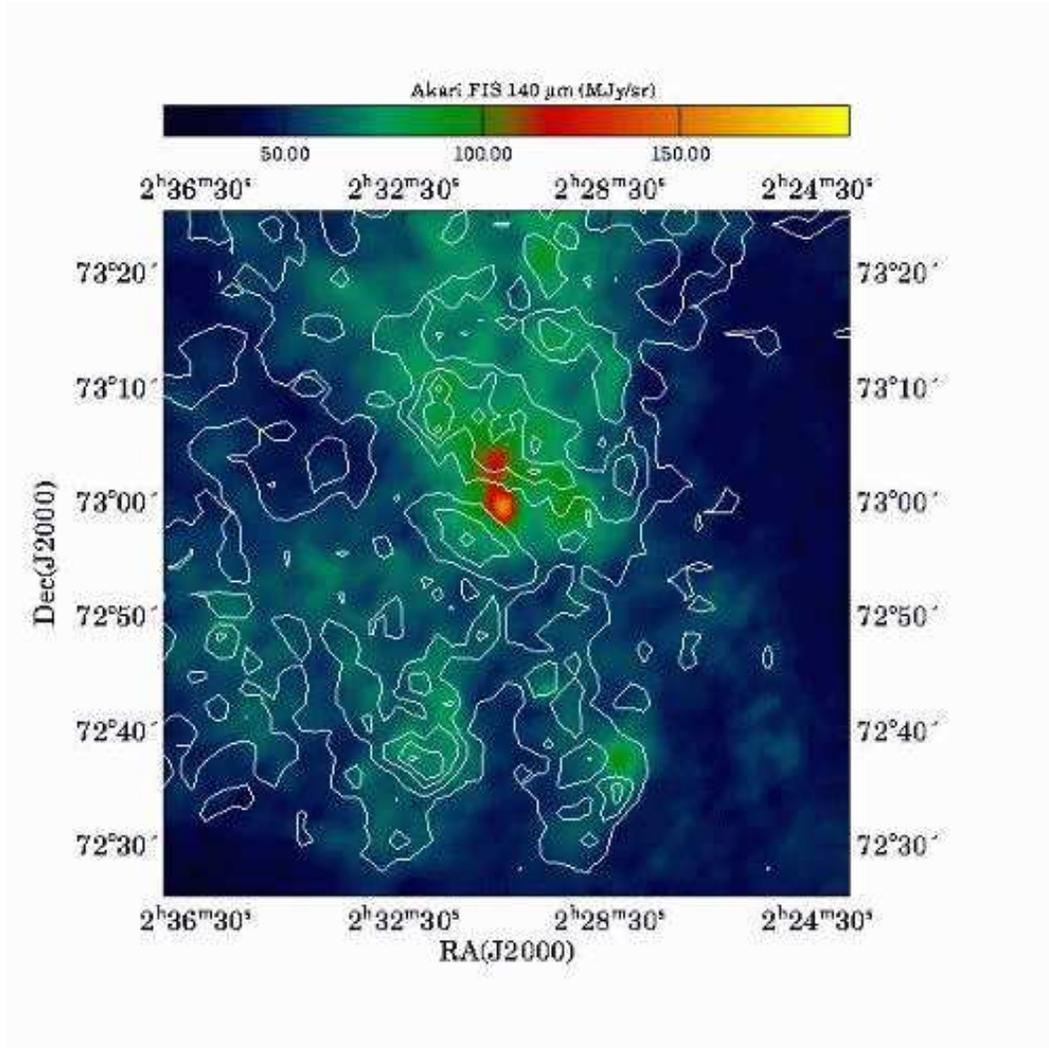}}
\caption{\textit{Akari Wide-L\/} map of L1340 (colors) with the visual extinction \citep[white solid contours,][]{Kun2016} overplotted. Both the lowest contour and increment for $A_\mathrm{V}$ are 1.0~mag. The center coordinates and size of the image are identical with those in Fig.~\ref{fig3}.}
\label{fig8}
\end{figure}

The \textit{Akari\/} Wide-L band image, centered on 140\,\micron, is displayed in Fig.~\ref{fig8}, with the contours of the visual extinction (Paper~III) overplotted. The Figure indicates that both the 140-\micron\ emission and the visual extinction trace the same component of the cloud. The lowest contour at $A_\mathrm{V}=1.0$~mag largely follows the 40--50~MJy\,sr$^{-1}$ level of the far-infrared emission. At a few positions, heated by embedded YSOs, the strong 140-\micron\ emission is not associated with high extinction. 

\placefigure{fig3}

\placefigure{fig4}

\placefigure{fig5}

\placefigure{fig6}

\placefigure{fig7}

\placefigure{fig8}

\section{YOUNG STELLAR OBJECTS IN L1340}
\label{Sect_yso}

\subsection{\textit{\textit{Spitzer\/} Sources}}
\label{Sect_spyso}

\subsubsection{SED-based Classification}

We classified the candidate YSOs, selected by the color criteria described in Sect.~\ref{Sect_spitzer}, based on the slope of their spectral energy distributions (SEDs), $\alpha=d\log(\lambda F(\lambda)) / d\log\lambda$. We derived  $\alpha$ both for the $K_\mathrm{s}$ --24\,\micron\ and the 3.6\,\micron--8.0\,\micron\ intervals (for 3.6\,\micron--5.8\,\micron\ when 8\,\micron\ observations were missing). We used the \wise\ 22\,\micron\ data when 24\,\micron\ MIPS data were missing. According to the canonical classification scheme \citep{Lada,Greene}, protostellar objects embedded in an envelope have $\alpha(2-24) > 0.3$, whereas $\alpha(2-24) < -0.3$ for pre-main sequence stars surrounded by accretion disks. Flat~SED sources with $-0.3 \le \alpha(2.0-24.0) \le 0.3$ represent the transition between the protostellar and pre-main sequence evolutionary phases. We classified 155 Class~II, 45 Class~I, and 25 Flat SED sources. We detected a further Class~I/Class~0 source in the 70-\micron\ MIPS image at $02^\mathrm{h}29^\mathrm{m}56\fs90$, $+73\degr02\arcmin17\farcs0$. This source is undetectable at shorter wavelengths, and coincides in position with an \textit{Akari FIS\/} source and with a submillimeter source. 

\subsubsection{Estimating Foreground Extinction}

Since the classification based on observed spectral slopes is biased by the extinction of the sources, we estimated the foreground extinctions of the candidate YSOs, and then reclassified them according to the extinction-corrected SED slopes.  
Foreground extinctions of Class~I and Flat~SED sources were estimated using the extinction map, derived from \textit{SDSS\/} star counts in Paper~III. We adopted the pixel value of the extinction map at the position of the source as the foreground extinction of an embedded source. On the one hand, the extinction obtained in this manner is an upper limit, since the sources may be situated at any depth within the dusty medium. On the other hand, small-scale, high-extinction cores, missed by the extinction mapping, may be present around embedded sources.  
For the Class~II sources we invoked \textit{SDSS\/} and \tm\  counterparts. We compared the optical and near-infrared side (from the {\it B\/} to the {\it J\/} band) of the SED with a grid of reddened photospheres, following the method described in Paper~III, and thus estimated the spectral type and extinction of the central star. Based on the slopes of the extinction-corrected SEDs, two sources, classified originally as Class~I, moved into the Flat class, and one Flat~SED source moved into the Class~II sample.
Tables~\ref{Tab1_class1}, \ref{Tab1_flat}, and \ref{Tab1_class2} list the SSTSL2 identifiers and \spitzer\ fluxes of the Class~0/I, Flat~SED, and Class~II sources of L1340, respectively.

The SEDs of Class~II sources can be divided into further subclasses by comparing the dereddened SED slopes with the median band of the Taurus pre-main sequence sample \citep{DAlessio,Furlan06}. The SED subclasses are indicative of the dust distribution in the circumstellar disks \citep{Evans09} of the classical T~Tauri stars, and may shed light on the processes governing disk evolution. 
We classified the infrared excesses of our candidate pre-main sequence stars into three groups: (1) the SED of primordial disks (II\,P subclass) does not drop below the Taurus median band; (2) the SED of the weak or \textit{anemic\/} disks (II\,A subclass) is below the Taurus band over the whole observed wavelength region, and (3) pre-transitional and transitional disks (II\,T) have SEDs below the Taurus median band at intermediate wavelengths, and start rising above 20\,\micron. For this latter group the spectral index $\alpha(8-24) > 0$.

\placetable{Tab1_class1}

\placetable{Tab1_flat}

\placetable{Tab1_class2}

\subsubsection{Submillimeter, Far-infrared, and Optical Counterparts}

Six \spitzer\  sources are associated with submillimeter sources listed in the \textit{JCMT SCUBA Fundamental Catalogue\/} \citep{DiFran}. Far-infrared counterparts of 17 candidate Class~0/I, and three Flat~SED YSOs were identified in the \textit{Herschel\/} PACS images. Table~\ref{Table_pacs} lists the SSTSL2 associations, 100\,\micron, and 160\,\micron\ fluxes of these \textit{Herschel\/} point sources. 

\placetable{Table_pacs}

Nine of the Spitzer-selected candidate YSOs coincide in position with far-infrared sources detected by the \textit{Akari/FIS} instrument \citep{Kawada07}. Four of them are included in the \textit{Akari/FIS\/} young stellar object catalog \citep{Toth2014}. A fifth catalog entry, \textit{Akari}~0232291+723855, has an associated mid-infrared point source, \textit{AllWISE\/} 023227.63+723841.4, within the half-maximum radius of the point-spread function of the FIS \citep{Arimatsu14}. Its fluxes, however, probably originate from more than one sources. Similarly, the far-infrared fluxes of \textit{Akari FIS}~0230333+725951, a bright candidate  YSO detected in each FIS band and associated with IRAS~02259+7246, are composed of several sources. An extended emission can be seen around this position in the \textit{Spitzer} 70-\micron\ image. 
We found \textit{SDSS\/} counterparts of all but seven Class~II infrared sources. 
A few Flat and Class~I sources also have \sdss\ counterparts. Most of these counterparts are classified as galaxies. The non-stellar appearance, however, may indicate their scattered light origin.

\sdss, \tm, \textit{AllWISE\/}, \textit{Akari\/}, and other identifiers of Class~I and Flat sources are listed in Tables~\ref{Tab2_class1} and \ref{Tab2_flat}, respectively. For the Class~II sample, excluded the 65 members common with the H$\alpha$ emission stars studied in Paper~III, we give the {\it UBVR}$_\mathrm{C}${\it I}$_\mathrm{C}${\it JHK}$_\mathrm{s}$ magnitudes in Table~A1 of the Appendix.

\placetable{Tab2_class1}

\placetable{Tab2_flat}

The SEDs of the candidate YSOs, constructed from all available data, are displayed in Fig.~\ref{fig9}, \ref{fig10}, and \ref{fig11} for the Class~I, Flat, and Class~II sources, respectively. Since the SEDs of the \ha\ emission stars, together with those of the best-fitting photospheres, have been presented in fig.~9 of Paper~III, Fig.~\ref{fig11} presents the results for the Class~II subsample not detected as \ha\ emission stars during our slitless spectroscopic \ha\ survey.  
The dereddened SEDs, as well as the best fitting photosphere \citep{Pecaut2013} are also plotted, and the derived spectral type and extinction are indicated in each plot. 

\placefigure{fig9}

\placefigure{fig10}

\placefigure{fig11}

\subsubsection{Bolometric temperatures and luminosities}

Bolometric temperatures and luminosities, as defined in \citet{ML93}, were derived from the dereddened SEDs for the Class~I and Flat SED objects, detected at least in one band beyond 24\,\micron, and for the Class~II sources, detected over the 0.36--24\,\micron\ region. \textit{Akari FIS\/}, \textit{Herschel PACS\/}, and \textit{JCMTSF\/} submillimeter data were included into the integration when available. Contribution of the spectral regions beyond the longest wavelength was estimated using the method described by \citet{Chavarria}. 
 The $L_\mathrm{bol}$ vs. $T_\mathrm{bol}$ diagram of the candidate YSOs is plotted in Fig.~\ref{fig12}.
The YSO Classes, defined by the spectral slopes, correspond to the $T_\mathrm{bol}$ intervals indicated in Fig.~\ref{fig12} \citep{Chen95}. It can be seen that both $\alpha(2-24)$ and $T_\mathrm{bol}$ are consistent with the Class~0/I identification. Flat SED sources overlap in $T_\mathrm{bol}$ with both Class~I and Class~II, whereas a significant part of the Class~II sample has $T_\mathrm{bol}$ above the theoretical boundary of 2800\,K. It is in accordance with the recent finding of \citet{Dunham2015}, that the extinction-corrected $T_\mathrm{bol}$ of a Class~II source depends on the $T_\mathrm{eff}$ of the central star, rather than on the disk properties. Figure~\ref{fig13} shows the histogram of bolometric luminosities of the candidate YSOs. The mean $L_\mathrm{bol}$ of the 28 Class~I sources, detected at $\lambda > 24$\micron,  is $\langle L_\mathrm{bol,ClassI}\rangle = 3.4~L_\sun$, and the same for the Class~II sample is  $\langle L_\mathrm{bol,ClassII}\rangle = 1.2~L_\sun$.

\placefigure{fig12}

\placefigure{fig13}

Tables~\ref{Tab3_class1} and \ref{Tab3_flat} present the derived extinctions, extinction-corrected SED slopes, bolometric temperatures and luminosities of Class~0/I and Flat~SED sources, respectively. Table~\ref{Tab3_class2}, in addition to the above quantities, lists the derived spectral types and luminosities for central stars of the Class~II sources, as well as the SED subclasses. 

\placetable{Tab3_class1}

\placetable{Tab3_flat}

\placetable{Tab3_class2}

\subsection{New Candidate Members of RNO\,7 in the Omega-Cass Data}

The three-color composite of the {\it J\/} (blue), {\it H\/} (green), and the {\it K\/} (red) Omega-Cass images is shown in the second panel of Fig.~\ref{fig14}. For comparison, we show in the first panel an optical three-color view of the same region, composed of the \sdss\ \textit{g\/} (blue), \textit{r\/} (green), and \textit{i\/} (red) images, whereas the third panel shows the \spitzer\ 3.6\,\micron\ (blue), 4.5\,\micron\ (green), and 8\,\micron\ (red) composite image. The high angular resolution Omega-Cass images reveal a few new objects, detectable neither in the optical nor in the IRAC images. Furthermore, they show  that the brightest member of RNO~7, SSTSL2 J022816.62+723732.6, associated with IRAS~02236+7224, has a faint companion at an angular distance of 1.12\arcsec\ (Fig.~\ref{fig14}, fourth panel), corresponding to some 760~AU at a distance of 825~pc.  

\placefigure{fig14}

\placefigure{fig15}

The magnitudes measured in the Omega-Cass images, and transformed into the \tm\ system, are compared with the \tm\ magnitudes of the same stars in the left panel of Fig.~\ref{fig15}. The right panel of Fig.~\ref{fig15} shows the {\it J\/}$-${\it H\/} vs.{\it H\/}$-${\it K}$_\mathrm{s}$ two-color diagram of the stars measured in each band. Twenty stars are located to the right of the band of the reddened normal main sequence and giant stars, indicating {\it K\/}$_\mathrm{s}$-band excess. Table~\ref{Table_oc} lists the derived magnitudes of these stars. All but two of them have 2MASS counterparts, but none of them has good (A or B) photometric quality in each band. Six of the 14 \ha\ emission stars, discovered by \citet{MMN03}, and seven \spitzer-identified candidate YSOs are found in this sample. Eight stars, marked with asterisks in Table~\ref{Table_oc}, are new candidate members of RNO~7. 
The SEDs of these eight stars, constructed from all available data, are presented in Fig.~\ref{fig16}. 

\placefigure{fig16}

\placetable{Table_oc}

\subsection{\textit{AllWISE\/} sources}

The 1~square degree area centered on RA(J2000) = 37\fdg625, Dec(J2000) = +72\fdg933 contained 954 sources, having signal to noise ratio greater than 5.0 in each band and not affected by upper-case contamination flag. We identified  seven new Class~II source candidates in the \wise\  database outside the area covered by the \spitzer\ images, but within the lowest significant $^{13}$CO contours of the molecular cloud. Each of them is located near the southern boundary of the cloud. Furthermore, two \wise\  sources without coinciding SSTSL2 entries, J022759.92+723556.4 and J023227.63+723841.4 were found within the field of view of the \spitzer\ observations. We measured their fluxes in the available bands, and added the sources to Tables~\ref{Tab1_flat} and \ref{Tab1_class1}, respectively. The selected \textit{AllWISE\/} sources are listed in Table~\ref{Table_wise}. The SEDs of the seven \wise\ sources, identified as candidate YSOs outside the field of view of the \spitzer\ observations, are displayed in Fig.~\ref{fig17}. Each of them is a Class~II source. Their $A_\mathrm{V}$ extinctions, spectral types, and luminosities derived from the photometric data, are listed in Table~\ref{Tab2_wise}.

\placetable{Table_wise}

\placefigure{fig17}

\placetable{Tab2_wise}

\subsection{Embedded Protostars and Herbig--Haro objects in L1340}
\label{Sect_class1}

\subsubsection{Candidate Class 0 Sources}

The extinction-corrected SED slopes revealed the presence of 45 Class~0/I and 27 Flat SED candidate YSOs. Eight sources have $T_\mathrm{bol} \lesssim 70$\,K, suggesting Class~0 evolutionary stage \citep{ML93}. These are as follows.
\begin{itemize}
\item[(1)] SSTSL2~J022808.60+725904.5 coincides with an \textit{Akari FIS\/} and a \textit{JCMTSF\/} submillimeter source (see Table~\ref{Tab2_class1}). Its SED, assembled from all available data (Fig.~\ref{fig9}), shows deep silicate absorption around 10\,\micron, suggesting a Class~0 protostar seen at high inclination \citep{Enoch09}. This object is associated with a parsec-scale outflow identified in H$_2$ 2.12-\micron\ observations (J. Walawender et al. 2016, in prep.). The three-color image of its environment, composed of IRAC 8\,\micron\ (red), 4.5\,\micron\ (green), and 3.6\,\micron\ (blue) images and displayed in Fig.~\ref{fig18}, shows 4.5-\micron\ emission, originating from shocked H$_2$. 
\item[(2)] SSTSL2~J022820.81+723500.5 lies outside the MIPS 70-\micron\ image. Its steeply rising SED is revealed by the \textit{Akari FIS\/} data. With  $L_\mathrm{bol} \approx 23$\,$L_{\sun}$ it is the most luminous protostar of L1340. This source, together with another nearby Class~I source 022818.51+723506.2, is located along the chain of Herbig--Haro objects HH~488 whose several knots were detected in optical \ha\ and \ion{S}{2} images by \citet{KAY03} and \citet{MMN03}. \citet{KAY03} suggested that the driving source was the brighter component of a binary star located at $2^\mathrm{h}28^\mathrm{m}00^\mathrm{s}, 72\degr35\arcmin58\arcsec$ (HH\,488\,S, Source~2 in Table~\ref{Tab1_flat}). The optical counterpart of HH\,488\,S is classified as a galaxy in the SDSS~DR9, and as an HH object by \citet{MMN03}. Our photometry suggests a Flat SED, although it results from the composite fluxes of the central objects. The positions of the two protostars with respect to the HH knots suggest that either of them is the probable driving source. The IRAC images reveal new knots of HH~488. In the upper panel of Fig.~\ref{fig19} we marked the known and new knots of HH~488 and the candidate driving sources. The lower panel shows a three-color composite image of HH~488, whose angular extension of 5.6\arcmin\ corresponds to a total length of some 1.3\,pc at a distance of 825\,pc.
\item[(3)] SSTSL J022931.98+725912.4 is associated with \textit{IRAS\/} and \textit{Akari\/} far infrared, and \textit{JCMTSF\/} submm source  (Table~\ref{Tab3_class1}). It is associated with a small fan-shaped reflection nebulosity, bright at 3.6\,\micron\ on the eastern side, and a jet-like feature, bright at  4.5\,\micron\ on the western side (Fig.~\ref{fig20}).
\item[(4)] The fourth candidate Class~0 protostar is the 70-micron source No.~22 in Table~\ref{Tab1_class1}. It is associated with the brightest submillimeter source of the region. 
\item[(5)] SSTSL~023256.14+724605.3 is an embedded eruptive young star in L1340\,C, discussed in \citep{Kun2014}. Its $T_\mathrm{bol}$ and $L_\mathrm{bol}$ were determined including the \textit{Herschel\/}  100\,\micron\ and 160\,\micron\ fluxes.
\item[(6)--(7)--(8)] SSTSL~023146.58+723729.4, 023237.90+723940.7, and 023330.92+724800.3 are low-luminosity sources, not detected in the 70\,\micron\ MIPS image. Their low bolometric temperatures were revealed by including the \textit{Herschel\/} 100\,\micron\ and 160\,\micron\ data into the SEDs. Their nature is uncertain: they may be either very low luminosity protostars or faint distant galaxies.
\end{itemize}

\placefigure{fig18}

\placefigure{fig19}

\placefigure{fig20}

\subsubsection{Class~I protostars associated with \textit{IRAS\/} sources}

Six \textit{IRAS\/} sources, listed in table~6 of Paper~I, are associated with Class~I \spitzer\ sources (see Table~\ref{Tab2_class1}). IRAS 02249+7230 in L1340\,A is the driving source of HH\,489 \citep{MMN03}. The \textit{Spitzer\/} data show it to be a wide binary, consisting of two Class~I sources, SSTSL2~J022943.01+724359.6 and SSTSL2~J022943.64+724358.6, separated by 2.8\arcsec. Their SEDs are shown in Fig.~\ref{fig9}, and the environment is displayed in the three-color image in Fig.~\ref{fig21}. HH~489\,A, identified in optical \ha\ and \ion{S}{2} images by \citet{MMN03} as well as a chain of faint HH knots to the south can clearly be seen. Their projected distribution suggests that both components of the binary and another nearby Class~I source, SSTSL2~J022950.37+724441.4 may contribute to their excitation.

\placefigure{fig21}

Figure~\ref{fig12} shows that $T_\mathrm{bol}$  of the second brightest Class~I object of L1340 falls into the Class~II regime near the Class~I/Class~II boundary. This ambiguous classification belongs to SSTSL2 J023032.44+725918.0, associated with \textit{IRAS\/}~02259+7246 and RNO~8. A faint optical star is visible at its position. Our low-resolution spectrum (Paper~III) reveals its late G spectral type with the Balmer lines in emission, and the optical color indices point to an unreddened star. The bolometric luminosity, determined from the \ic\ or {\it J\/} magnitudes, places this star near the ZAMS. All these data suggest the high inclination of the disk of this star. The optical and infrared images confirm this statement. Three-color images, shown in Fig.~\ref{fig22}, reveal the connections between various components of the circumstellar environment of RNO~8.
The gap between the star and the nebulosity in the optical three-color image (first panel of Fig.~\ref{fig22}) suggests a huge shadow of the circumstellar disk on the dusty envelope, stretching far beyond the disk. The image composed of the optical {\it g\/} (blue), IRAC 3.6\,\micron\ (green), and IRAC 8\,\micron\  (red), and shown in the second panel of Fig.~\ref{fig22}, reveals streaks of 8-\micron\  emission overlapping with the reflected starlight. The image in the third panel is composed of the 4.5\,\micron\ (blue), 8\,\micron\ (green), and 24\,\micron\ (red) images. The overplotted contours of the 70\,\micron\ emission reveal a cloud core associated with RNO~8. 

\placefigure{fig22}

In L1340\,C a J-shaped chain, consisting of five Class~I and four Class~II YSOs, can be seen close to the extinction peak (see Fig.~\ref{fig29}). \textit{IRAS\/}~02276+7225 and \textit{Akari FIS\/} 0232291+723855 are situated in the same area, but neither of them can be unambiguously associated with mid-infrared sources.
Similarly, \textit{IRAS\/}~02267+7226 and \textit{Akari FIS\/} 0231270+724015 coincide with the Class~I source SSTSL2~J023127.34+724012.9 within the position uncertainties, but other nearby sources may contribute to their catalogued fluxes. 

SSTSL2~J023302.41+724331.2, coinciding with \textit{IRAS\/}~02283+7230, is the Class~I companion of the eruptive star V1180~Cas. This protostar drives a jet, detected by \citet{Anton14} in [\ion{S}{2}] and \ha\ narrow-band images. The IRAC 4.5\,\micron\ image also clearly show the jet (Fig.~\ref{fig23}), as well as several faint HH objects.  The 8-\micron\ image reveals a probable third component of the system, located at $4.8\arcsec$ towards the north-northwest from V1180~Cas.

\placefigure{fig23}

\textit{IRAS\/}~02240+7259, detected at 100~\micron\ only by \textit{IRAS\/} and thus not listed in Paper~I, coincides with a faint candidate protostar SSTSL2~J022855.69+731333.1, not detected at 70\,\micron. Taking into account the \textit{IRAS\/} 100-\micron\ flux the SED suggests a Class~0/I source with $T_\mathrm{bol} \approx 75$\,K. The nature of this source, however, is uncertain: it may be a distant galaxy. 

\subsection{Classical T Tauri stars}
\label{Sect_ttau}

The \spitzer, \wise, and Omega-Cass data resulted in 170 Class~II young stars in the region of L1340. These stars represent the classical T~Tauri star (CTTS) population of L1340. Sixty-five of the 77 \ha\ emission stars, presented in Paper~III, are members of this sample. These stars are marked with asterisks in Tables~\ref{Tab1_class2} and \ref{Tab3_class2}. Histograms of their {\it K\/}$_\mathrm{s}$ magnitudes, derived $A_\mathrm{V}$ and $T_\mathrm{eff}$ values are shown in Fig.~\ref{fig24}, together with those of the \ha\  emission subset (Paper~III). It can be seen that \ha\ emission was detected in brighter and hotter Class~II stars. Only five of the Class~II stars brighter than $K_\mathrm{s}=11.5$ were not detected during the \ha\ survey, and only one \ha\ emission star has spectral type later than M2. The derived extinctions of the Class~II sources peak between $2~mag <$ {\it A\/}$_\mathrm{V} < 3~mag$. 

\placefigure{fig24}

After estimating their spectral classes and extinctions we plotted the positions of all candidate pre-main sequence stars in the $\log T_\mathrm{eff}$--$\log L$ plane. The intermediate-mass young main sequence stars, identified in Paper~III, are also plotted. Effective temperatures of the spectral types were adopted from \citet{Pecaut2013}. Bolometric luminosities were derived from the extinction-corrected \ic\ and {\it J\/} magnitudes, separately, using the bolometric corrections and color indices tabulated for pre-main sequence stars by \citet{Pecaut2013}, and adopting the distance of 825\,pc. Finally the results obtained from the \ic\ and {\it J\/} magnitudes were averaged. Figure~\ref{fig25} shows the Hertzsprung--Russell diagram. Evolutionary tracks and isochrones for the 0.1\,\msun $\le M_\mathrm{star} \le 5.0$\,\msun\ interval are from \citet{Siess}, and the track for 0.07\,\msun\  from \citet{BHAC2015} is also plotted. Most of the candidate YSOs are located between the 1 and 10 million-year isochrones, confirming their pre-main sequence star nature at 825\,pc from us. Exceptions are a few Class~II objects close to or below the ZAMS. The SEDs of these stars suggest that their disks have high inclinations, and thus most of the optical fluxes arise from scattered light (see Paper~III for further details). The HRD suggest a mass range between 0.07\,\msun\ (M5 type) and 2.5\,\msun\  (G--early K type stars, evolving towards higher T$_\mathrm{eff}$). 

\placefigure{fig25}

We examined whether the average properties of stars surrounded by primordial (SED subtype II\,P in Table~\ref{Tab3_class2}), weak (II\,A), and transitional (II\,T) disks can be distinguished or not. Table~\ref{Table_sedcomp} show the mean {\it K\/}$_\mathrm{s}$ magnitudes, and derived mean {\it A\/}$_\mathrm{V}$, $T_\mathrm{eff}$, $L_\mathrm{star}$, $T_\mathrm{bol}$, and $L_\mathrm{bol}$ values of the three groups. The Table shows that most of the candidate CTTSs of L1340 have weak (anemic) disks. We find that the central stars of primordial disks are brighter in each photometric bands, and have higher average $T_\mathrm{eff}$ than the others, in accordance with the findings of Paper~III. The bright \ha\ emission stars of Flat SED (Table~\ref{Tab1_flat}, Paper~III) fit into this trend.

\placetable{Table_sedcomp}

A most prominent member of the T Tauri star population of the region is the \ha\ emission star associated with \textit{IRAS\/}~02236+7224. Its early G spectral type (Paper~III) suggests a mass of $\sim 2$\,\msun. The Omega-Cass images reveal a faint companion at an angular distance of $1.12\arcsec$ ($\sim$ 760\,AU). The IRAC 8-\micron\ image shows a further companion at 4.3\arcsec\  (3550~AU) to the northwest, and another one at 4.9\arcsec\  (4040~AU) to the southeast from the primary star (Fig.~\ref{fig14}, lower right panel).

\section{SURFACE DISTRIBUTION OF THE YOUNG STELLAR POPULATION}
\label{Sect_cluster}

The positions of all candidate YSOs, identified by infrared color indices, are overplotted on the extinction map of the region, together with $^{13}$CO and C$^{18}$O contours (from Paper~I), in Fig.~\ref{fig26}. More detailed maps of the central regions of the L1340\,A, L1340\,B, and L1340\,C clumps are presented in Figs.~\ref{fig27}, \ref{fig28}, and \ref{fig29}, respectively.

\placefigure{fig26}
\placefigure{fig27}
\placefigure{fig28}
\placefigure{fig29}

The surface distribution of the candidate YSOs  reveals a rich population of young stars clustered over the clumps L1340\,A, B, and C. We constructed surface density maps of the YSOs following the method described by \citet{Gutermuth2005}. We determined the $r_N(i,j)$ distance of the $N$th nearest star at each \textit{(i,j)} position of a uniform grid, and obtained the local surface density of YSOs at the grid point as 
$\rho(i,j) =N/\pi r_N^2(i,j)$. The surface density contour plot, shown in the left panel of Fig.~\ref{fig30}, was constructed using a 30\arcsec\  grid and $N = 6$, and shows the surface densities of Class~I+Flat (red dot-dashed contours) and Class~II (blue solid contours) sources separately, overlaid on the \textit{WISE\/} 12-\micron\ image of L1340. The contour labels indicate the surface densities in star\,pc$^{-2}$ units. The YSO groups associated with the cloud clumps are apparent. Like the three clumps, the associated YSOs show diverse surface structures. The surface distribution in L1340~A suggests a west-to-east progression of star formation. Similarly, in L1340\,C, Class~I and Class~II sources are apparently separated from each other. The largest clump L1340\,B is associated with an extended, low surface density population. The right panel of Fig.~\ref{fig30} shows a composite surface density distribution of all YSO classes, derived at the same grid points, and using the distance of the 20th nearest YSO. The three clusterings, associated with the three clumps, remain apparent in the smoothed surface density map. The area of each YSO group and the number of stars within the surface density contour 2~stars\,pc$^{-2}$ are listed in Table~\ref{Table_clump}. 

\placefigure{fig30}

\placetable{Table_clump}

\subsection{Young Clusters in L1340}

To find and characterise clusters in the YSO population of L1340 we examined the projected distances between the stars within the three groups seen in Fig.~\ref{fig30}. Figure~\ref{fig31} shows the histograms of the nearest neighbor separations for the three groups, separately. The histograms of groups associated with L1340\,A and L1340\,C  show peaks at short spacings, similarly to other nearby star-forming regions \citep{Gutermuth09}. On the contrary, no preferred spacing range can be seen in the histogram of L1340\,B. The median separations of the YSOs are 0.117\,pc, 0.243\,pc, and 0.141\,pc in L1340\,A, B, and C, respectively. 
 
\placefigure{fig31}

Figure~\ref{fig32} show the YSO distribution overplotted on the extinction map, and the stars having a neighbor closer than 0.15~pc ($\approx$42\arcsec) are marked by underlying black dots. In L1340\,A, 75 percent of the YSO population belong to this clustered subsystem, while 66\,\% of the YSOs in L1340\,C and 34\,\% in L1340\,B have neighbors within this distance. We identified four small clusters encircled by the overplotted ellipses. This criterion reveals 56 members of the RNO~7 cluster in L1340\,A, including the K-band excess stars identified in the Omega-Cass data. The RNO~9 cluster in L1340\,C consists of 22 Class~II, three Flat, and one Class~I sources, whereas six of the 12 members of the cluster associated with IRAS~02276+7225, are Class~I/Flat sources. The only small clustering in clump~B consists of eight stars, including the bright Class~I source RNO~8.
The coordinates and sizes of the clusters, identified by the nearest neighbor spacings, and the number of stars within them are listed in Table~\ref{Table_cluster}. The sizes are described by the major axis (a) and aspect ratio (AR) of the smallest ellipse encircling the members. The sampling of the members was not homogeneous, since part of L1340\,A was not covered by the MIPS observations, whereas the \textit{IRAS\/}~02276+7225 cluster is outside of the 4.5\,\micron\  and 8\,\micron\ IRAC images. Moreover, eight members of the central core of RNO~7 comes from the Omega-Cass observations. For comparison, the last row of Table~\ref{Table_cluster} lists the median values derived for the young cluster sample in our 1-kpc Galactic environment \citep{Gutermuth09}. 

\placefigure{fig32}

\placetable{Table_cluster}

\subsection{YSO Distribution and the Cloud Structure}

The four small, compact clusters identified above comprise nearly half of the candidate YSOs. The distributed population consists of Class~II stars scattered widely over low-extinction regions, and small groups of a few closely spaced YSOs. An example is the small aggregate marked by a red circle in Fig.~\ref{fig32}, consisting of Class~I, Flat, submillimeter, and strongly reddened Class~II sources, and similar in angular size to the knots seen in the extinction map. To demonstrate the connection between the cloud structure and YSO distribution, we present in Fig.~\ref{fig33} a multi-wavelength view of L1340\,B, revealing various aspects of interactions between the cloud and embedded stars. The upper panel of Fig~\ref{fig33} suggests that Class~0/I sources of L1340\,B are associated with small-scale dust clumps. The morphology of this image suggests that the filamentary structure, detected at 850\,\micron, might have been created by past and present winds of the nearby young B- and A-type stars. The middle panel demonstrates the interactions of the intermediate-mass stars with the gas and dust, and reveals a diversity of the embedded YSOs. The lower panel reveals that a chain of Class~0/I/Flat sources and two ammonia cores (Paper~II) are lined up along a ridge of 850\,\micron\ emission, starting with the Class~0 source J022808.60+725904.5 at the south-western side, and stretching over a projected length of some 3~pc to the Flat~SED source 023042.36+730305.1 at the north-eastern end. The average separation of the protostars/bright knots along the submillimeter filament, $\sim$1.6\arcmin, corresponds to  $\sim$0.4\,pc at 825\,pc. 

\placefigure{fig33}

Linear configurations in the distribution of protostars are thought to result from fragmentation of dense molecular filaments \citep[e.~g.][]{Teixeira06}. The separation of protostars along the filament is of the order of the Jeans length. Temperatures and densities derived from the ammonia mapping of L1340 (from Paper~II) suggest a Jeans length of $\sim 0.14$\,pc for the dense cores of L1340. The wide separation of the protostars along the submillimeter filament of L1340\,B, as well as the large average spacing of the nearest neighbors throughout the clump suggest higher temperature of the ambient medium. The NH$_3$ cores are probably the coldest regions of the cloud, embedded in a warmer gas, heated by the nearby B and A-type stars.

Another conspicuous linear feature is the J-shaped configuration of YSOs in L1340\,C (the IRAS~02276+7225 cluster, Fig.~\ref{fig29}). The average separation of the objects within that chain is 27.9\arcsec, corresponding to 0.11~pc at a distance of 825\,pc. The total length of the chain is some 0.9\,pc, suggesting that these stars have been formed from cores of $\sim 0.1$\,pc in diameter. This coincides with the average size of the ammonia cores studied in Paper~II, and is same as the Jeans length at $T_\mathrm{kin} \approx 12.5$\,K and $n_{H_2}\approx 1.29\times10^{4}$\,cm$^{-3}$, resulted from the NH$_{3}$ observations. The cloud structure, underlying the observed distribution of the protostars, can be seen in the distribution of the cold dust, revealed by the \textit{Herschel SPIRE\/} images (Fig.~\ref{fig7}).

 \section{STAR FORMATION IN L1340}
 \label{Sect_5}

At   a distance of 825~pc and a latitude of 11\fdg5 L1340 is situated some 160~pc above the Galactic plane, in a low-density outer region of the molecular disk of our Galaxy. (The Swan is floating on the surface of the Galactic molecular disk). The average hydrogen column densities of the three molecular clumps are about $2.5\times10^{21}$\,cm$^{-2}$, slightly lower than the mean column density of  $4.38\times10^{21}$\,cm$^{-2}$, obtained by \citet{Lund15} for IM\,SFRs in the outer Galaxy. The extinction map of L1340 (Paper~III) reveals a shallow molecular cloud, spotted with dense knots of a few arcminutes ($\sim 0.5$\,pc) characteristic size. YSOs are grouped on similar angular scales, and Class~0/I--Flat sources appear closely associated with extinction knots (see Fig.~\ref{fig33}), suggesting that star formation occurs in small groups, consisting of a few stars, and scattered over the surface of the cloud.
The most massive star in L1340 is an optically visible B4-type star of some 5\,M$_{\sun}$, whereas the YSOs revealed by our present survey are in the $0.07 \la M/M_{\sun} \la 2.5$ mass interval. 

The number of embedded sources, and their ratio to the more evolved pre-main sequence stars in a star-forming region is an indicator of evolutionary state. \citet{Myers12} established relations between Class~II/Class~I number ratios, as well as ages and birthrates of young stellar clusters, assuming a constant protostellar birthrate. The Class~II/Class~I ratio (Table~\ref{Table_clump}) suggests an age of 1 million years and a birthrate of 200--300 protostars/Myr. The three clumps of L1340 differ from each other in several respects. The effects of young intermediate-mass stars on the environment are conspicuous in L1340\,B. The wispy structure of the 8-\micron\ emission, the bow-shock like structure associated with an A2-type star, and the double-peaked CO lines at the positions of the \textit{Planck\/} Cold Clumps \citep{Wu2012} associated with L1340, suggest violent swirling of the gas in this region. The low surface density of YSOs, compared to the other clumps of the cloud, indicates that the prestellar gas in L1340\,B had higher temperature and lower density than in L1340\,A and L1340\,C, due to the heating from the ambient B-type stars. The higher proportion of protostars in L1340\,B (NII/(NI+NFlat)=1.47) suggests that the average age of the YSO sample is lower in this clump than in the others. Star-forming regions like L1340\,B are probably more transient structures than centrally condensed young embedded clusters. \citet{Pfalzner15} have found that only clusters and associations with initial central surface densities exceeding a few 1000\,\msun\,pc$^{-2}$ will be detected as clusters at ages longer than 5~Myr. 

Assuming an average mass of 0.5\,\msun\ for each candidate YSO, and including the intermediate-mass stars, discussed in Paper~III, we find the star formation efficiencies (SFE = $M_{star} / (M_{star}+M_{cloud}$) listed in Table~\ref{Table_clump} for the three clumps of L1340. It can be seen that while some 17\% of the gas turned into star in L1340\,A, the SFE is only 3\% for L1340\,B and also for the whole cloud.  
The actual SFEs are probably somewhat higher, since the low-mass diskless YSO population of L1340 is still unknown.   

\subsection{Comparison with Other Star-Forming Regions}
\label{Sect_comp}

Comparison of our target cloud with IM\,SFRs, located in similar environments, may help to understand the interstellar processes, leading to star formation near the outer boundaries of the Galactic molecular disk. The short expected lifetime of L1340 (probably $\ll 5$~Myr) suggests that similar star-forming regions may be rare in our Galactic neighborhood. A sample 50 IMSFRs, studied by \citet{Arvidsson10}, contains objects similar in stellar content and total mass to L1340. Most of them are, however, more distant and thus their detailed structures are still unrevealed. The \spitzer\ sample of young clusters in our Galactic neighborhood \citep[SSYSC,][]{Gutermuth09} also contains several IMSFRs. Comparison of our results with several properties of this sample of young clusters is shown in Table~\ref{Table_cluster}. It suggests that the clusters, identified in the YSO population of L1340 are similar in size, shape, and stellar content to the SSYSC average. 
The distribution of the projected YSO separations, however, suggests that the mode of star formation in L1340\,B is quite atypical. The median nearest neighbor separations are significantly smaller in each of the SSYSC clusters than in L1340\,B. Another atypical feature of L1340 is that conspicuous YSO groups are being formed in the smaller clumps, whereas the largest clump, associated with the highest luminosity stars of the region, has a fragmented structure, associated with tiny groups of YSOs, scattered over the area of the clump. 

A few IM\,SFRs in our 1-kpc Galactic environment are also located at latitudes around 10\degr\ or higher, and are apparently not associated with giant molecular clouds. Well-known examples are NGC~7023 and NGC~7129, both located more than 100~pc above the Galactic plane, and forming small clusters with the brightest stars of B3 type. These regions may have star-forming histories similar to L1340. Expanding supershells could create conditions of star formation intermediate Galactic latitudes. Apparently none of these star-forming regions are associated with supershells, thus some other process, such as infall of high velocity clouds, or Kelvin--Helmholtz instabilities arising at the shearing surface between gas layers of different velocities might have compressed the gas.

\section{Conclusions}
\label{Sect_sum}

We identified some 250 candidate YSOs associated with the moderate-mass ($\sim 3700$\,\msun\, Paper~III) molecular cloud L1340, based on \textit{Spitzer\/}, \textit{AllWISE\/} mid-infrared, and Omega-Cass near-infrared data, using various published color criteria. Supplemented with our measurements on the \textit{Herschel PACS\/} 100\,\micron\ and 160\,\micron\ images and publicly available photometric data we constructed spectral energy distributions, and classified 8 candidate Class~0, 37 Class~I, 27 Flat~SED, and 170 Class~II sources. Based on the SEDs we derived extinctions and spectral types for the Class~II sources, and plotted them onto the Hertzsprung--Russell diagram. The HRD suggests a mass interval of 0.07--2.5\,\msun\ for the CTTS of our sample. 

We identified new Herbig--Haro objects, associated with the Class~0 protostar SSTSL2 022808.60+725904.5 in the \textit{Spitzer\/} images.
The \spitzer\ data reveal that the bright IRAS source 02249+7230 is a binary protostar associated with Herbig--Haro objects. The \spitzer\ data also suggest that the probable driving source of HH~488 is a Class~0 protostar, SSTSL2 J022820.81+723500.5. The projected length of HH~488 is some 1.3~pc. 

The Omega-Cass \textit{JHK\/} data resulted in 8 new candidate members of RNO~7, and revealed a close companion of its brighest member, IRAS~02236+7224. The \spitzer\ 8-\micron\ image revealed two further wide companions of this intermediate-mass T~Tauri star.

The surface density distribution of young stellar objects shows three groups, associated with the three major molecular clumps of L1340, each consisting of $\lesssim 100$ members, including both pre-main sequence stars and embedded protostars. Based on the distribution of nearest neighbor separations we identified four small clusters in the cloud, the RNO~7 cluster in L1340\,A, RNO~8 in L1340\,B, and RNO~9 and IRAS~02276+7225 in L1340\,C. Filamentary configurations of the protostars follow the distribution of the cold dust, traced by \textit{SCUBA\/} and \textit{Herschel\/} observations. The efficiency of the star formation in L1340 is some 3\,\%.
Our results demonstrate that L1340 is a prolific star-forming region of our Galactic environment in which several specific properties of the intermediate-mass mode of star formation can be studied in detail.
The distribution of dense gas and YSOs suggest that star-forming regions like L1340 are short-lived, transient objects.

\begin{acknowledgements}
This work is based on observations made with the \textit{Spitzer Space Telescope}, which is operated by the Jet Propulsion Laboratory, California Institute of Technology under a contract with NASA.  This research utilized observations collected at the Centro Astron\'omico Hispano Alem\'an (CAHA) at Calar Alto, operated jointly by the Max-Planck Institut f\"ur Astronomie and the Instituto de Astrof\'{\i}sica de Andaluc\'{\i}a (CSIC). This research has made use of the NASA/ IPAC Infrared Science Archive, which is operated by the Jet Propulsion Laboratory, California Institute of Technology, under contract with the National Aeronautics and Space Administration. Our research has benefited from the VizieR catalogue access tool, CDS, Strasbourg, France. Financial support from the Hungarian OTKA grant K81966 and K101393 is acknowledged. This work was partly supported by the Momentum grant of the MTA CSFK Lend\"ulet Disk Research Group. 
\end{acknowledgements}

\appendix

\section{{\it UBVR}$_\mathrm{C}${\it I}$_\mathrm{C}${\it JHK}$_\mathrm{s}$ PHOTOMETRIC DATA OF THE CLASS II YOUNG STELLAR OBJECTS}
 We list {\it UBVR}$_\mathrm{C}${\it I}$_\mathrm{C}$ magnitudes, transformed from the \sdss\ data, and \textit{2MASS\/} \textit{JHK$_s$} magnitudes of the color-selected candidate Class~II young stars associated with L1340 in Table~A1, excluded the H$\alpha$ emission stars, whose data are given in Paper~III.

\clearpage

\newpage

\clearpage

\clearpage

\clearpage

\begin{figure*}
\centering{\includegraphics[width=16cm]{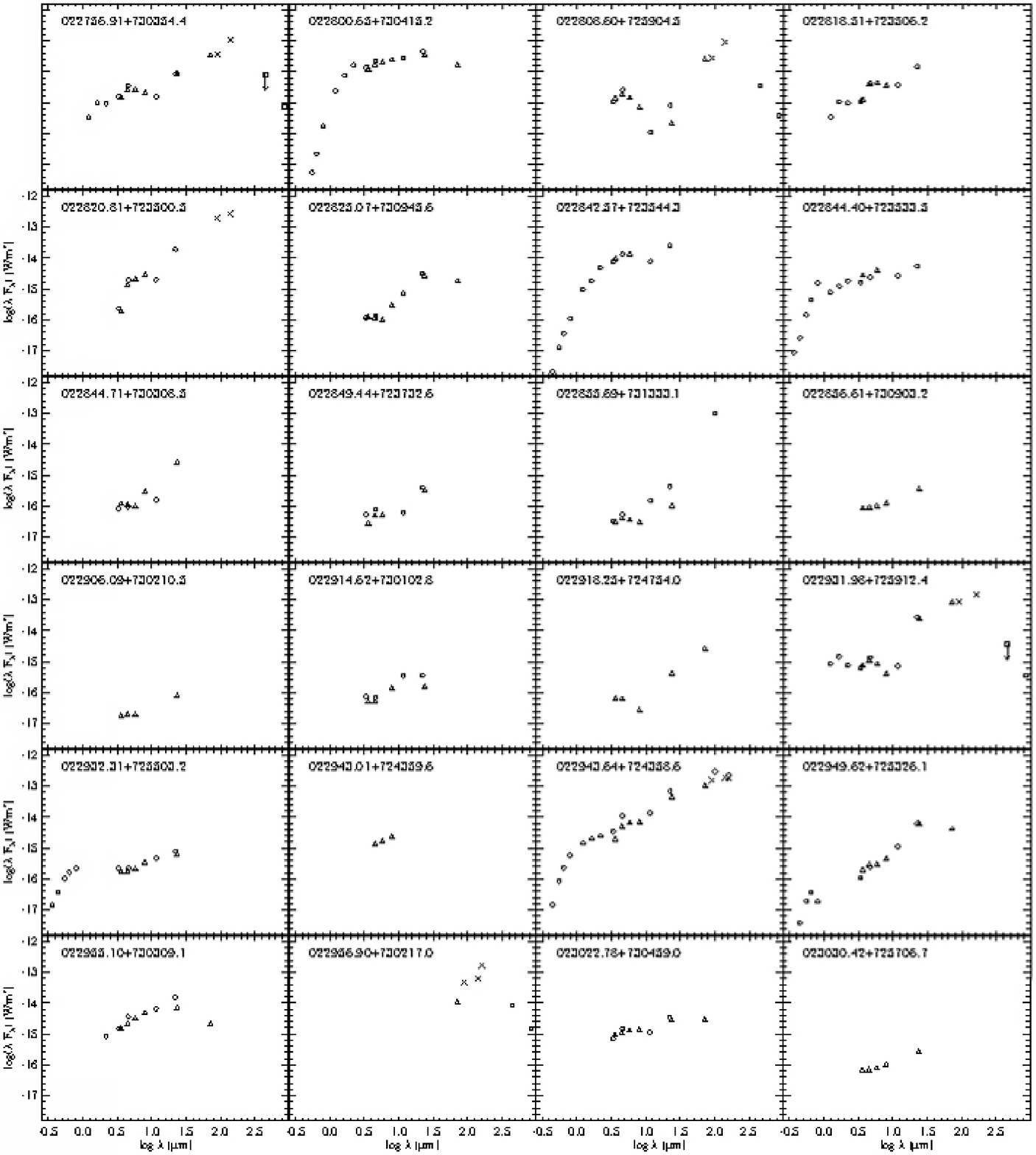}}
\caption{Spectral energy distributions of Class~I YSOs. Open circles show the \textit{SDSS}, \textit{2MASS}, and \textit{WISE} data, and triangles indicate the \spitzer\ data. Crosses are for \textit{Akari FIS\/} data, diamonds show the \textit{Herschel\/} data points, and squares indicate \textit{SCUBA\/} submillimeter fluxes.}
\label{fig9}
\end{figure*}

\addtocounter{figure}{-1}
\begin{figure*}
\centering{\includegraphics[width=16cm]{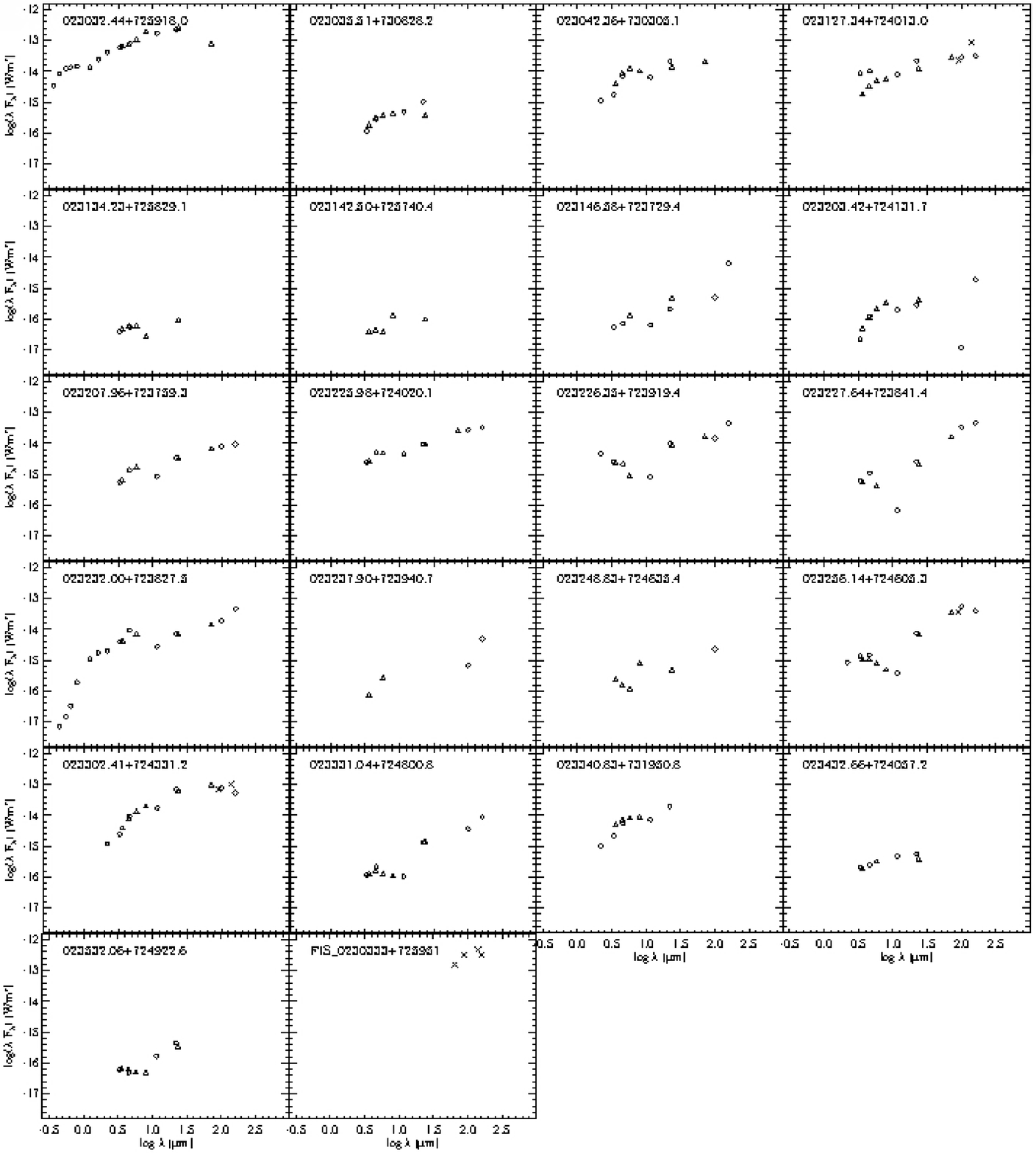}}
\caption{SEDs of the Class~I YSOs (continued).}
\label{fig9}
\end{figure*}

\begin{figure*}
\centering{\includegraphics[width=16cm]{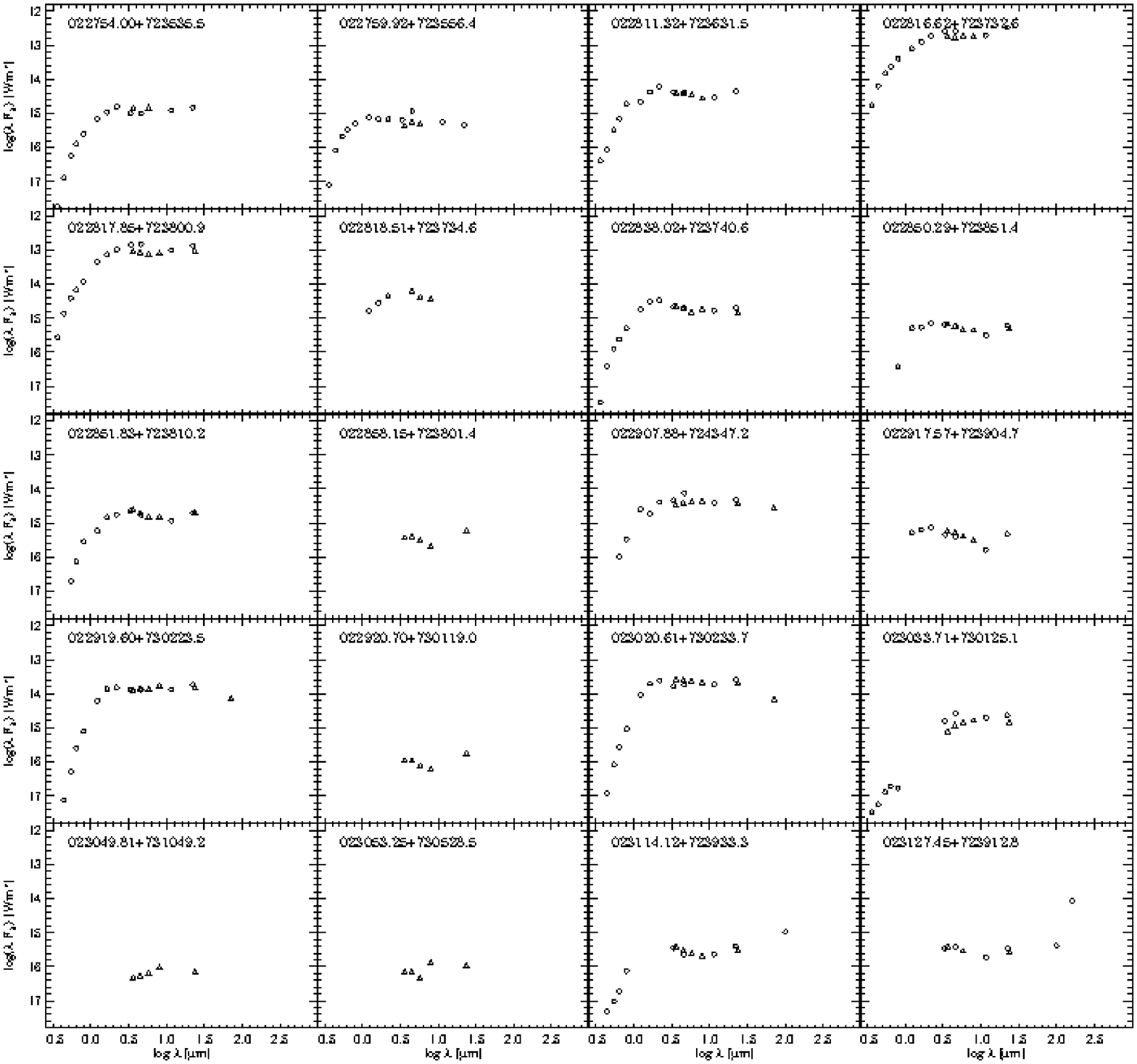}}
\caption{SEDs of the  Flat SED infrared sources. Symbols are same as in Fig.~\ref{fig9}.}
\label{fig10}
\end{figure*}

\addtocounter{figure}{-1}
\begin{figure*}
\centering{\includegraphics[width=16cm]{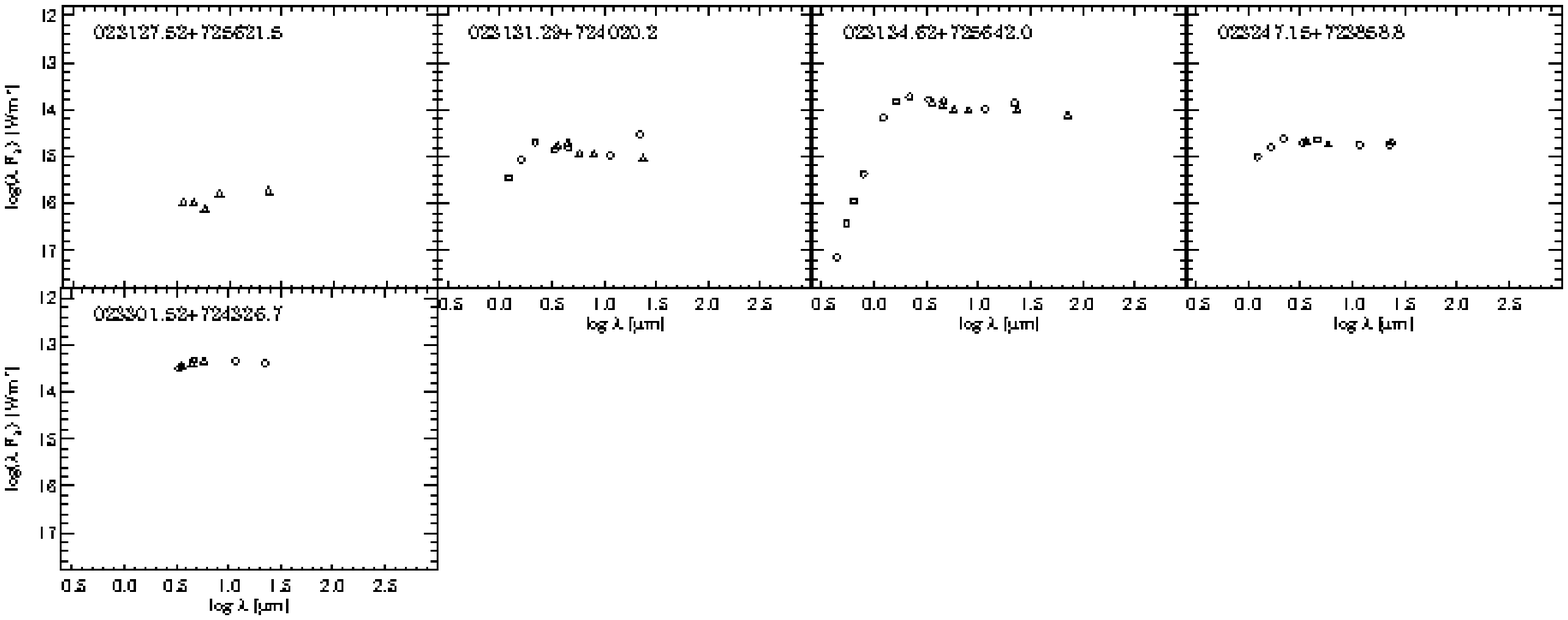}}
\caption{SEDs of the  Flat SED infrared sources (continued).}
\label{fig10}
\end{figure*}

\begin{figure*}
\centering{\includegraphics[width=16cm]{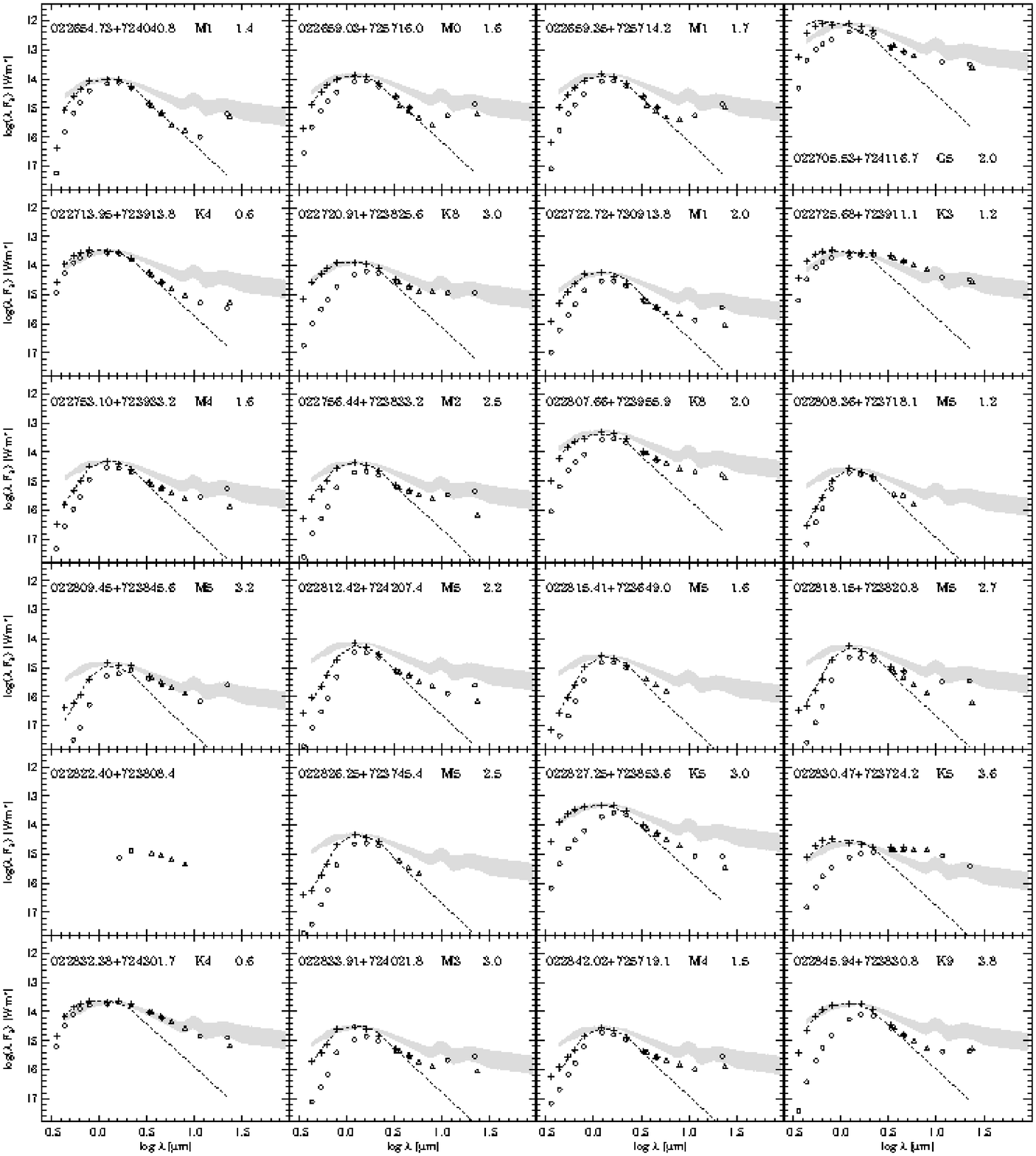}}
\caption{SEDs of the Class~II candidate YSOs identified in the SEIP Source List data, and not identified as \ha\ emission stars. The SEDs of \ha\ emission stars can be seen in fig.~9 of Paper~III. Open circles show the \textit{SDSS}, \textit{2MASS}, and \textit{WISE} data, and triangles are for \spitzer\ data. Plusses indicate the dereddened SED, and the dashed line shows the photospheric SED of the spectral type, obtained by fitting a model to the data. The gray shaded area indicates the median SED of the T~Tauri stars of the Taurus star-forming region \citep{DAlessio}. SSTSL2 identifiers, spectral types, and $A_\mathrm{V}$ extinctions, derived from photometric data are indicated at the top of each panel.}
\label{fig11}
\end{figure*}

\addtocounter{figure}{-1}
\begin{figure*}
\centering{\includegraphics[width=16cm]{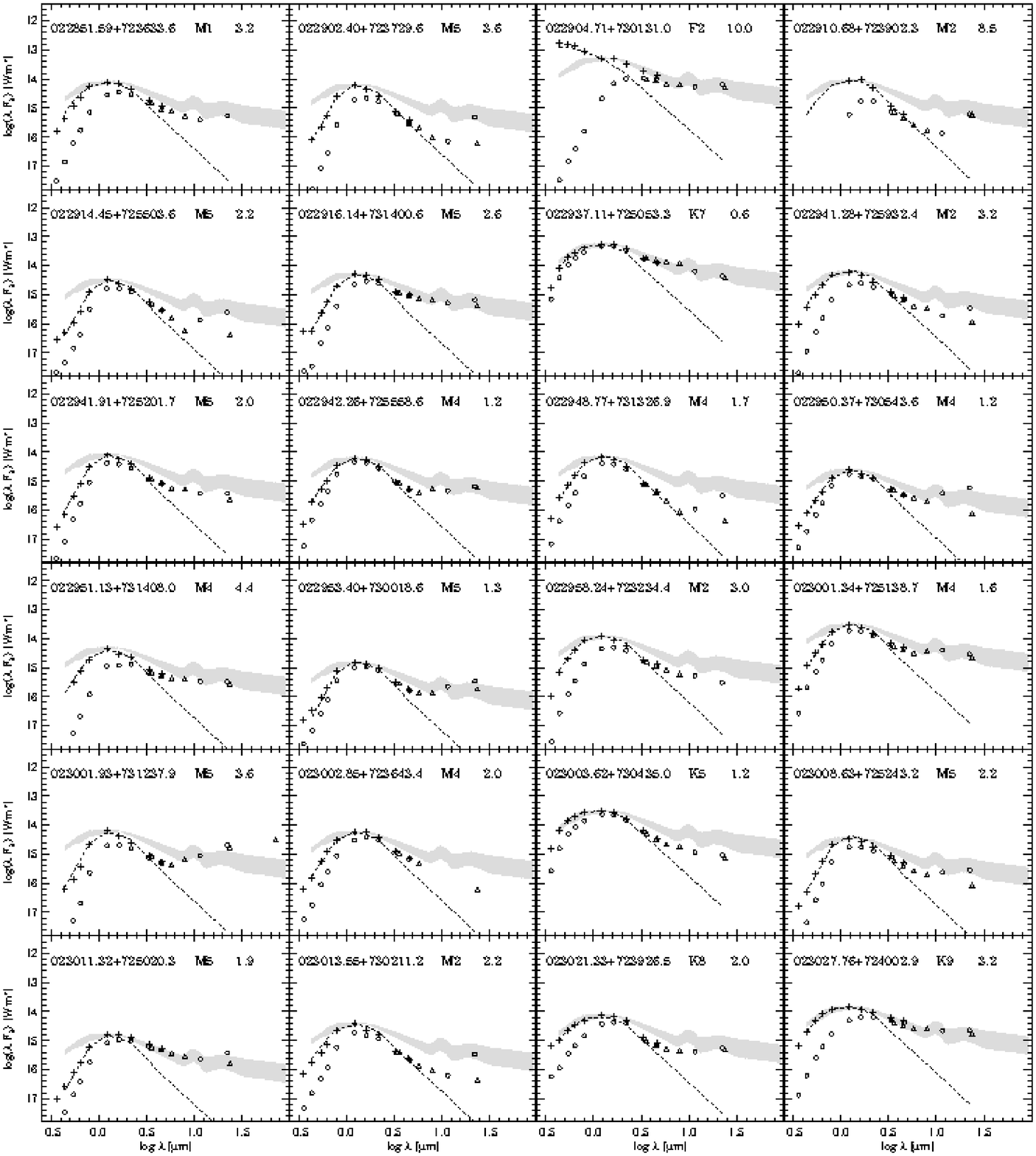}}
\caption{SEDs of the Class~II candidate YSOs (continued).}
\label{fig11}
\end{figure*}

\addtocounter{figure}{-1}
\begin{figure*}
\centering{\includegraphics[width=16cm]{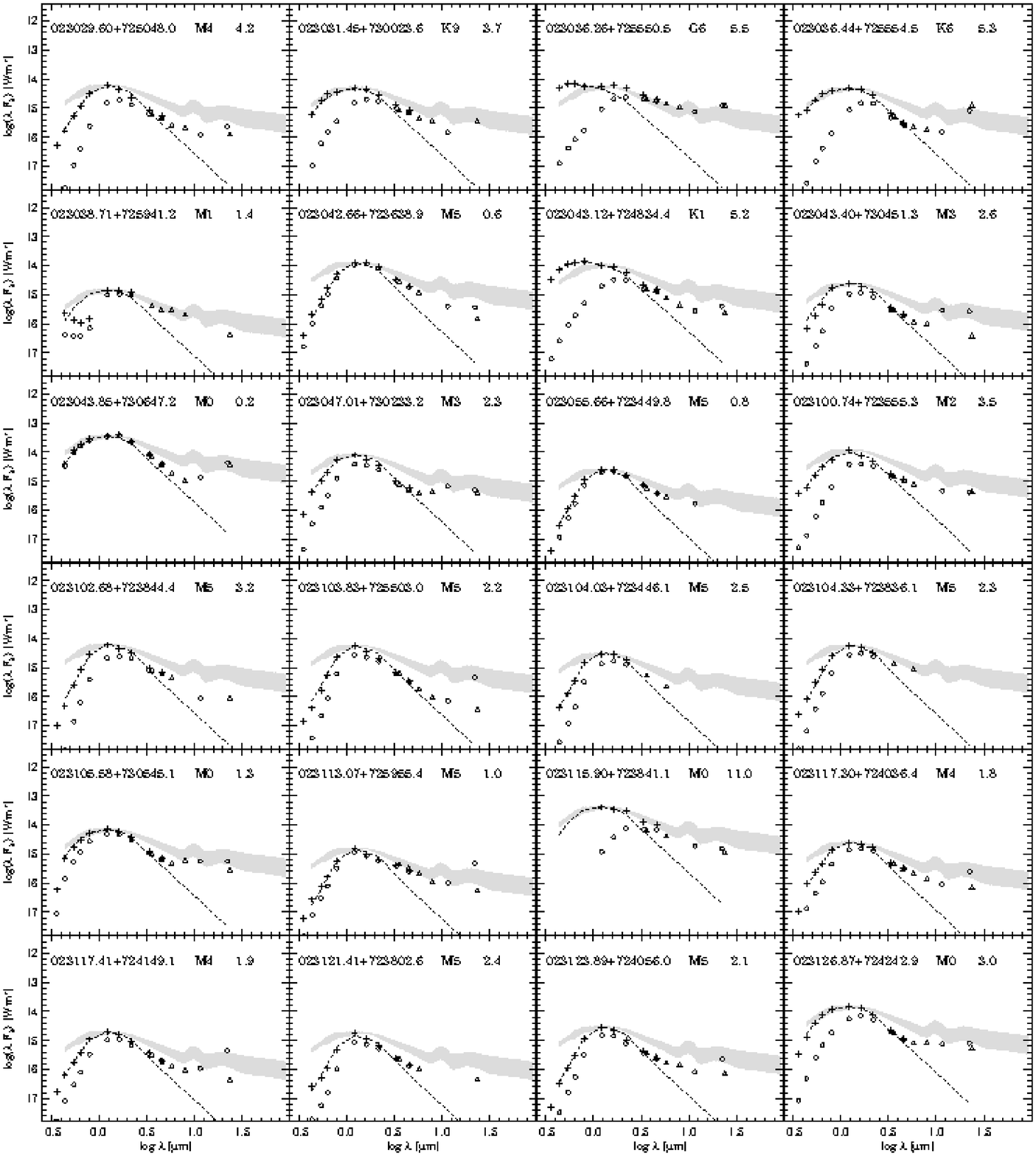}}
\caption{SEDs of the Class~II candidate YSOs (continued).}
\label{fig11}
\end{figure*}

\addtocounter{figure}{-1}
\begin{figure*}
\centering{\includegraphics[width=16cm]{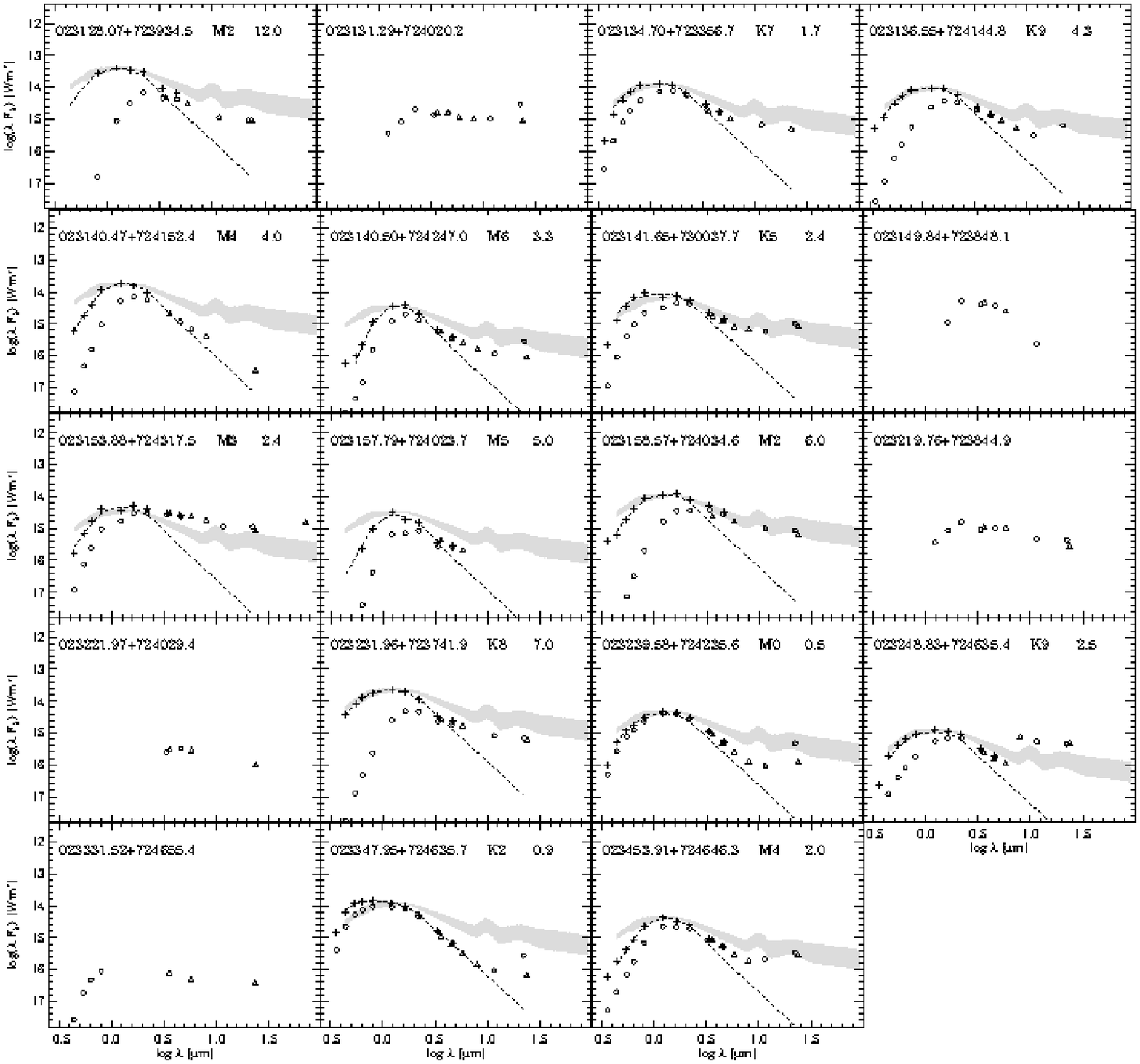}}
\caption{SEDs of the Class~II candidate YSOs (continued).}
\label{fig11}
\end{figure*}

\begin{figure}
\centering{\includegraphics[width=12cm]{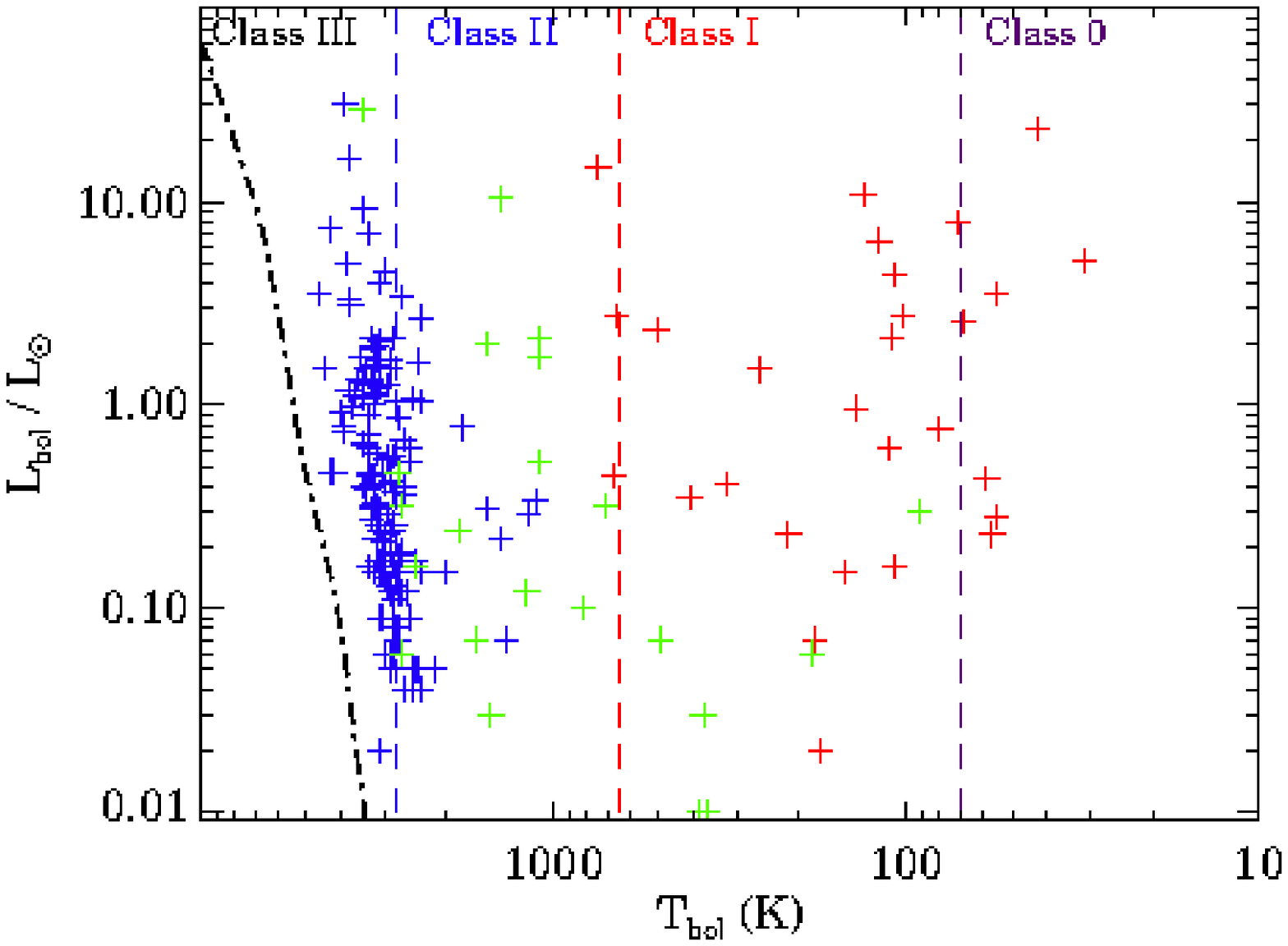}}
\caption{Extinction-corrected bolometric luminosities plotted against extinction-corrected bolometric temperatures. Red crosses indicate Claas~0/I sources, green symbols are for Flat, and blue for Class~II SED slopes. Vertical dashed lines indicate the boundaries between the Classes, and the dash-dotted line indicates the position of the zero-age main sequence \citep{Siess}. }
\label{fig12}
\end{figure}

\begin{figure}
\centering{\includegraphics[width=8cm]{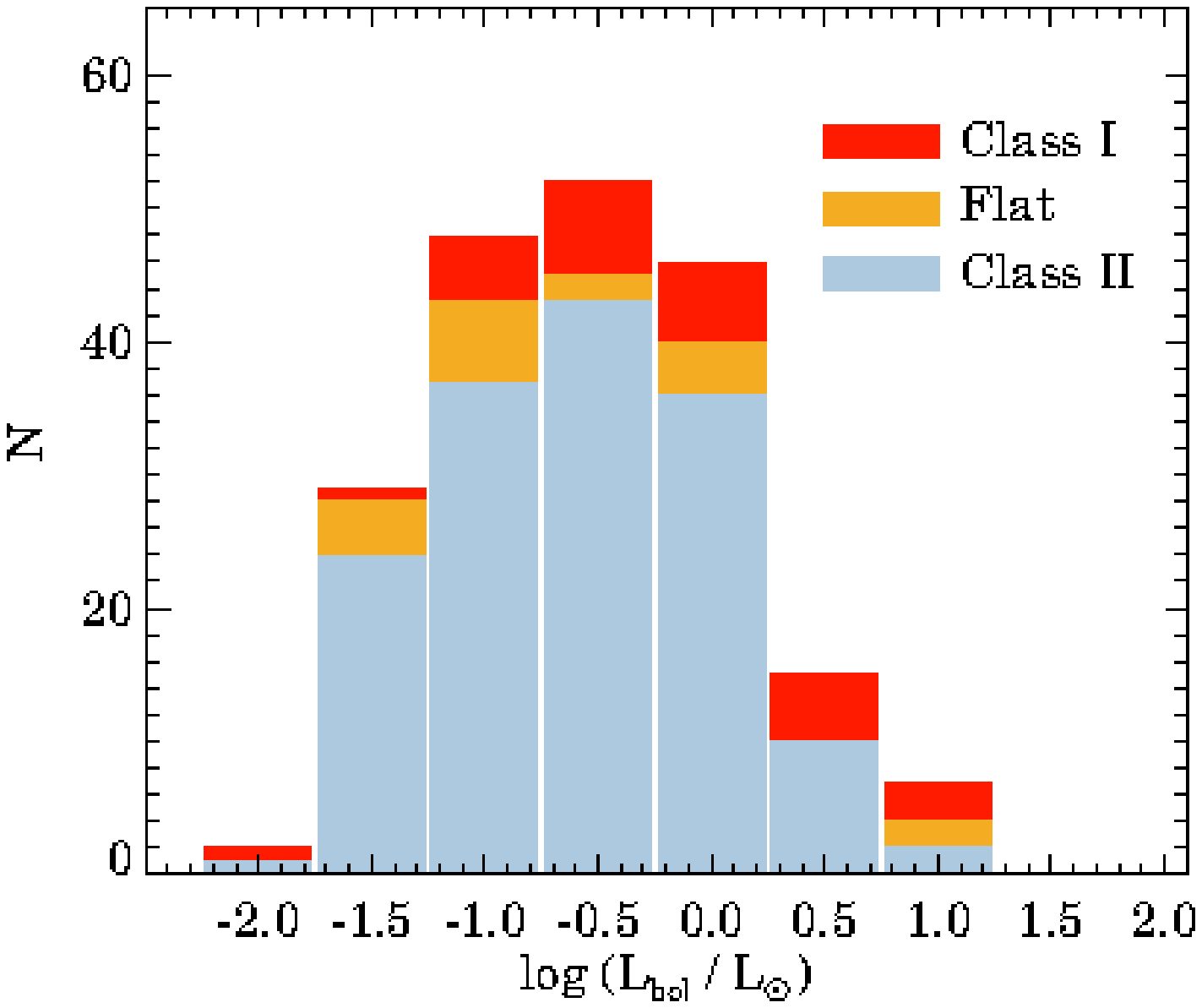}}
\caption{Histogram of bolometric luminosities of Class~0/I/Flat protostars and Class~II pre-main sequence stars.}
\label{fig13}
\end{figure}

\clearpage

\begin{figure*}
\centerline{\includegraphics[width=5cm]{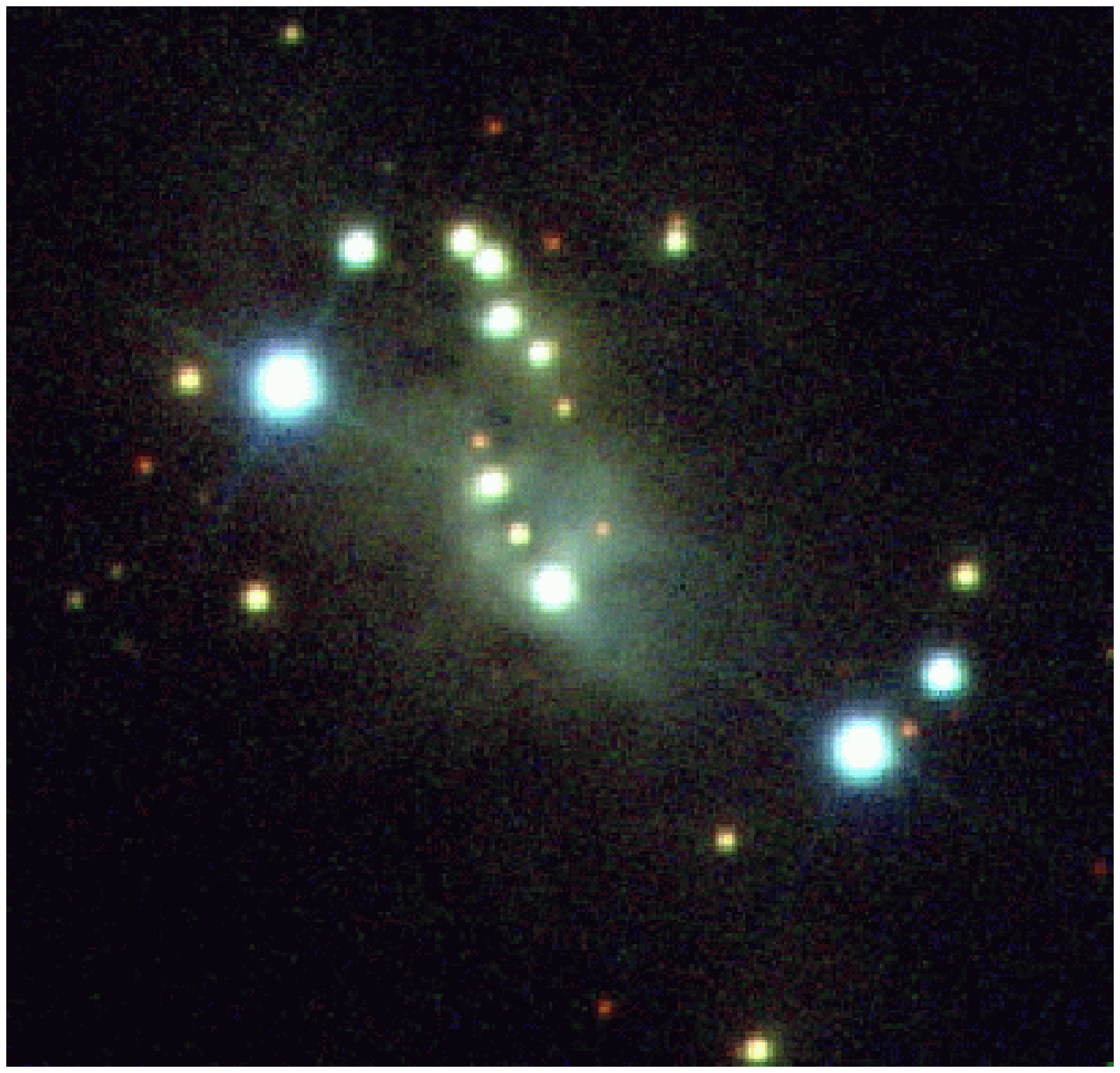}\hskip2mm\includegraphics[width=5cm]{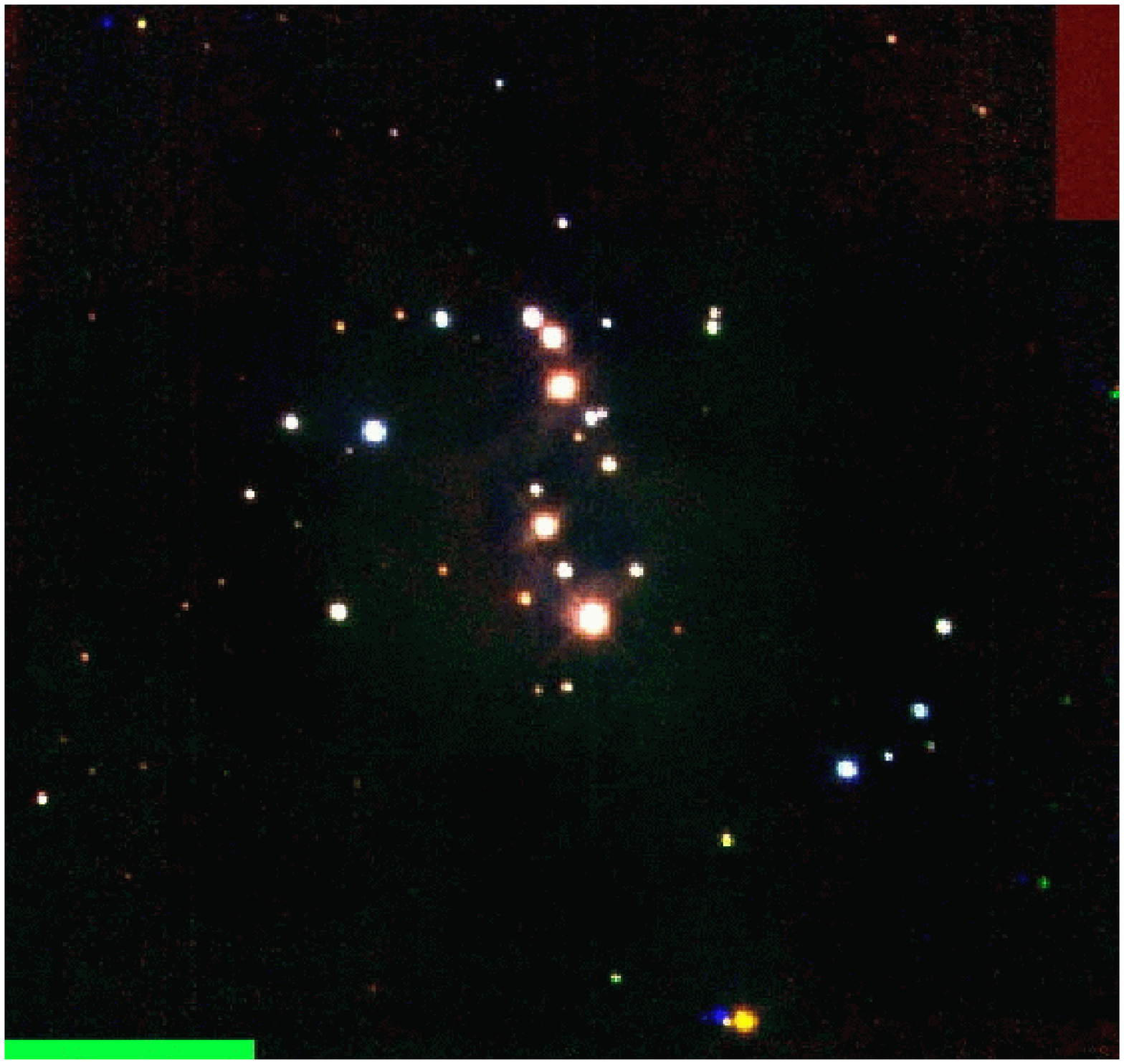}}
\centerline{\includegraphics[width=5cm]{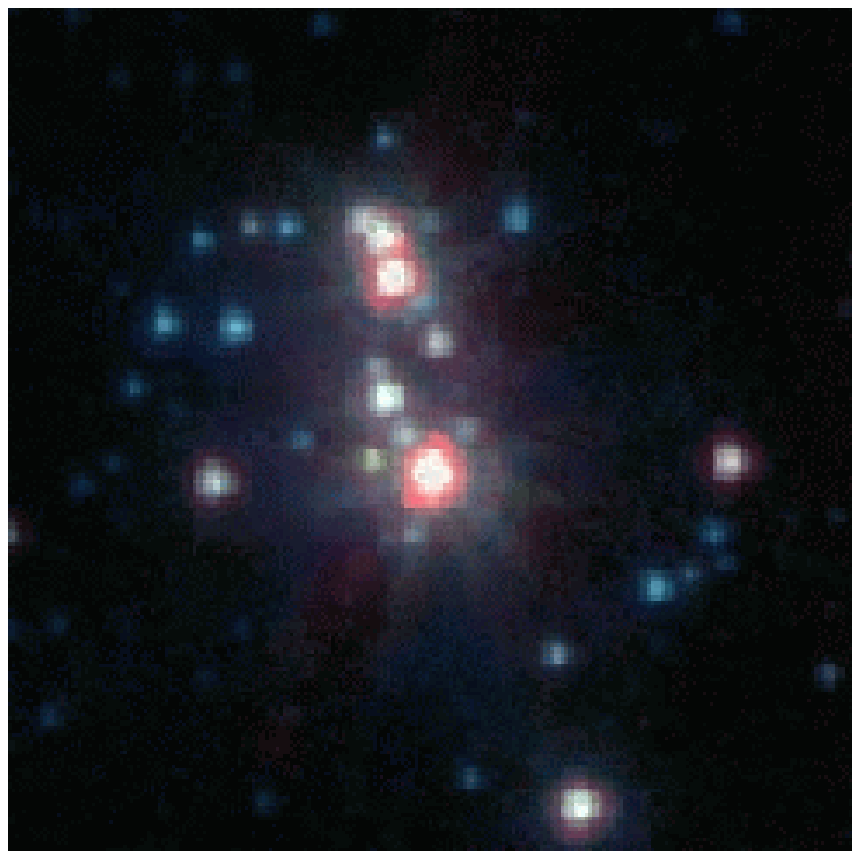}\hskip2mm\includegraphics[width=5cm]{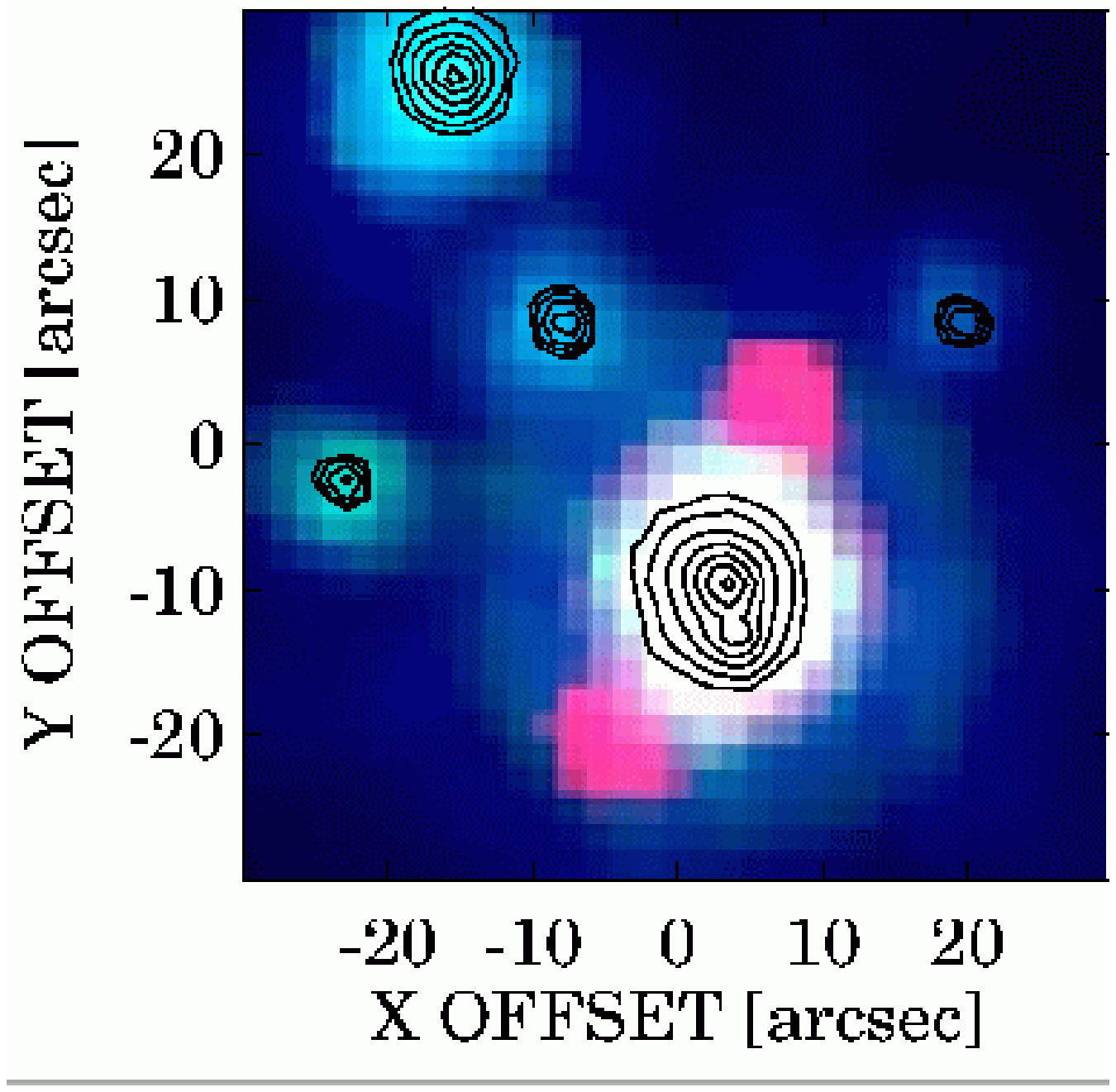}}
\caption{Optical, near-infrared, and mid-infrared three-color composite images of RNO~7. The upper left panel is composed of the \sdss\ {\it g\/} (blue) {\it r\/} (green), and {\it i\/} (red) images, the upper right panel resulted from our high angular resolution {\it J\/} (blue), {\it H\/} (green), and {\it K\/} (red) observations, and the lower left panel presents a composite of the \spitzer\ IRAC 3.6\,\micron\ (blue), 4.5\,\micron\ (green), and 8.0\,\micron\ (red) images. The lower right panel magnifies a $30\arcsec\times30\arcsec$ area of the previous image around the brightest member of the cluster. The IRAC composite image is scaled to show the two faint, 8-\micron\ companions, and Omega-Cass $K$-band contours, revealing a close near-infrared companion, are overplotted.}
\label{fig14}
\end{figure*}

 \begin{figure}
 \centering{
 \includegraphics[width=8cm]{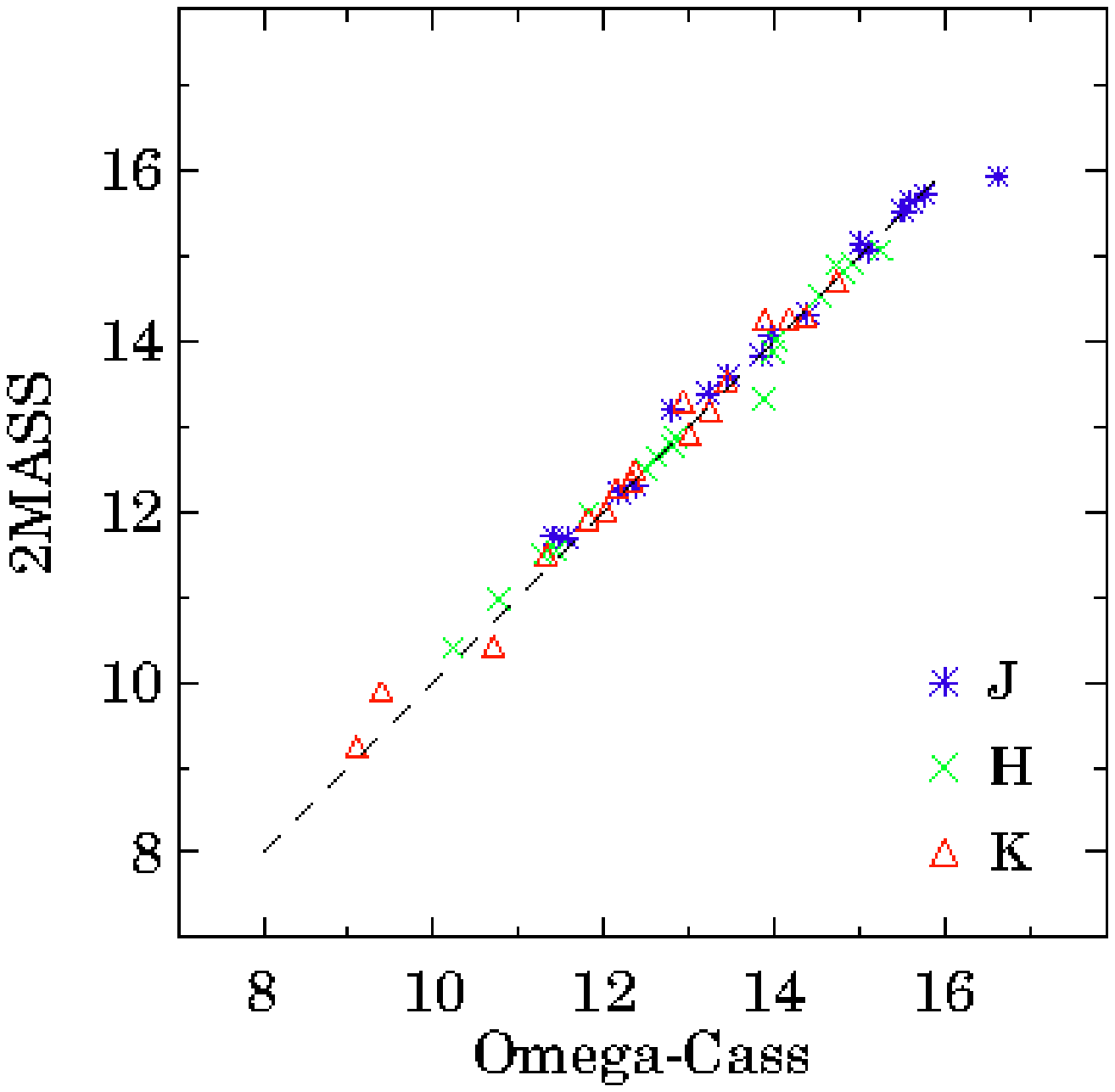}\includegraphics[width=8cm]{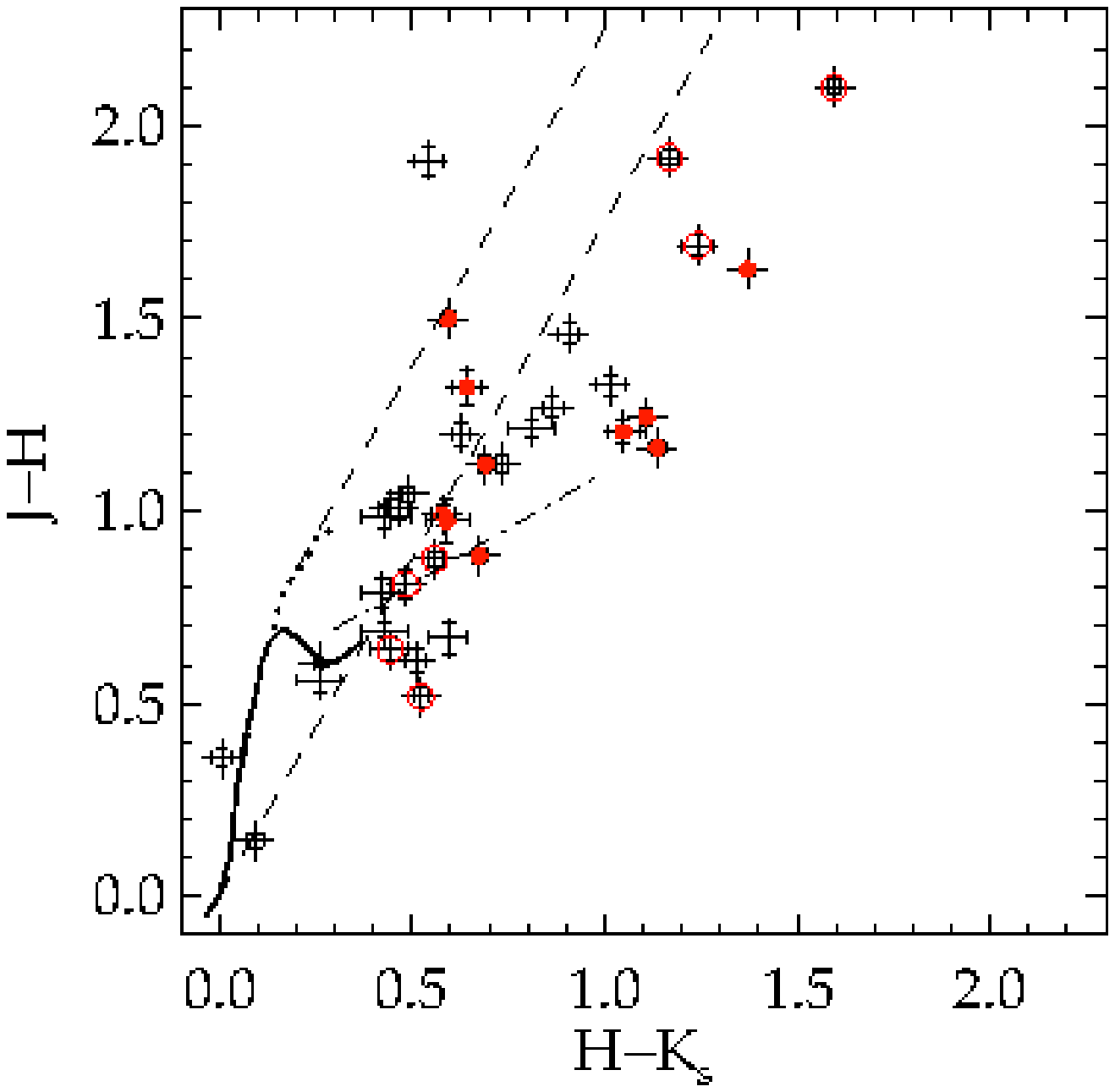}}
 \caption{{\it Left\/}: \tm\ magnitudes of 17 stars within the field of view of the Omega-Cass, plotted against those measured in the Omega-Cass images. {\it Right\/}: $J-H$ vs. $H-K_\mathrm{s}$ two-color diagram of the stars in the region of RNO~7, detected in each band in the Omega-Cass images, and listed in Table~\ref{Table_oc}. Solid line indicates the colors of the main sequence stars, and dotted line shows those of the giants \citep{Bessell}. Dashed lines border the band of the reddened main sequence and giant stars, and the dash-dotted line is the locus of T~Tauri stars \citep{Meyer97}. Red dots indicate the known \ha\ emission stars, and open circles show the stars selected as Class~II infrared sources in the \spitzer\ data.} 
\label{fig15}
 \end{figure}

\begin{figure*}
\centering{\includegraphics[width=16cm]{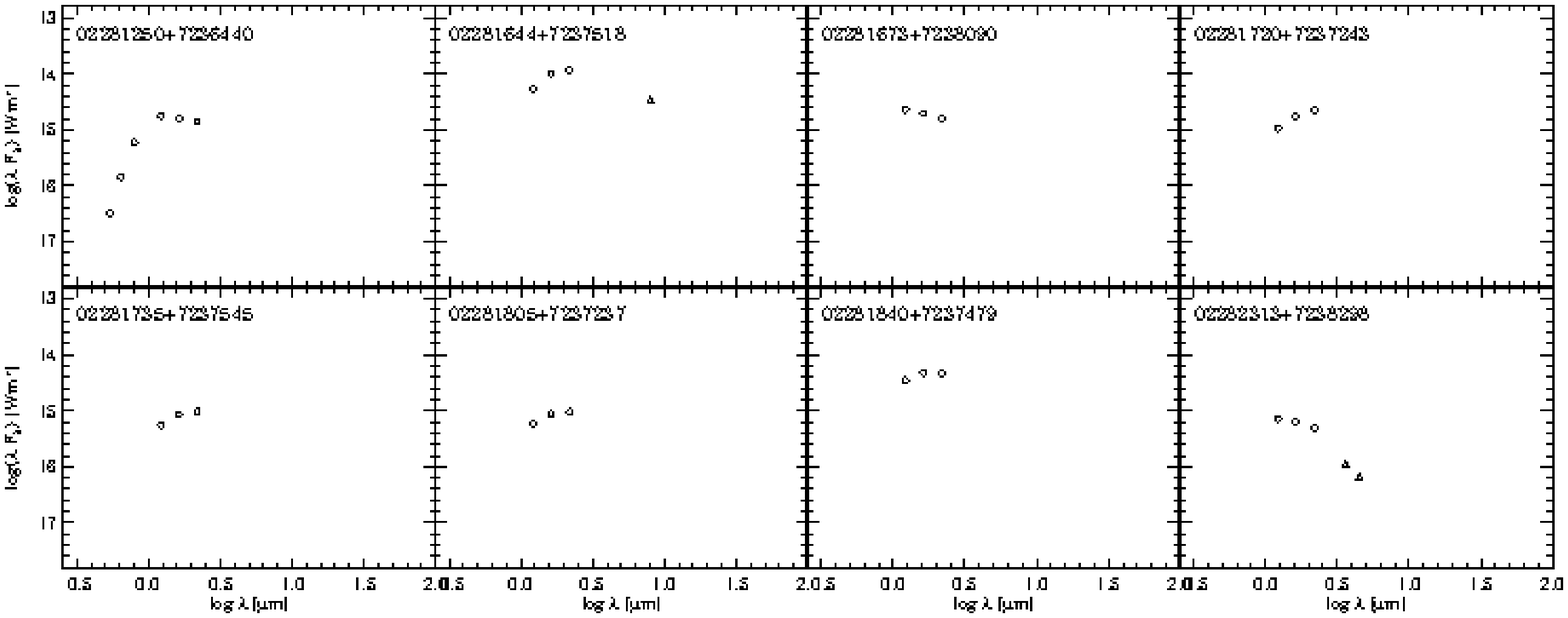}}
\caption{SEDs of the candidate young stars identified in the Omega-Cass \textit{JHK\/} data.}
\label{fig16}
\end{figure*}

\begin{figure*}
\centering{\includegraphics[width=16cm]{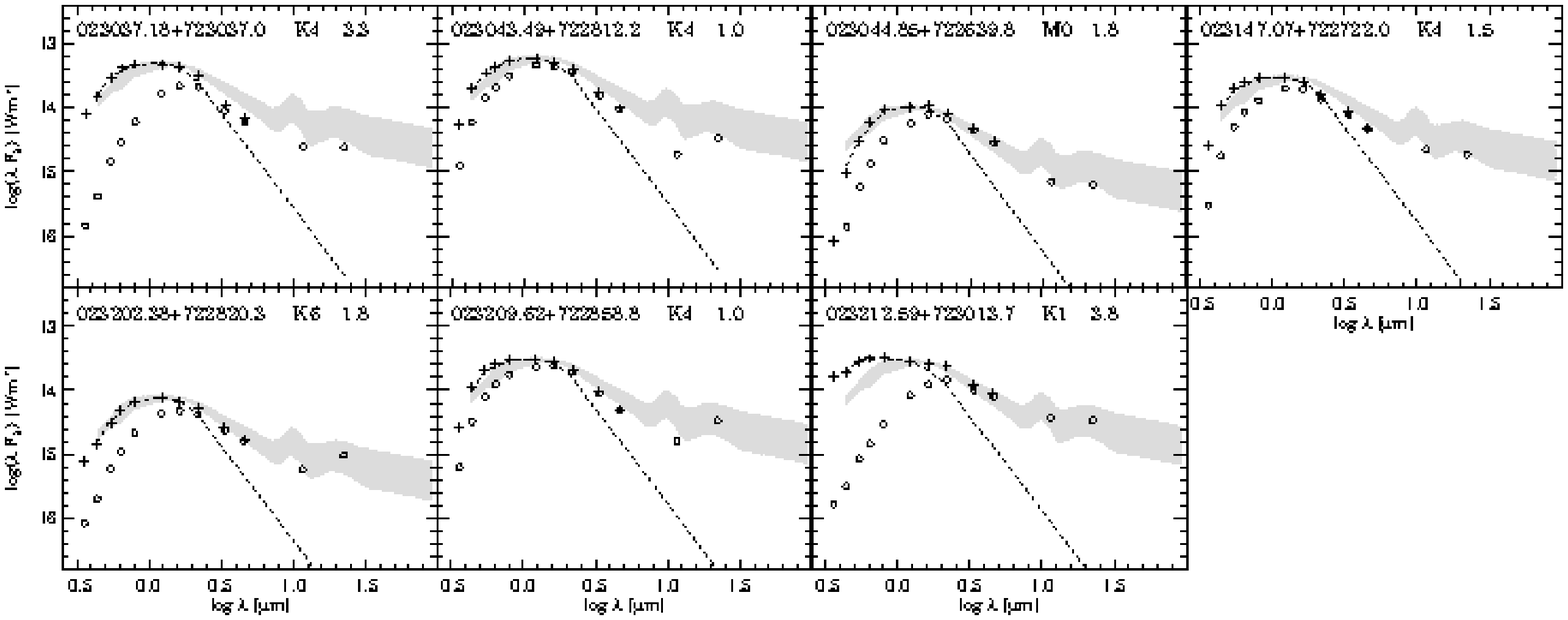}}
\caption{SEDs of the candidate young stars identified in the  \textit{AllWISE\/} data, located outside of the field of view of the \spitzer\ observations.}
\label{fig17}
\end{figure*}

\clearpage

\begin{figure*}
\centering{\includegraphics[width=8cm]{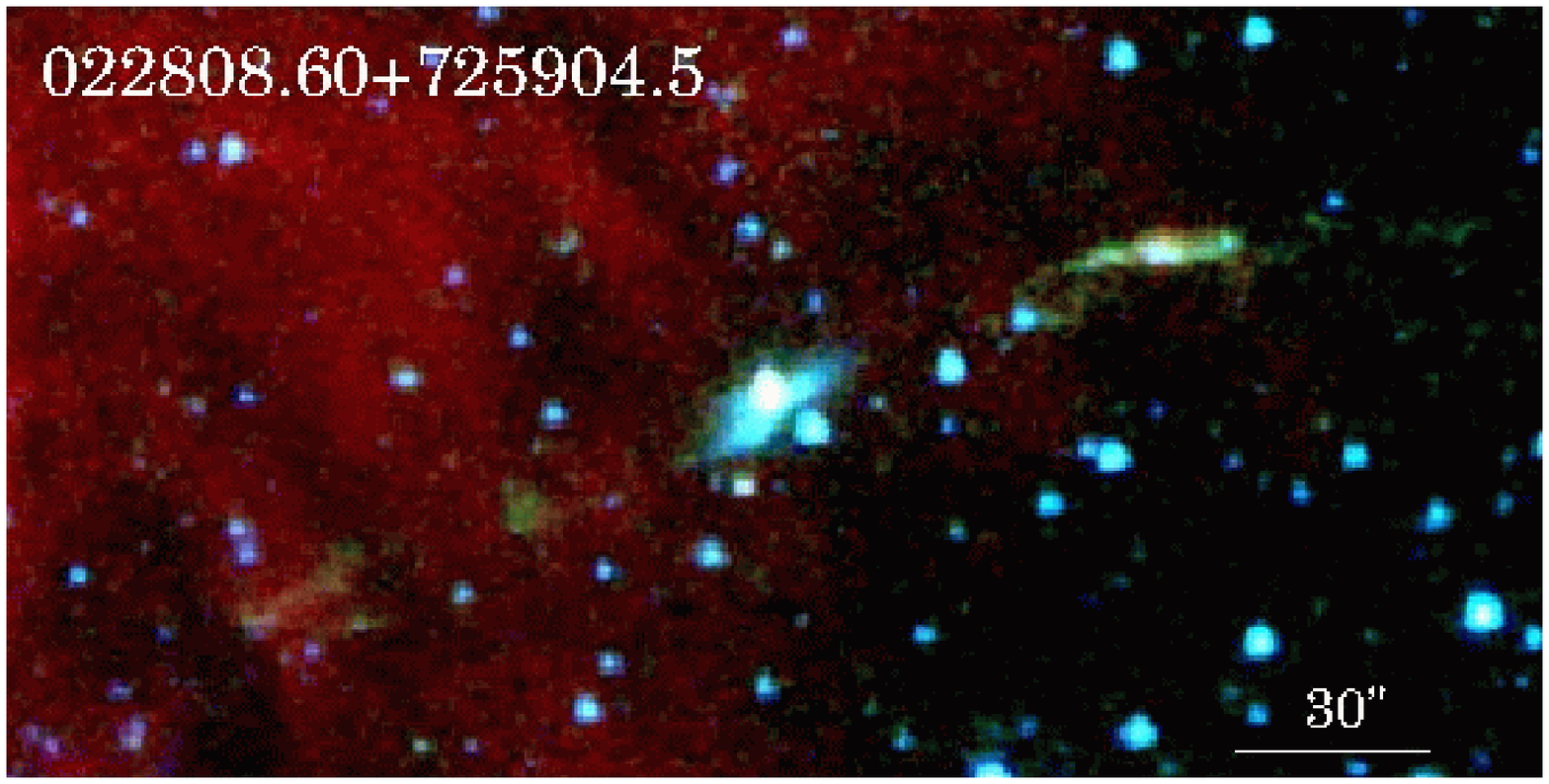}}
\caption{RGB image composed from IRAC 8\,\micron\ (red),  4.5\,\micron\ (green), and 3.6\,\micron\ (blue) images of the candidate Class~0 protostar SSTSL2~J022808.60+725904.5. The 4.5\,\micron\ knots at both side of the nebulous source indicate HH objects driven by the Class~0 protostar, and corresponding to the protostellar outflow reported in Walawender et al. 2016.}
\label{fig18}
\end{figure*}

\begin{figure*}
\centering{\includegraphics[width=8cm]{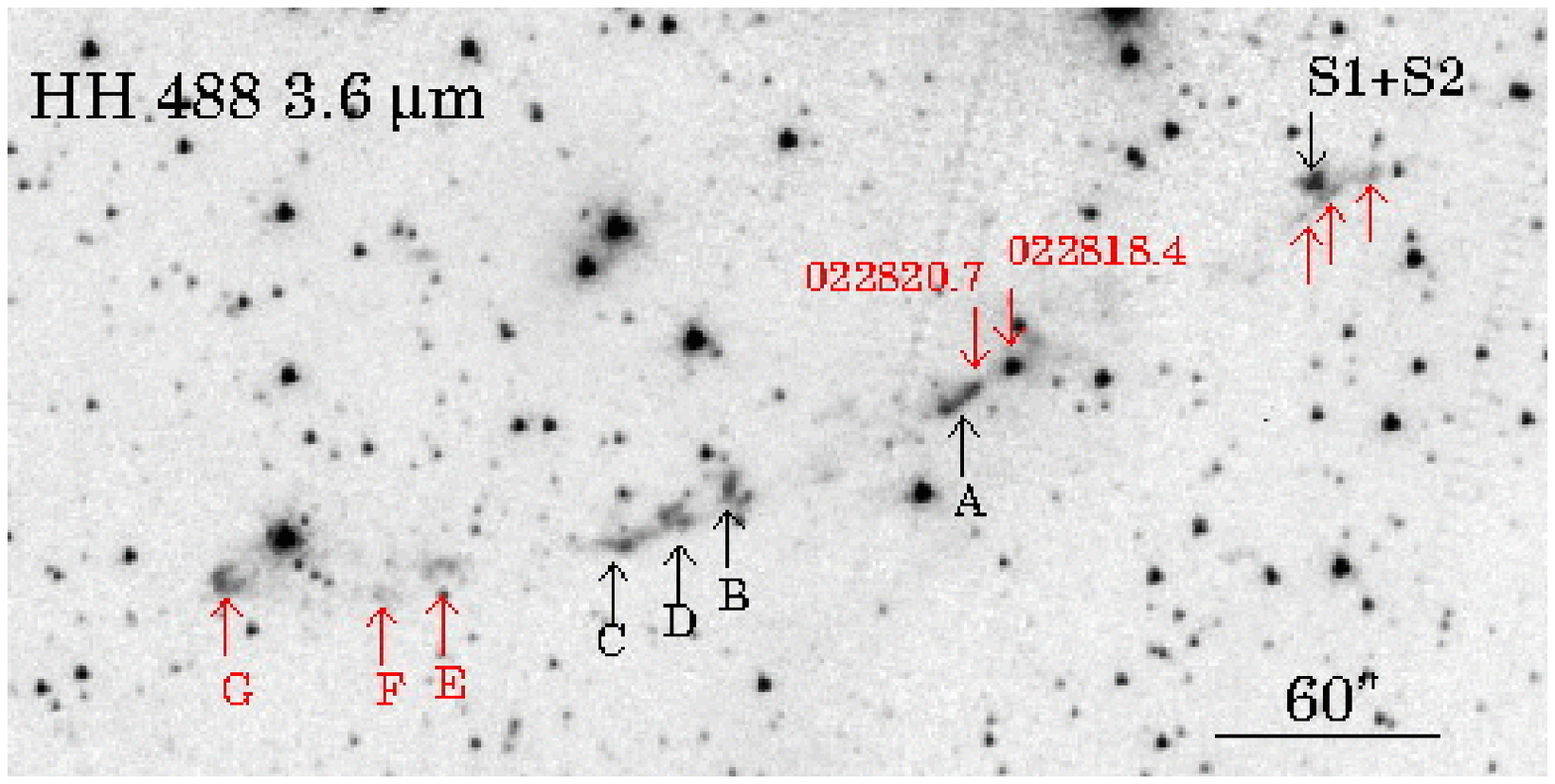}}\vskip2mm
\centering{\includegraphics[width=8cm]{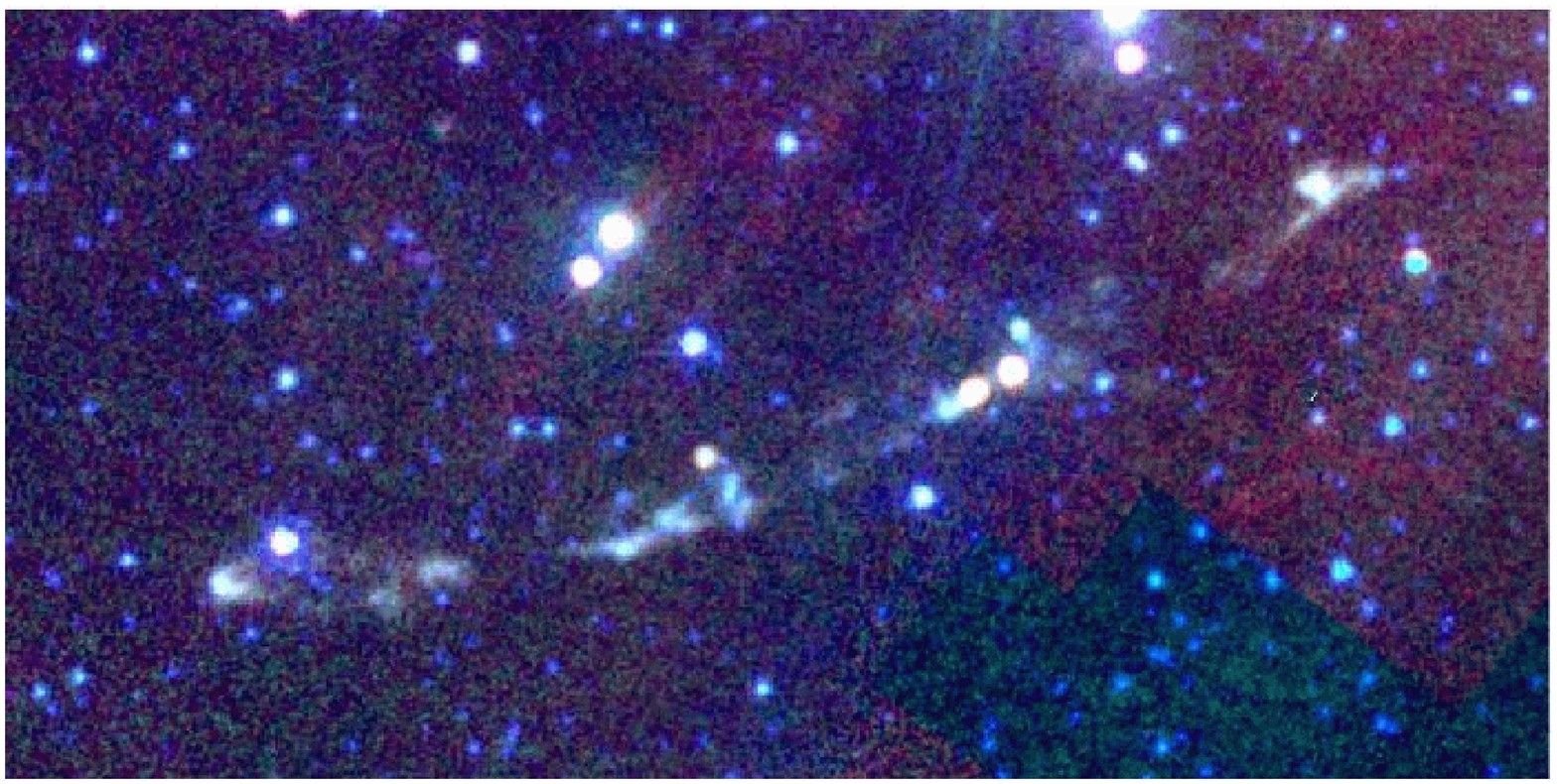}}
\caption{Top: HH 488 in the 3.6\,\micron\  IRAC band. The HH knots A, B, C, D, and the binary star S1+S2, driving source candidate identified by \citet{KAY03} are indicated by the black characters, and red characters mark the new protostars SSTSL2~022818.51+723506.2 and SSTSL2~022820.81+723500.5, and HH knots E, F, G, revealed by the \spitzer\ images. Bottom: Three-color image of the same region, composed from IRAC 8\,\micron\ (red),  5.8\,\micron\ (green), and 3.6\,\micron\ (blue) images.}
\label{fig19}
\end{figure*}

\begin{figure*}
\centering{\includegraphics{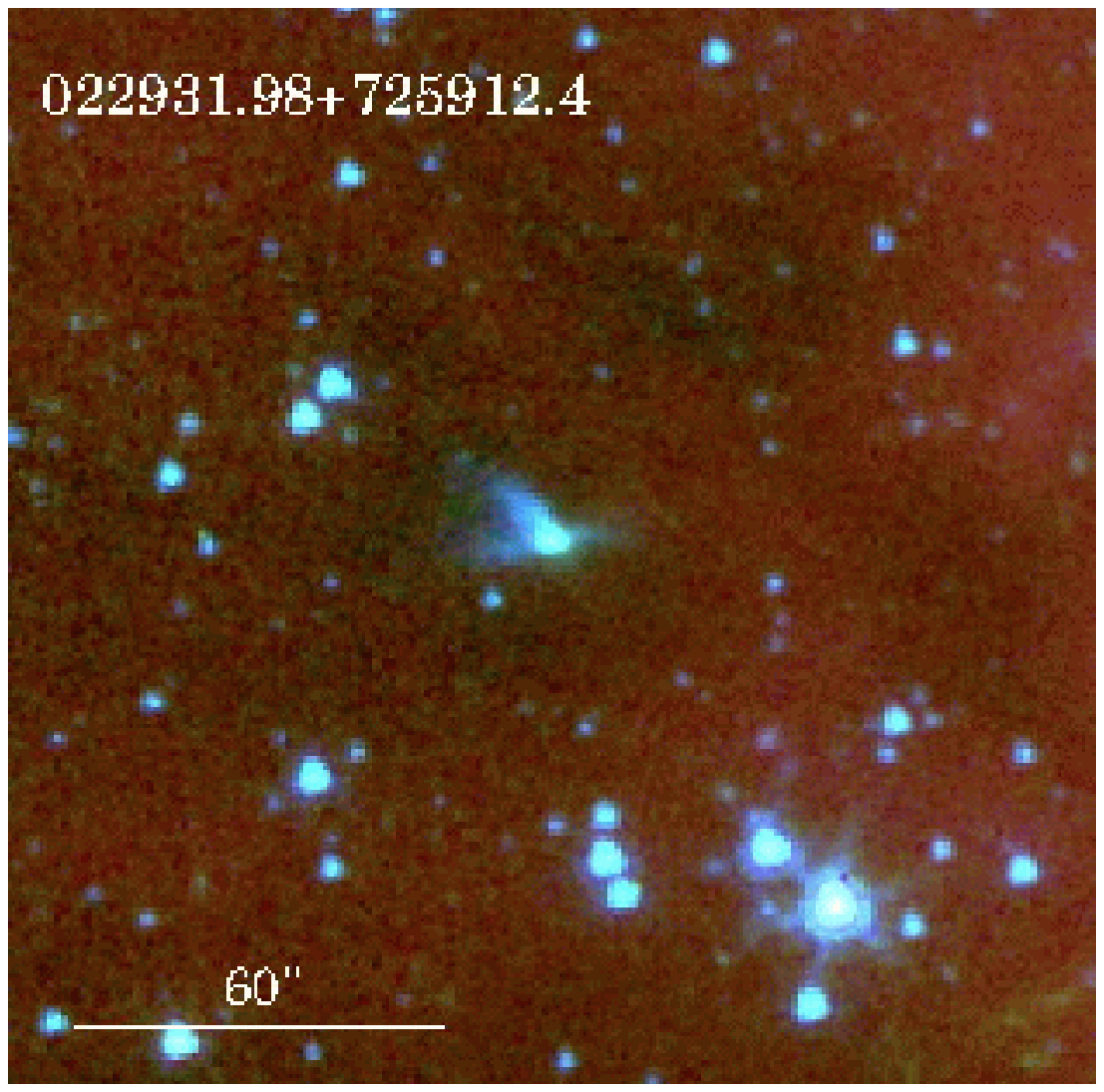}}
\caption{Three-color image, composed of the IRAC 3.6\,\micron\ (blue), 4.5\,\micron\ (green), and 8\,\micron\ (red) images of the environment of the candidate Class~0 protostar SSTSL2~J022931.98+725912.4. Notice the color difference between the eastern and western nebulosities.} 
\label{fig20}
\end{figure*}

\begin{figure*}
\centering{\includegraphics{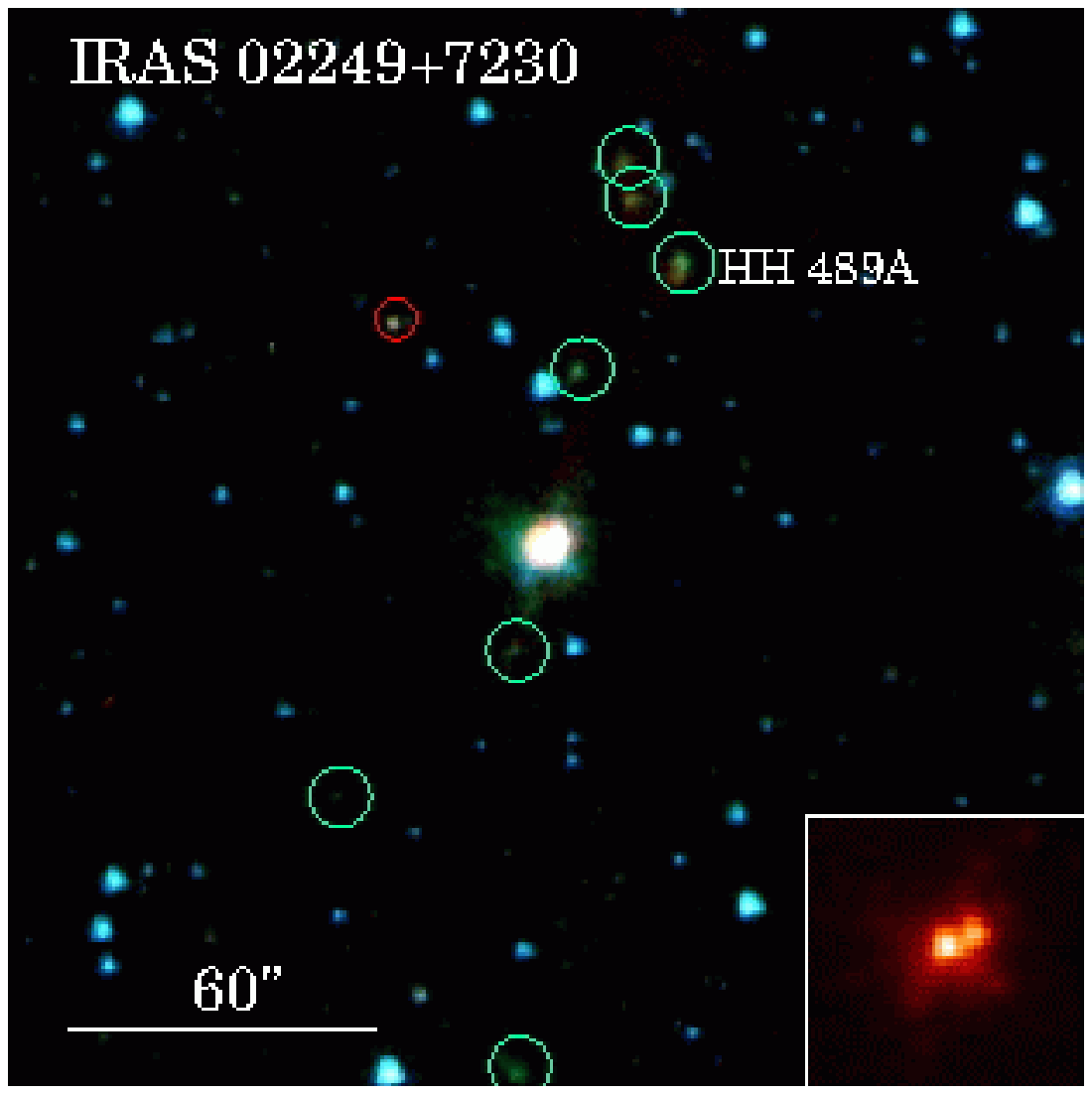}}
\caption{Three-color image, composed from the IRAC 3.6\,\micron\ (blue), 4.5\,\micron\ (green), and 8\,\micron\ (red) images of the environment of the binary protostar \textit{IRAS\/}~02249+7230, the driving source of HH~489. HH\,489\,A, and further HH objects revealed by the 4.5-\micron\ image are marked by the green circles. The faint red object within the red square is another Class~I object SSTSL2~J022950.37+724441.4. The inset in the lower right corner shows the 30\arcsec\ environment of the central object, magnified and scaled for better visibility. }
\label{fig21}
\end{figure*}

\begin{figure*}
\centering{\includegraphics[width=8cm]{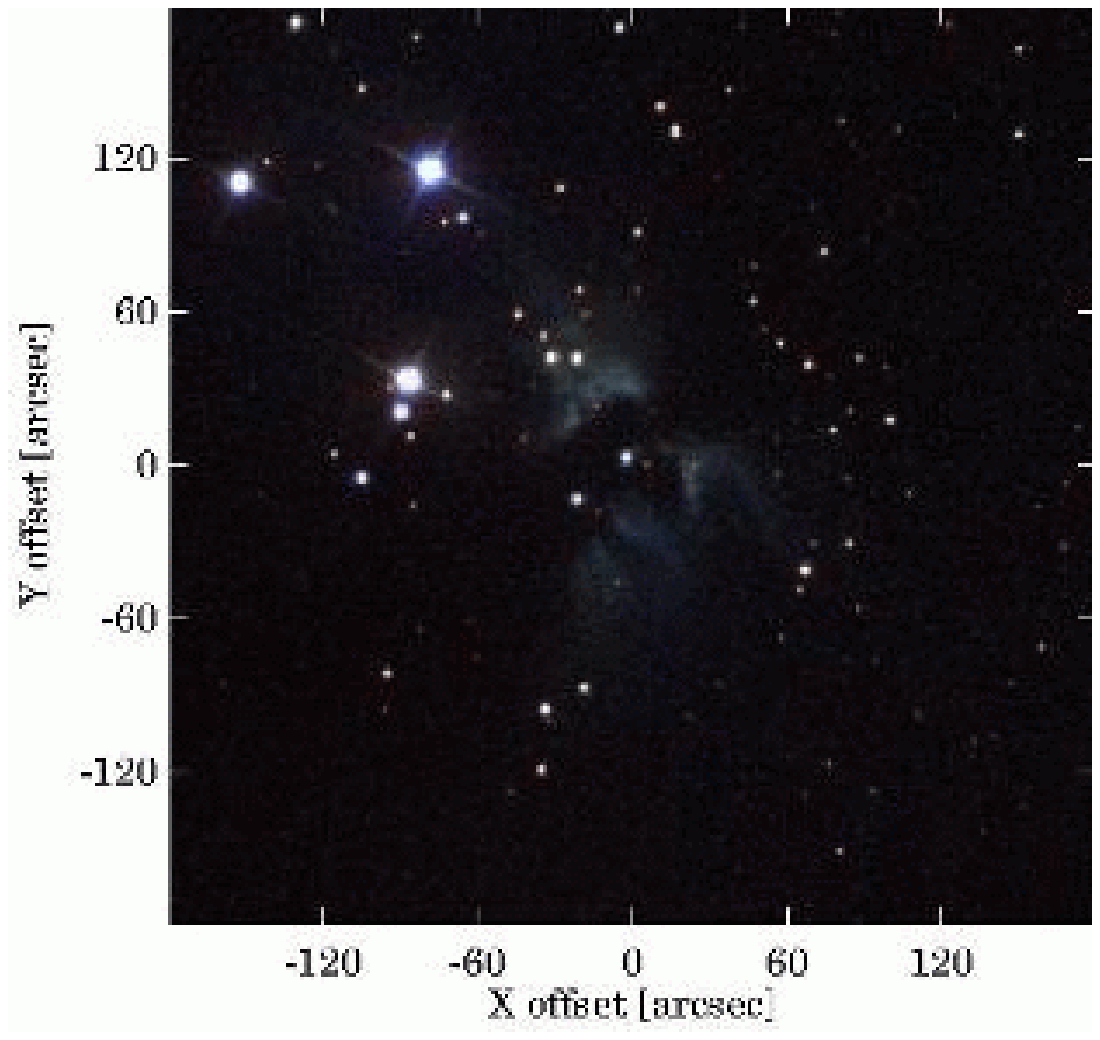}}
\centering{\includegraphics[width=8cm]{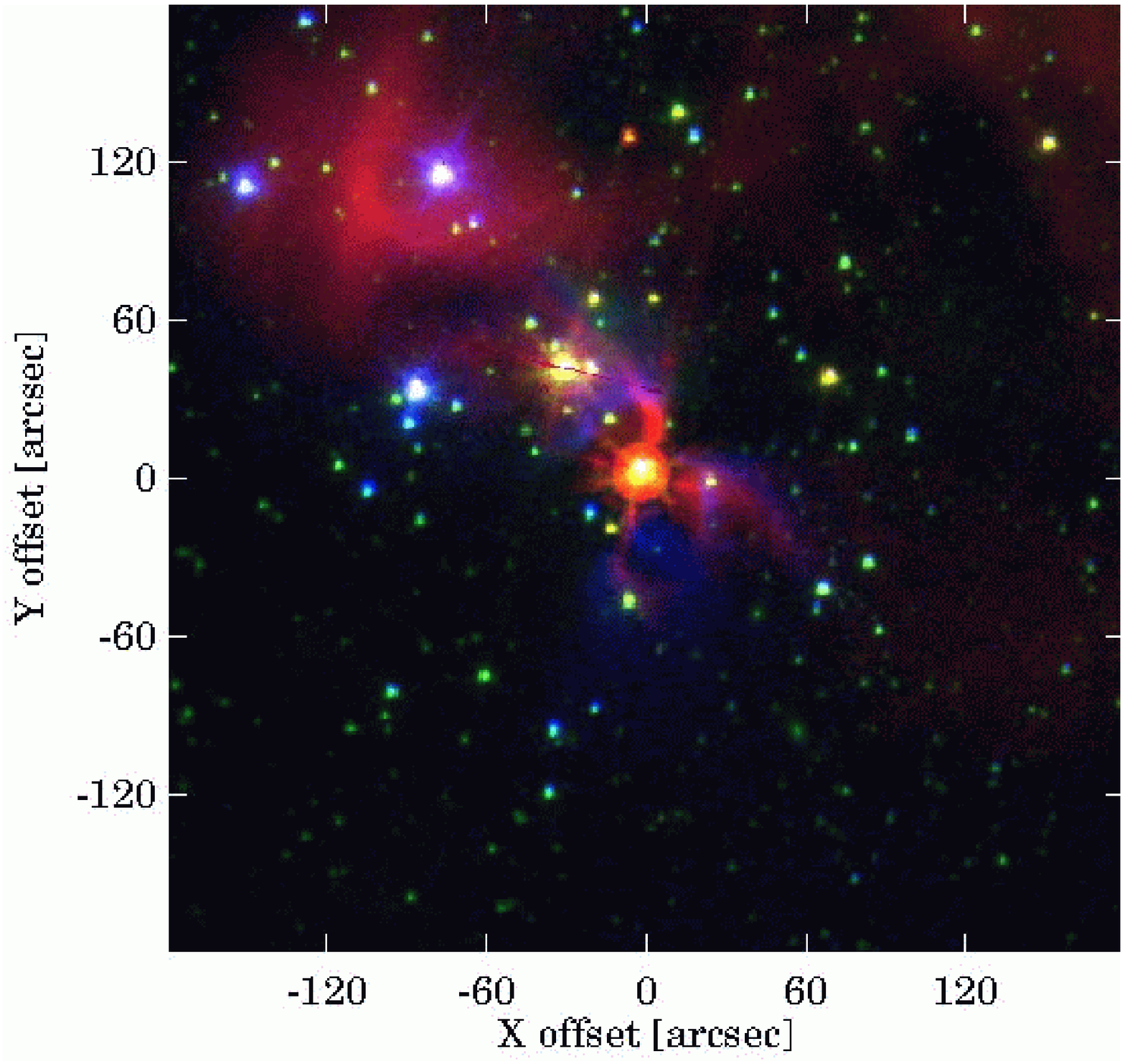}}
\centering{\includegraphics[width=8cm]{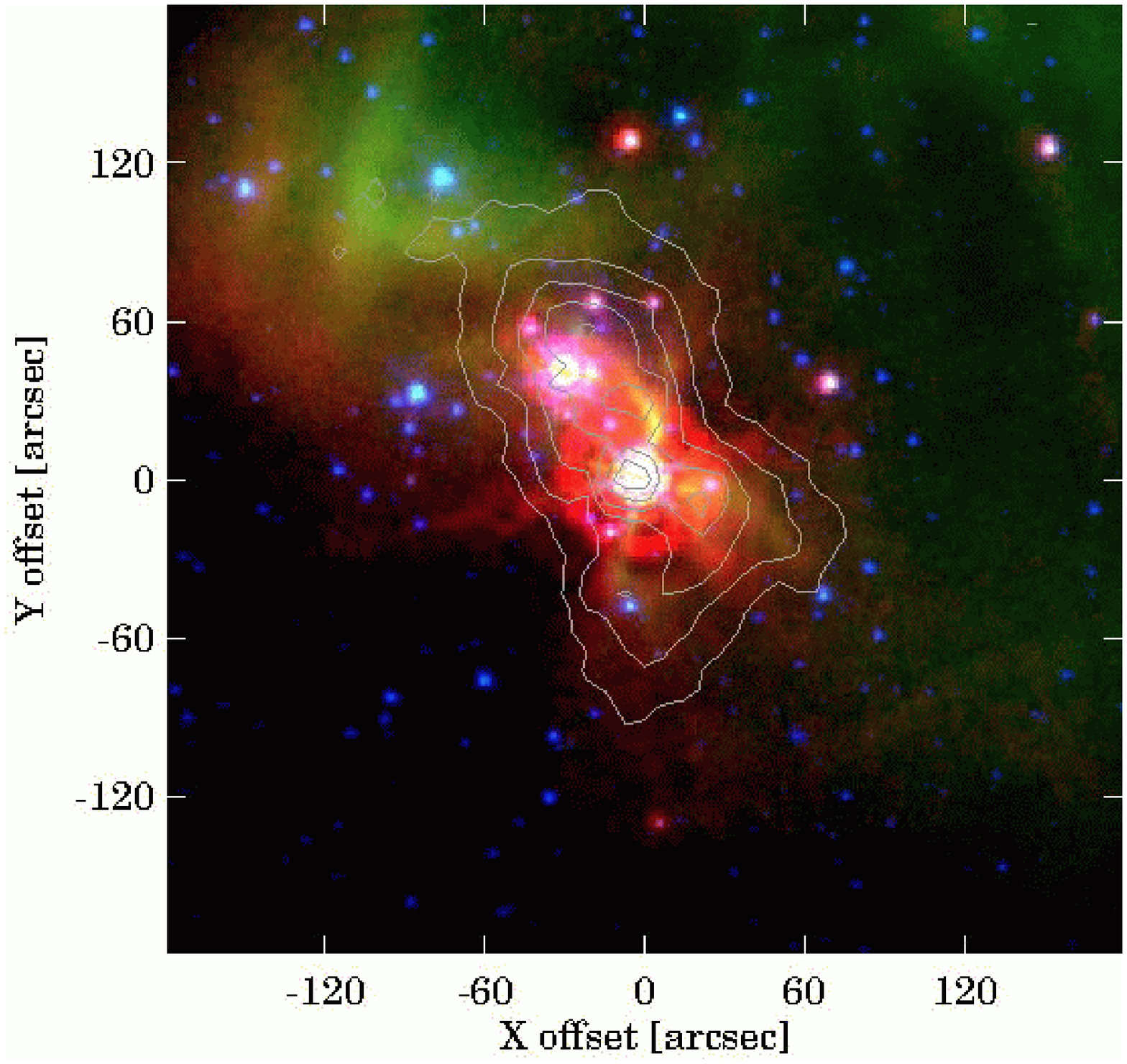}}
\caption{\textit{First panel\/}: Optical three-color image of RNO~8, composed from the \sdss\ {\it g\/} (blue), {\it r\/} (green), and {\it i\/} (red) images. \textit{Second panel\/}: three-color image of the same region composed of \sdss\ {\it g\/} (blue), IRAC 3.6\,\micron\ (green), and IRAC 8\,\micron\ (red) images. \textit{Third panel\/}: Three-color image composed of the IRAC 4.5\,\micron\ (blue), 8.0\,\micron\ (green), and MIPS 24\,\micron\ (red) images. Light grey contours show distribution of the 70\,\micron\  emission. }
\label{fig22}
\end{figure*}

\begin{figure*}
\centering{\includegraphics{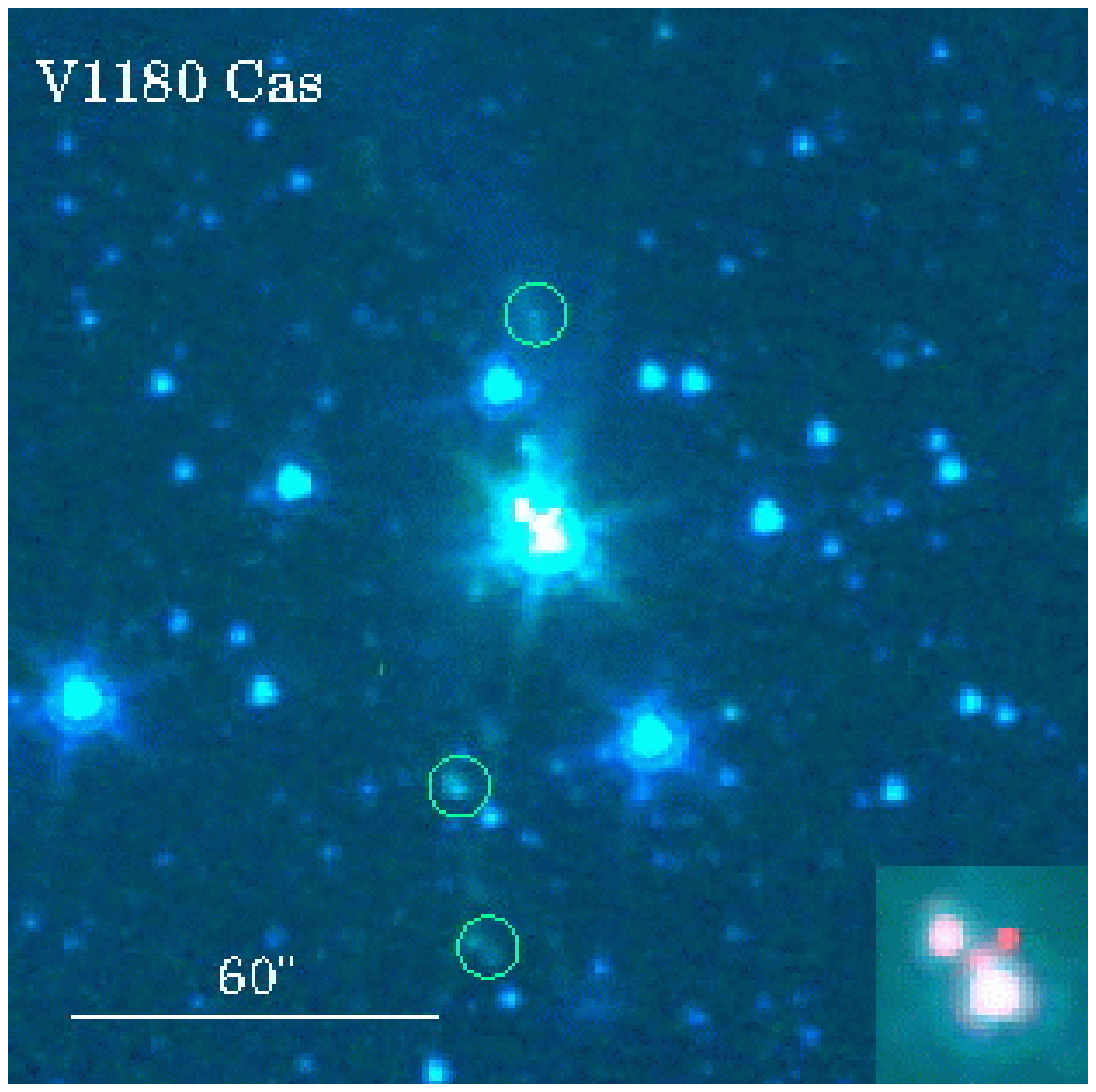}}
\caption{Same as Fig.~\ref{fig20}, for the environment of V1180~Cas. In addition to the jet, emanating northward from the Class~I component \citep{Anton14}, several fainter HH knots can be detected. A third component of the system emerges in the 8-\micron\ image. The inset shows a $18\arcsec\times18\arcsec$ area of the central objects.}
\label{fig23}
\end{figure*}

\begin{figure*}
\centering{\includegraphics[width=5cm]{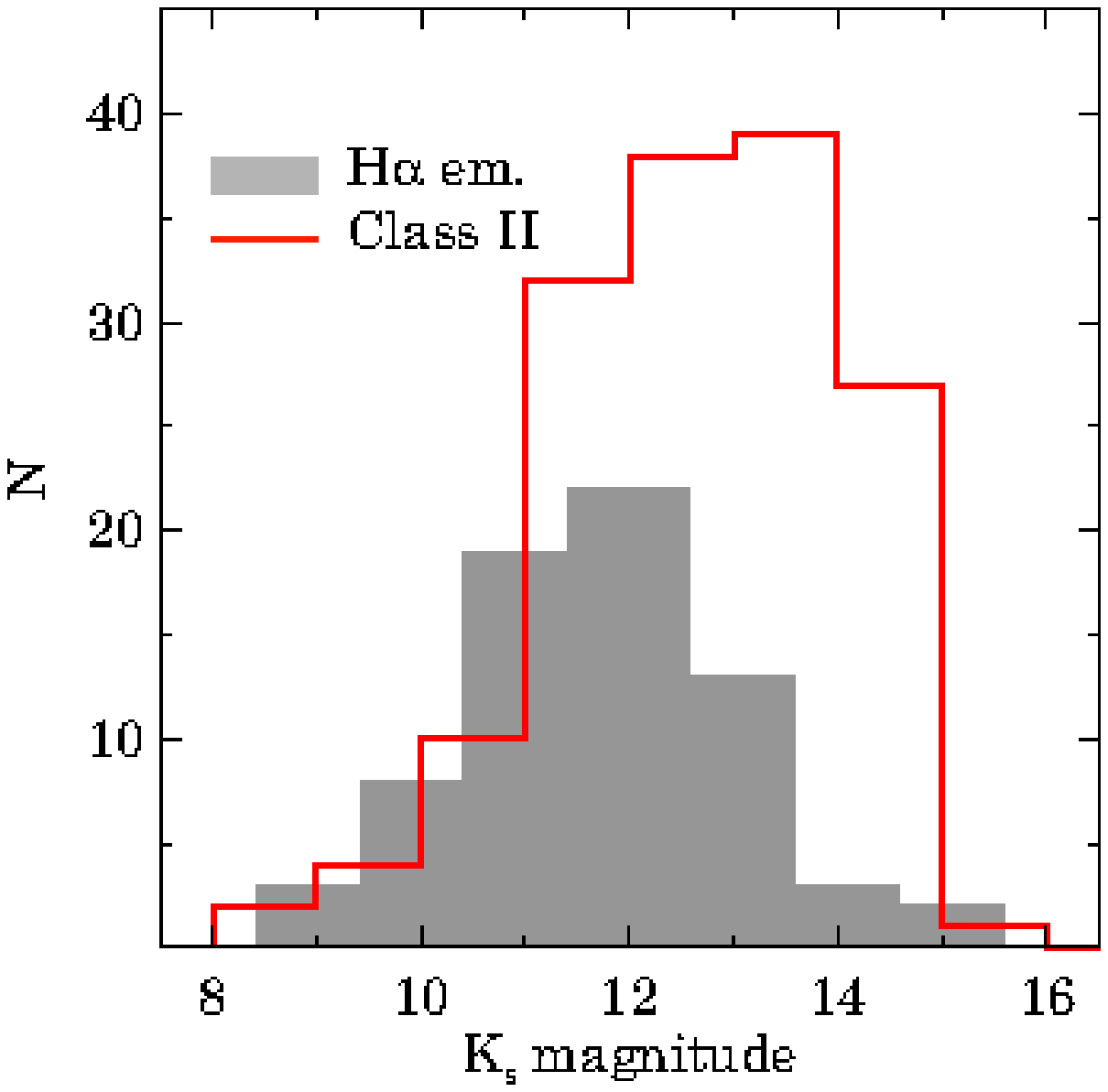}}
\centering{\includegraphics[width=5cm]{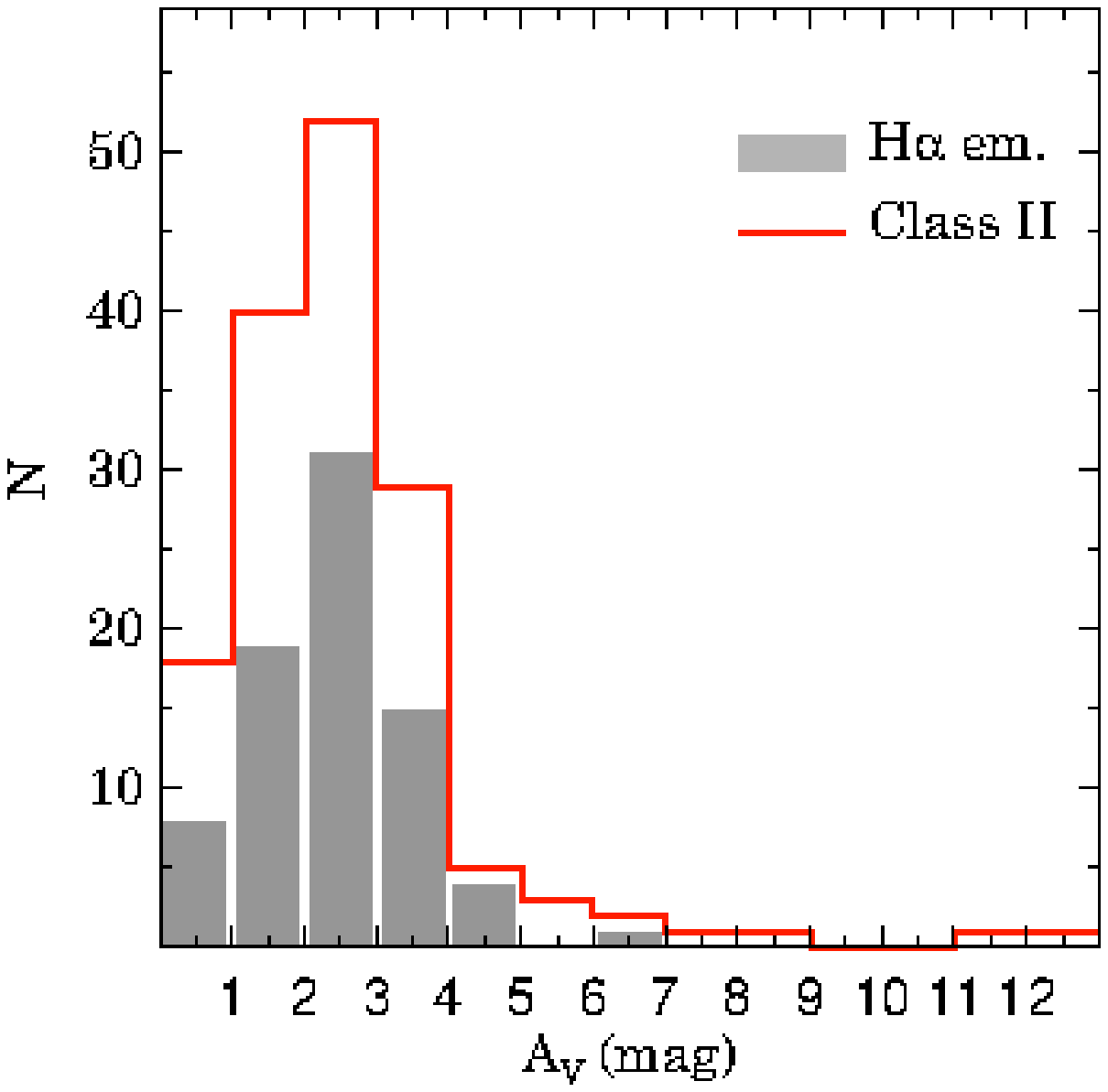}}
\centering{\includegraphics[width=5cm]{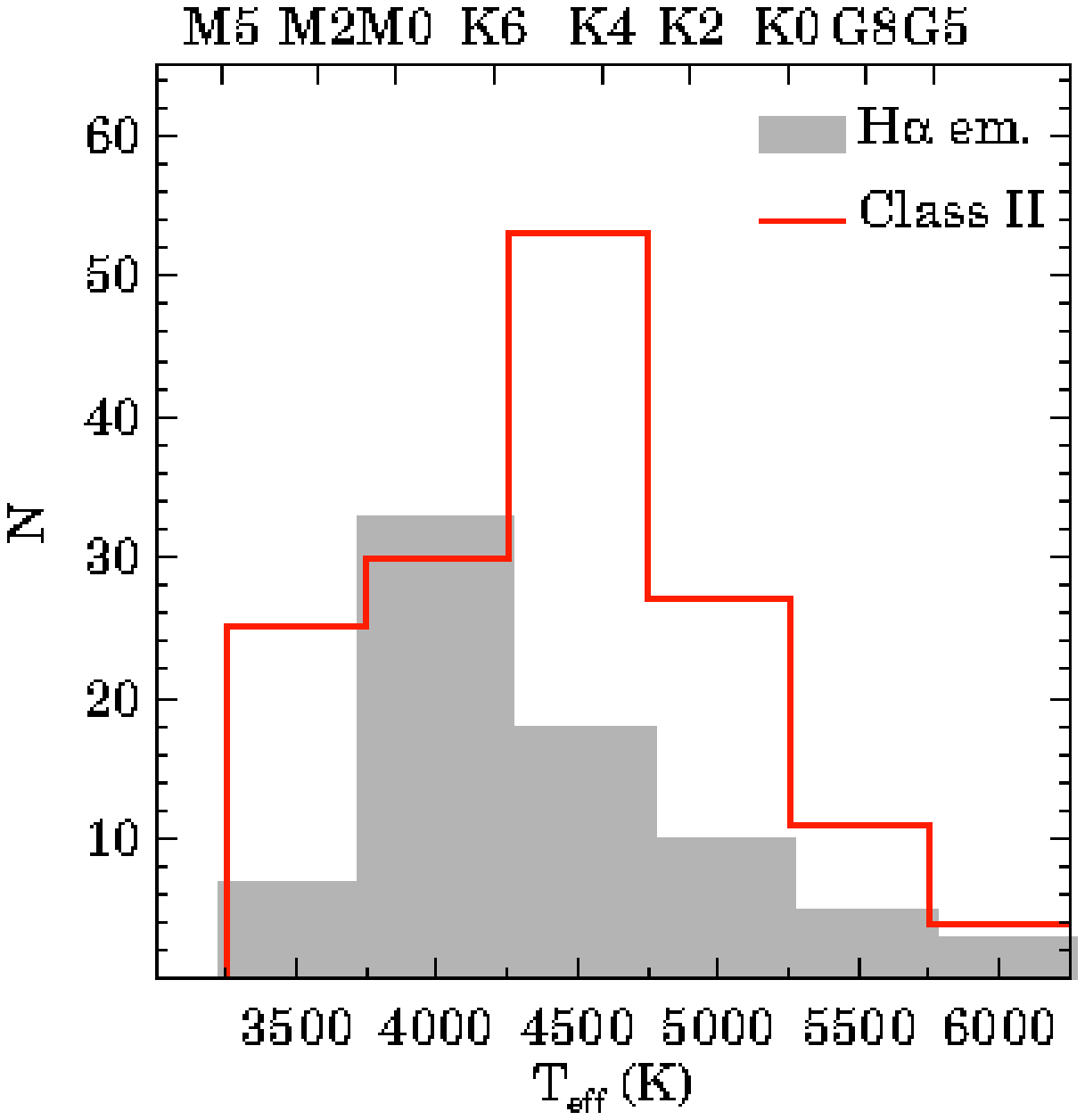}}
\caption{Histogram of the {\it K\/}$_\mathrm{s}$ magnitudes (left), visual extinctions (middle), and effective temperatures/spectral types (right), derived from photometric data of the candidate pre-main sequence stars of L1340.}
\label{fig24}
\end{figure*}

\begin{figure*}
\centering{\includegraphics[width=14cm]{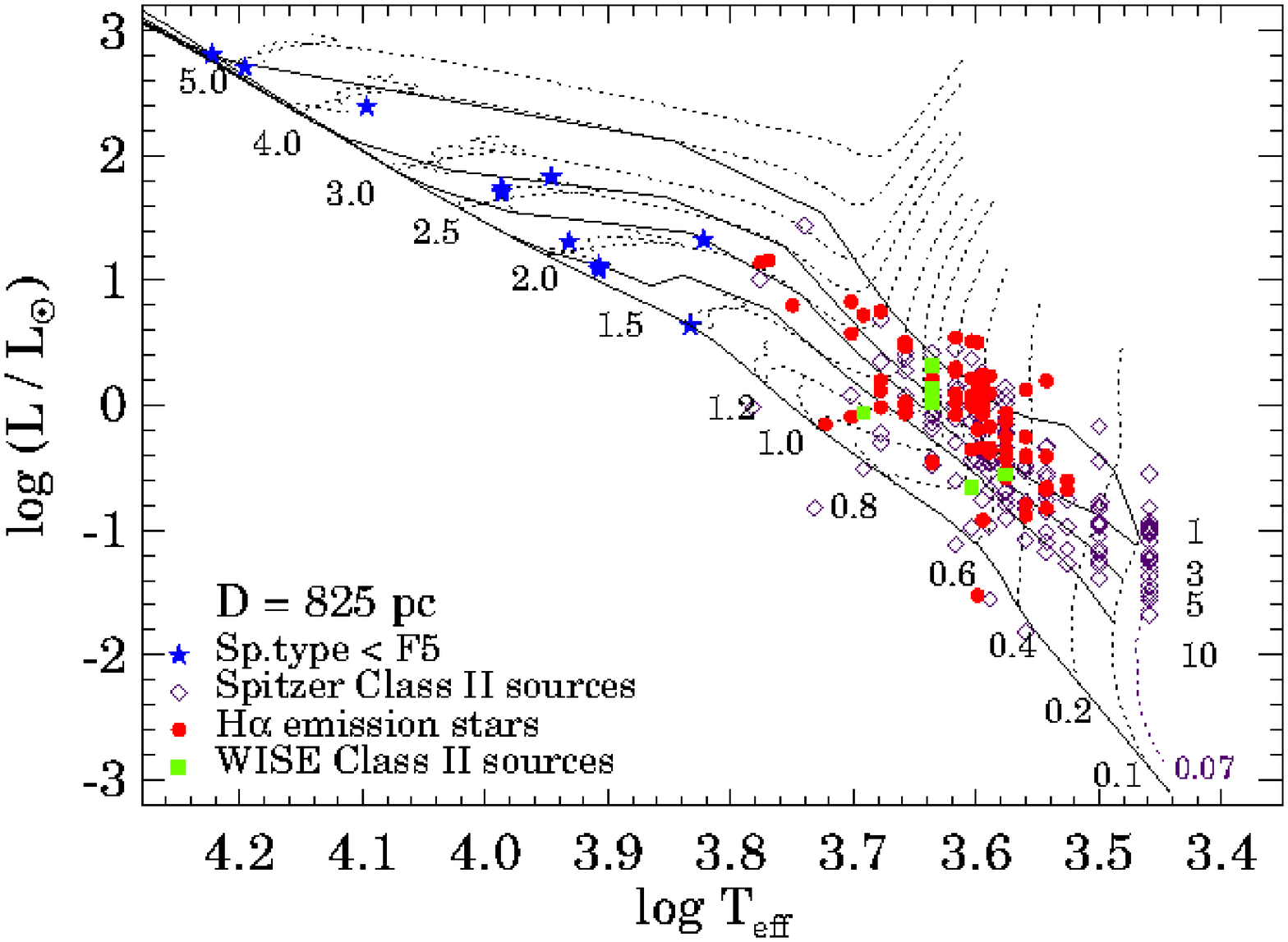}}
\caption{Hertzsprung--Russell diagram of the candidate young population of L1340. Blue star symbols indicate the stars earlier than F5 (Paper~III), red filled circles show Class~II sources with \ha\ emission detected (Paper~III). Open diamonds indicate Class~II sources not detected during our \ha\ survey. Black dotted lines indicate evolutionary tracks, and thin solid lines mark the 1, 3, 5, 10 million year isochrones from \citet{Siess}. The track for 0.07\,\msun\ (purple dotted line) is from \citet{BHAC2015}. The bolometric luminosities of the selected sources were calculated for a distance of 825\,pc.}
\label{fig25}
\end{figure*}

\clearpage

\begin{figure*}
\centering{\includegraphics[width=14cm]{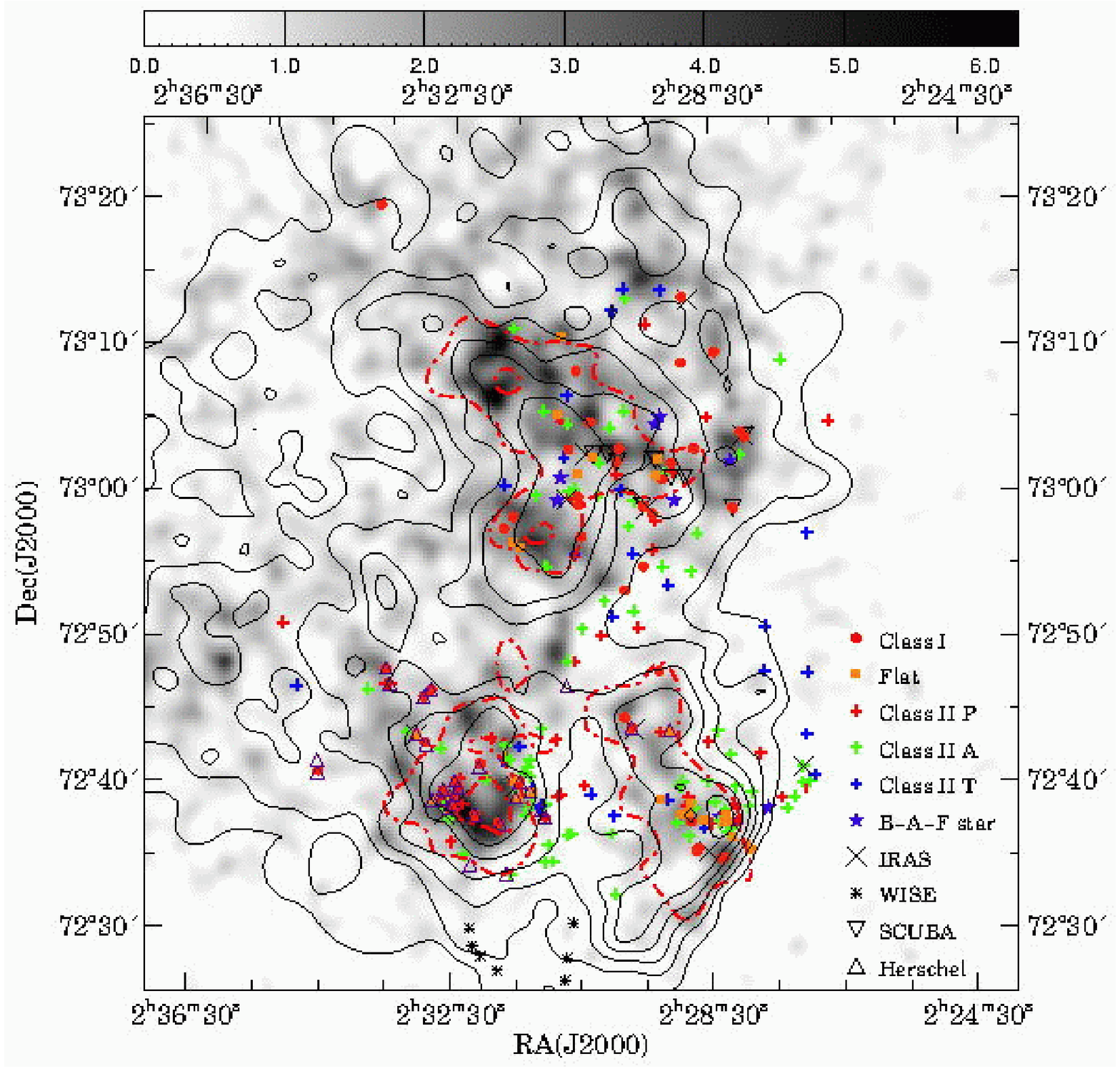}}
\caption{Surface distribution of the candidate YSOs, identified by infrared colors, overplotted on the visual extinction map of L1340, determined from star counts in the \textit{SDSS} DR9 (Paper~III). Black contours show \co\ integrated intensities, drawn at 0.6, 1.2, 1.8...K\,km\,s$^{-1}$, and red dash-dotted contour show the C$^{18}$O contour at 0.35~K\,km\,s$^{-1}$. Meaning of the symbols are shown in the lower right corner. }
\label{fig26}
\end{figure*}

\begin{figure}
\centering{\includegraphics[width=12cm]{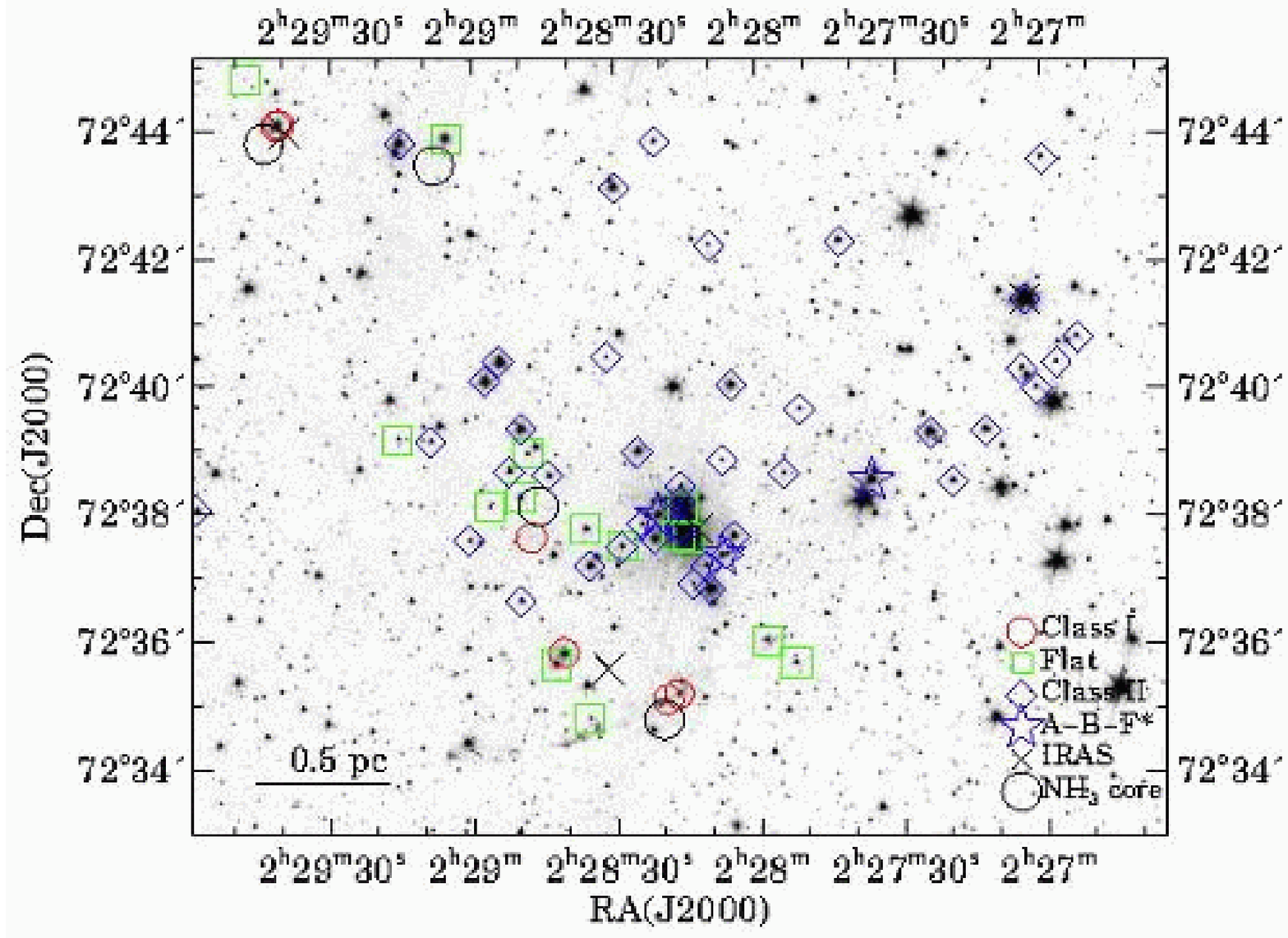}}
\caption{Surface distribution of the candidate YSOs and dense NH$_3$ cores in a $15.6\arcmin \times 12\arcmin$ area of L1340\,A, plotted on the IRAC 3.6\,\micron\ image of the field.} 
\label{fig27}
\end{figure}

\begin{figure}
\centering{\includegraphics[width=14cm]{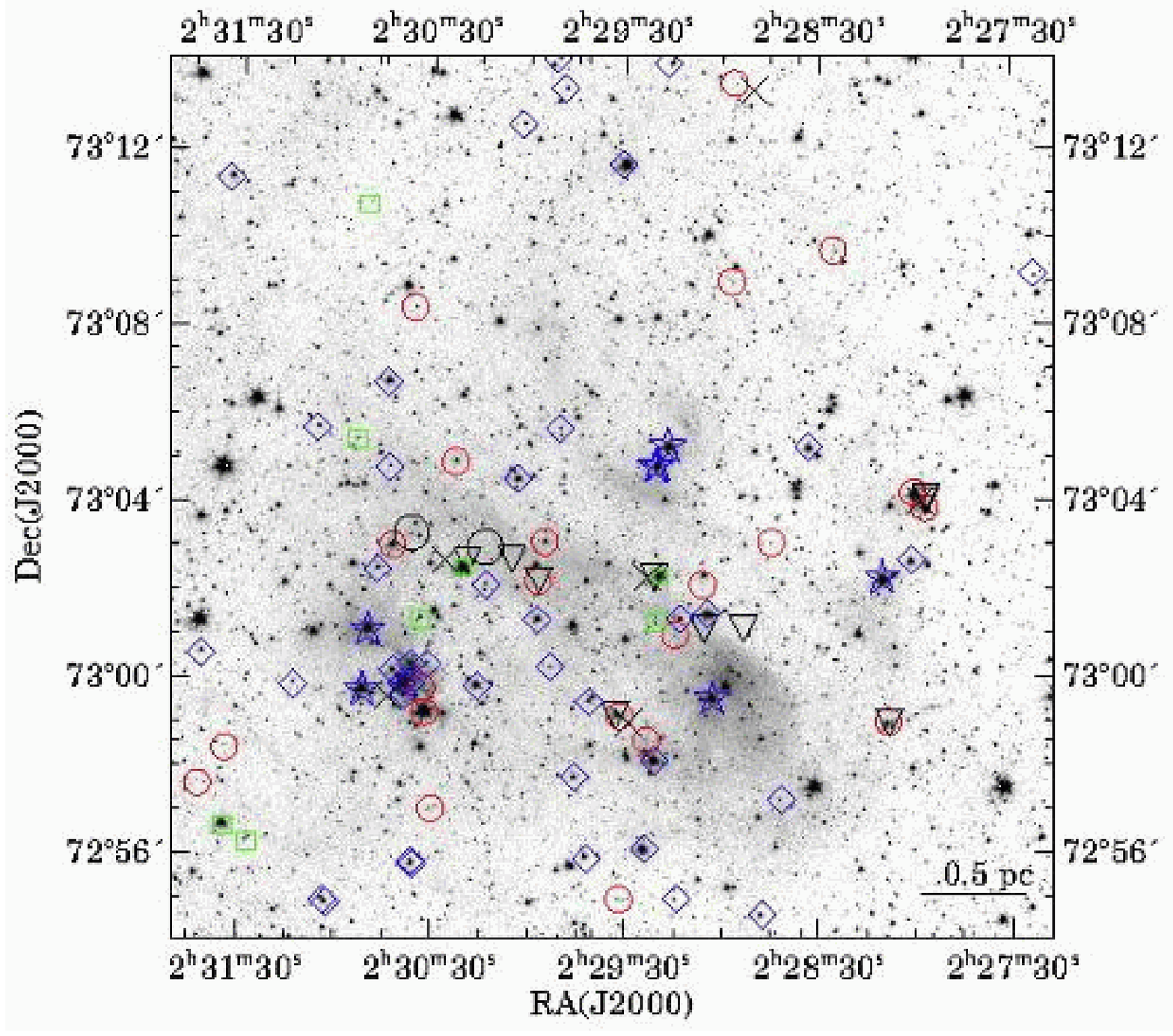}}
\caption{Surface distribution of the candidate YSOs and dense NH$_3$ cores in a $20\arcmin \times 20\arcmin$ area of L1340\,B. Symbols are same as in Fig.~\ref{fig27}, plus downward triangles mark submillimeter sources.} 
\label{fig28}
\end{figure}

\begin{figure}
\centering{\includegraphics[width=10cm]{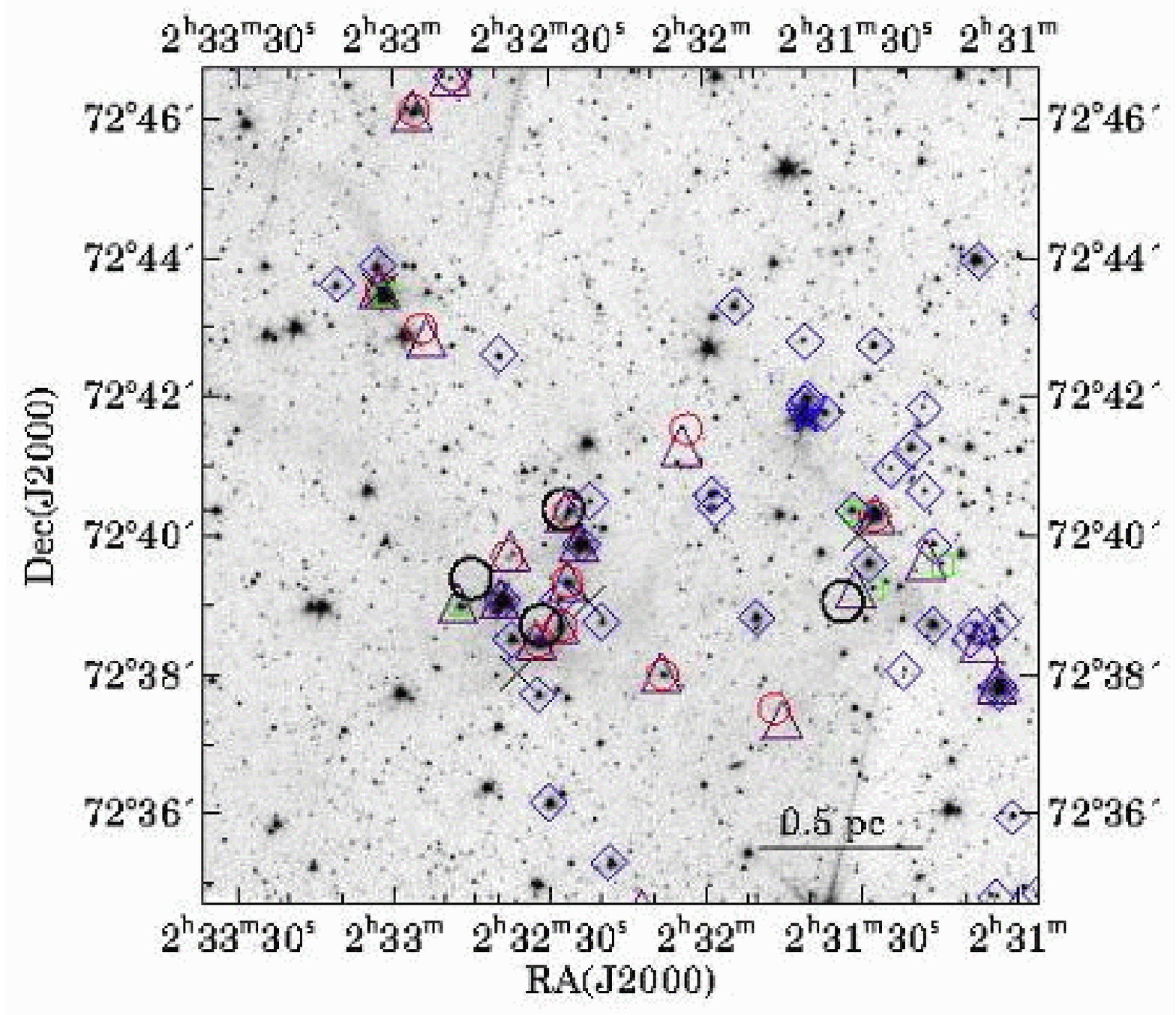}}
\caption{Surface distribution of the candidate YSOs and dense NH$_3$ cores in a $12\arcmin \times 12\arcmin$ area of L1340\,C. Symbols are same as in Fig.~\ref{fig27}, and triangles indicate far-infrared sources detected in the \textit{Herschel PACS\/} images.} 
\label{fig29}
\end{figure}

\begin{figure*}
\centering{\includegraphics[width=8cm]{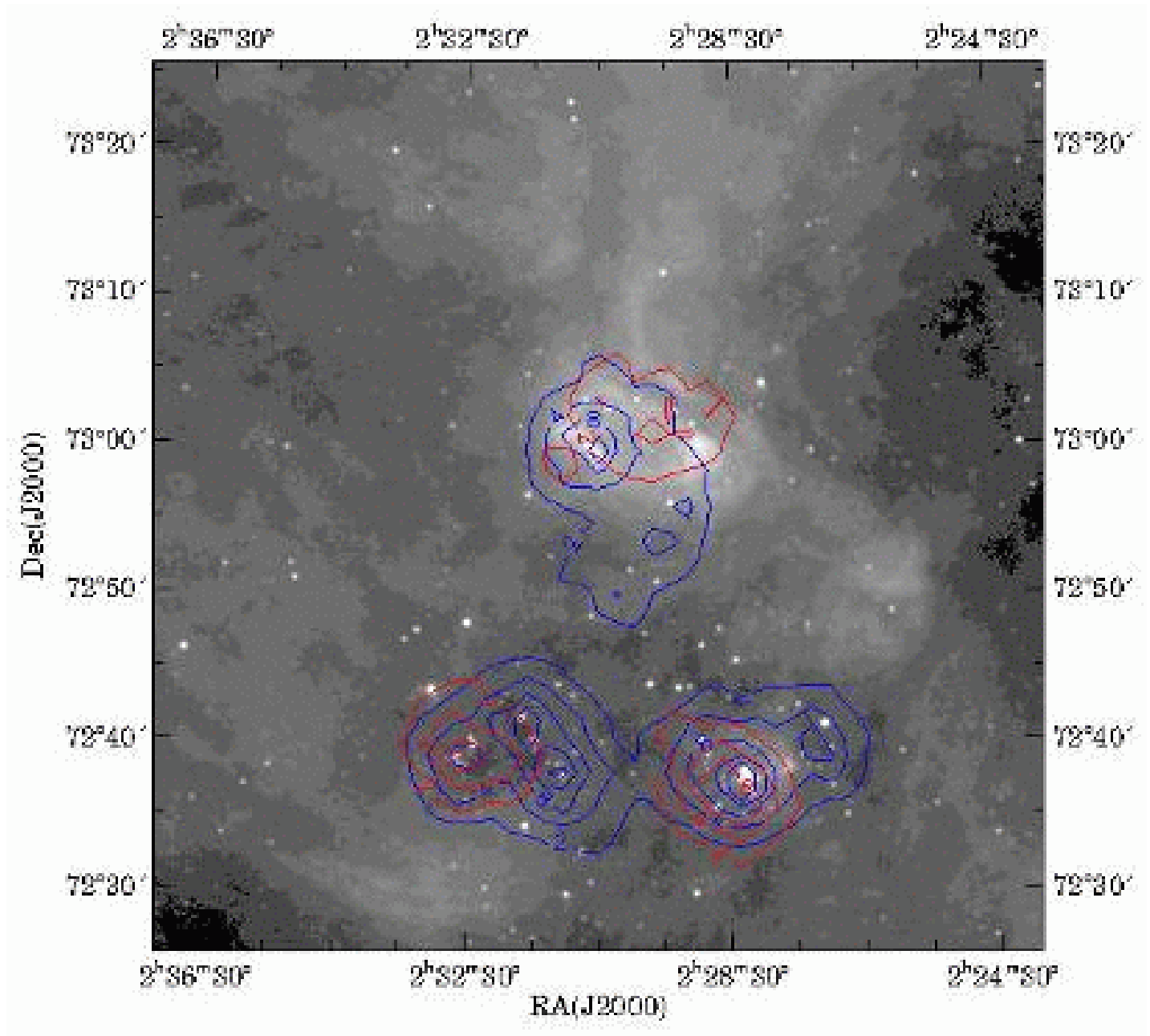}
\includegraphics[width=8cm]{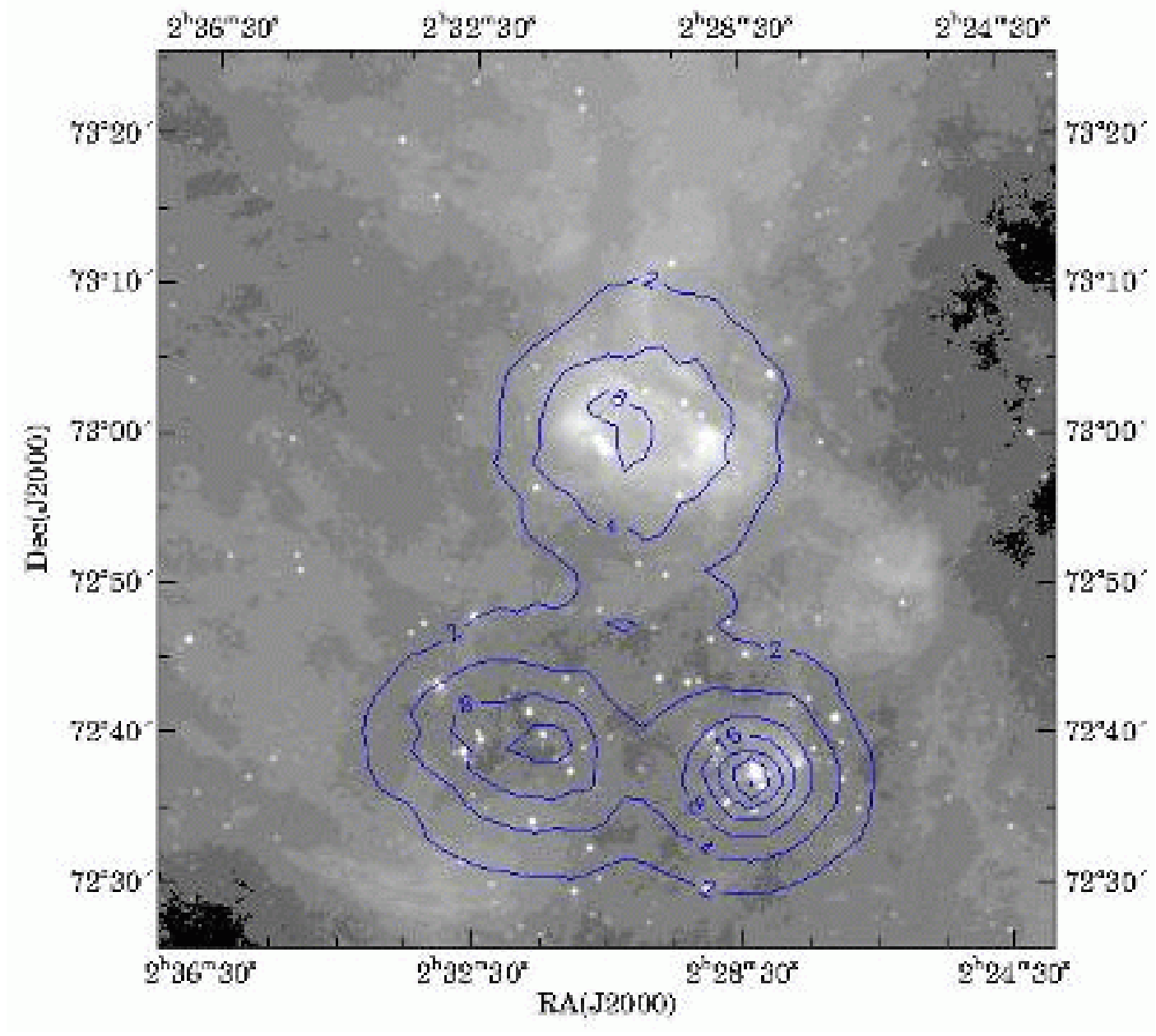}}
\caption{{\it Left\/}: Surface density distribution of the young stars, overplotted on the \textit{WISE\/} 12-\micron\ map of the cloud.  Blue contours show the surface density of pre-main sequence stars, and red contours show that of the Class~I + Flat SED sources. Each distribution was computed from the distances of the sixth nearest stars to the grid points. {\it Right\/}: The smoothed, composite surface density distribution of all candidate YSO classes, derived from the distance of the 20th nearest YSO to the grid points.}
\label{fig30}
\end{figure*}

\newpage

\begin{figure}
\centering{\includegraphics[width=8cm]{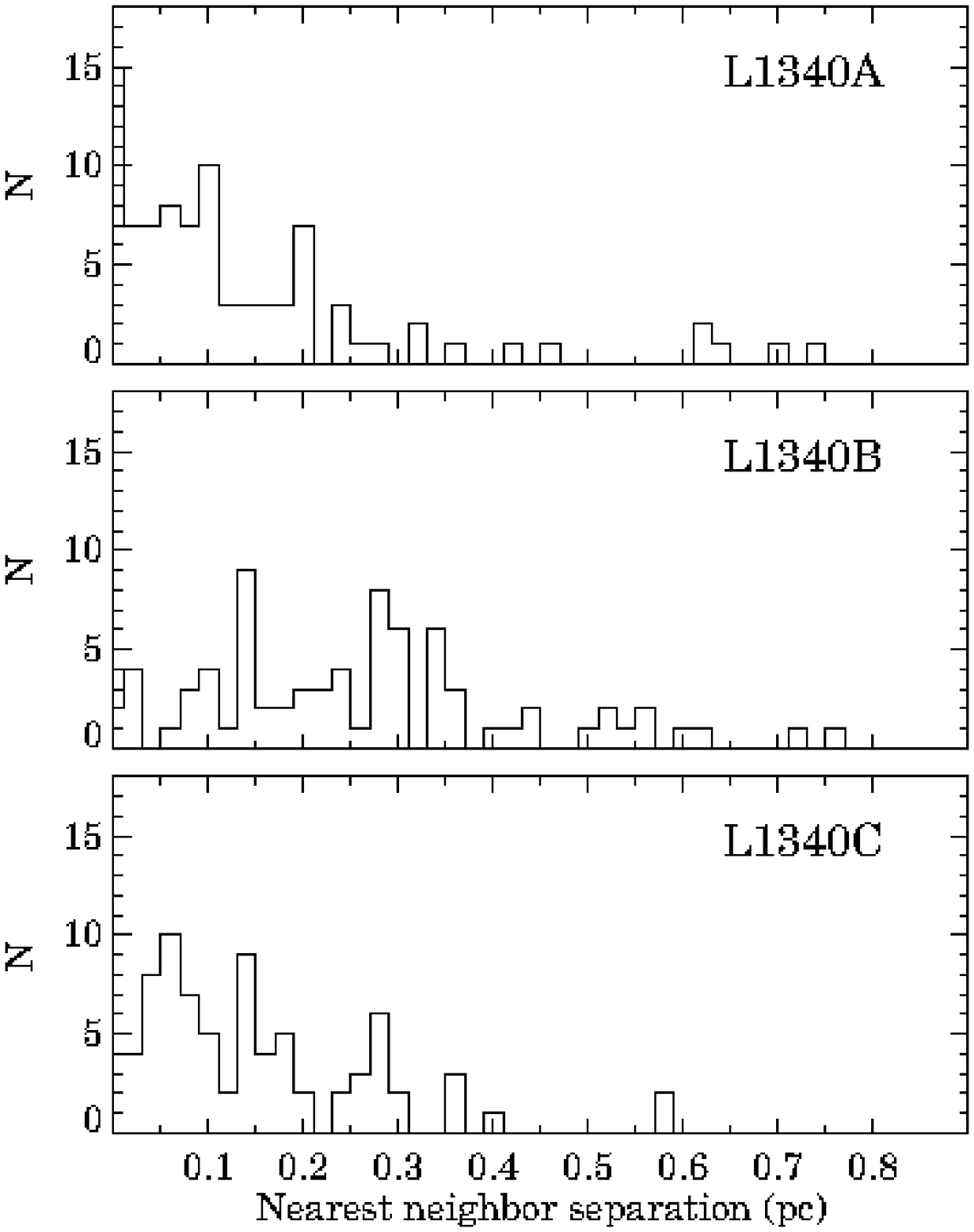}}
\caption{Histograms of the nearest neighbor separations for the three clumps of L1340.}
\label{fig31}
\end{figure}

\begin{figure}
\centering{\includegraphics[width=12cm]{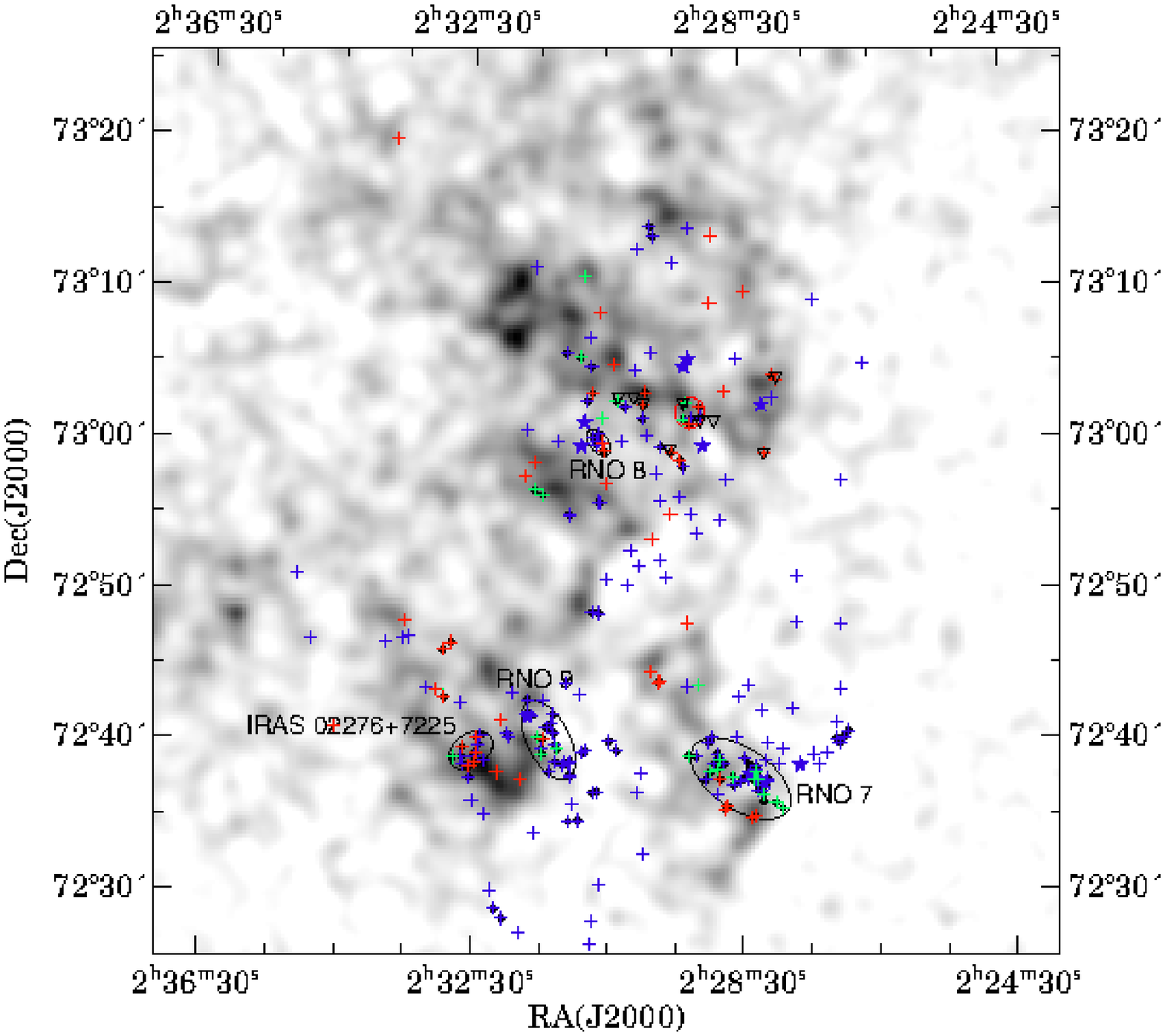}}
\caption{Distribution of candidate YSOs overplotted on the extinction map of L1340. Red symbols indicate Class~I sources, green is for Flat, blue for Class~II YSOs, and triangles indicate the submillimeter sources. Sources having a neighbor at a projected separation smaller than 0.15\,pc are marked with underlying small black circles. Ellipses encircle the most prominent clusterings identified during this procedure.}
\label{fig32}
\end{figure}

\begin{figure}
\centering{\includegraphics[width=8.5cm]{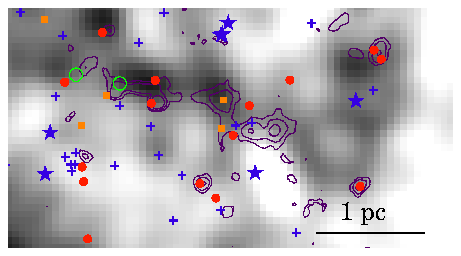}}
\centering{\includegraphics[width=8.5cm]{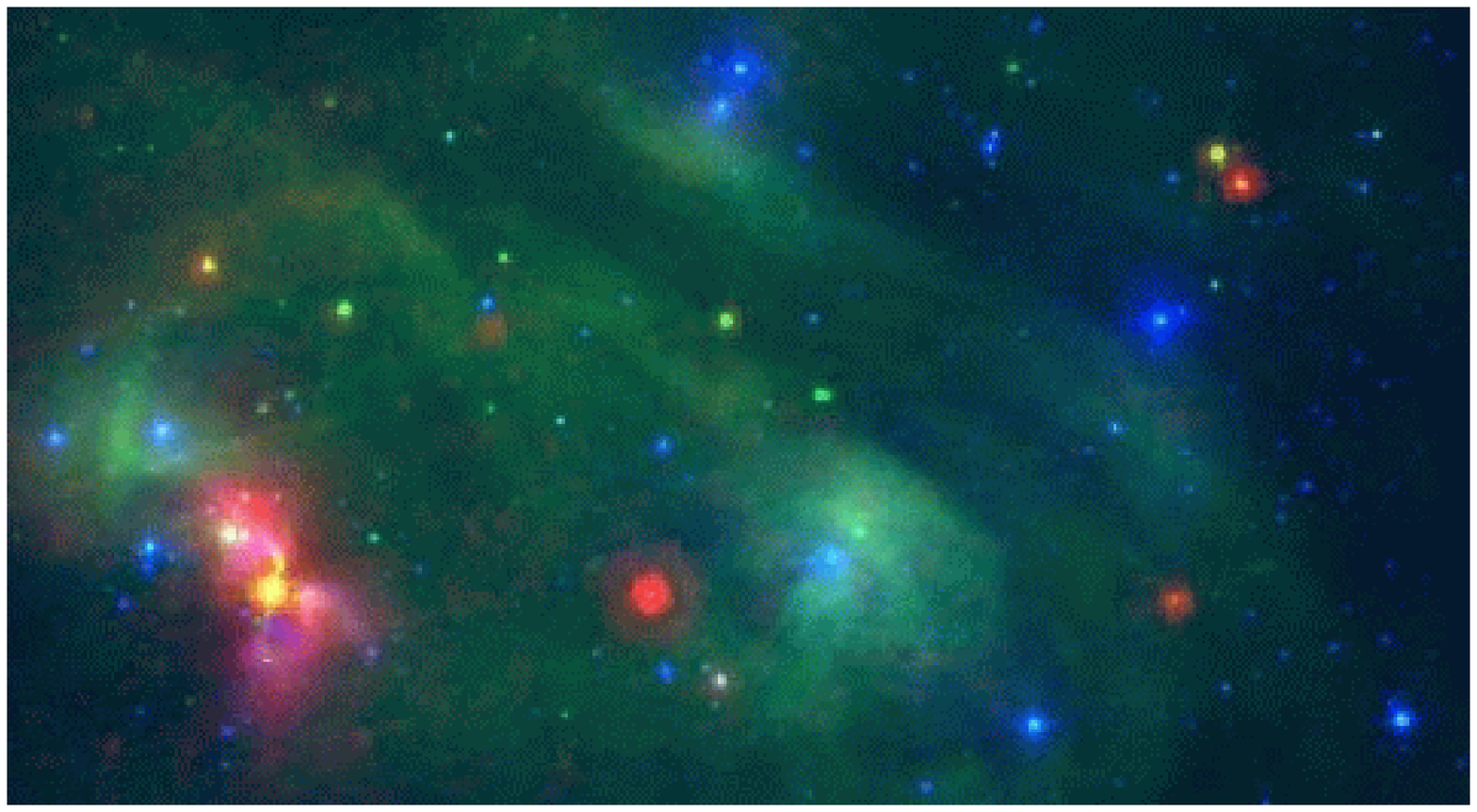}}
\centering{\includegraphics[width=8.5cm]{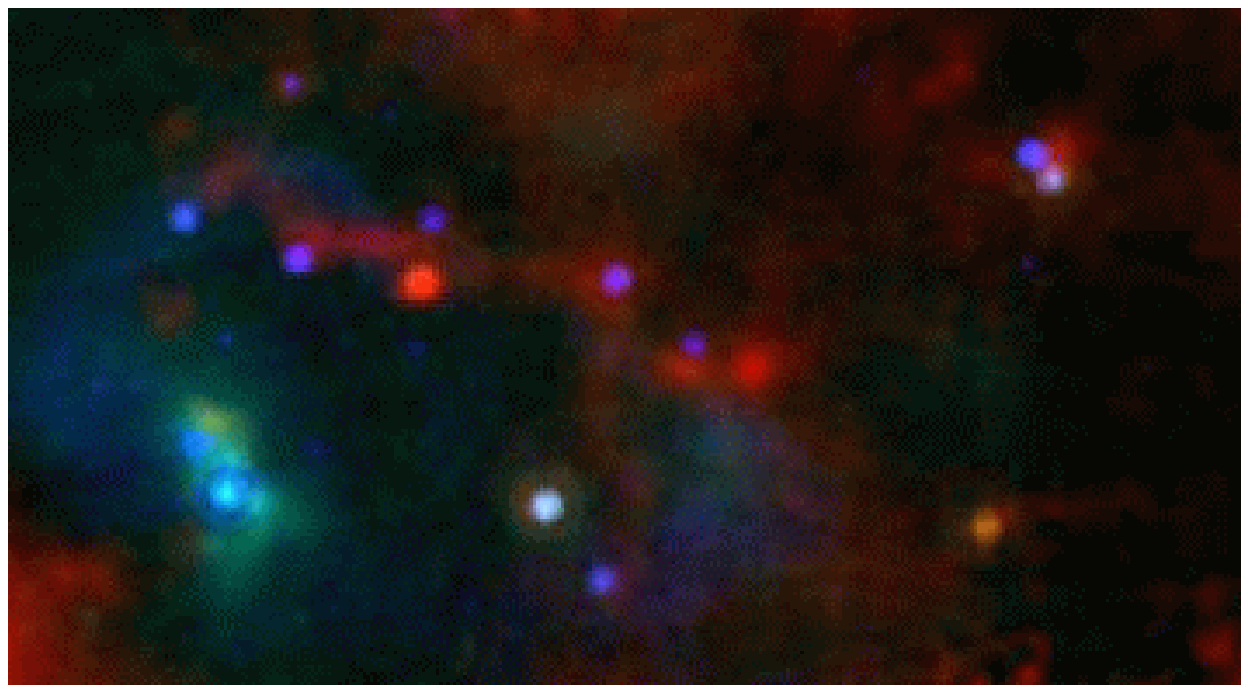}}
\caption{A multi-wavelength view of the central, $17.16\arcmin \times 9.3\arcmin$ area of the molecular clump L1340\,B. The upper panel shows the positions of the B--A type stars (blue star symbols), Class~0/I infrared sources (red dots), Flat SED (orange squares), Class~II YSOs, ammonia cores (Paper~II, green circles) and the contours of 850-\micron\ emission (purple contours, plotted at 30, 60, and 120 Jy\,beam$^{-1}$), overplotted on the extinction map. Class~I sources associated with tenth-of-parsec scale density enhancements, and a bubble-like  cavity, bordered by submm emission around an A-type star can be seen. The middle panel is a three-color image of the same area, composed of DSS2 blue (blue), \textit{Spitzer\/} IRAC 8\,\micron\ (green), and MIPS 70\,\micron\ (red) images. It reveals a bow-shock like structure in the 8-\micron\ emission around an A-type star, and the wispy structure of the diffuse 8-\micron\ emission. Brightness and color diversities of the embedded stars reveal their diverse evolutionary stages and luminosities. Lower panel: Three-color image of the same region, composed of the MIPS 24\,\micron\ (blue), MIPS 70\,\micron\ (green), and SCUBA 800\,\micron\ (red) images. The infrared sources lined up along $\sim$ 2\,pc long, cold filamentary structure.}
\label{fig33}
\end{figure}

\clearpage

%\documentclass[12pt,preprint]{aastex}
%\begin{document}
{\samepage
\begin{deluxetable}{rccccccc}
\tabletypesize{\scriptsize}
\rotate
\tablecolumns{8}
\tablewidth{0cm}
\tablecaption{SEIP SSTSL2 Fluxes and uncertainties  of Class~0/I sources in L1340 \label{Tab1_class1}}
\tablehead{
\colhead{No.} & \colhead{SSTSL2} & \colhead{$F_{3.6}$} &  \colhead{$F_{4.5}$} & \colhead{$F_{5.8}$} & \colhead{$F_{8.0}$} & \colhead{$F_{24}$} &
\colhead{$F_{70}$\tablenotemark{a}} \\
& & \colhead{(mJy)} & \colhead{(mJy)} & \colhead{(mJy)} & \colhead{(mJy)} & \colhead{(mJy)} & \colhead{(mJy)} }
\startdata
 1 & 022756.91+730354.4 &    1.915\,$\pm$\,0.005 &   3.782\,$\pm$\,0.007 &   5.412\,$\pm$\,0.016 &   6.104\,$\pm$\,0.013 &   73.620\,$\pm$\,0.164 & 828.038\,$\pm$\,58.375	\\  
 2 & 022800.65+730415.2 &   14.510\,$\pm$\,0.014 &  25.430\,$\pm$\,0.018 &  40.200\,$\pm$\,0.044 &  67.940\,$\pm$\,0.035 &  289.700\,$\pm$\,0.263 & 404.928\,$\pm$\,28.732    \\     
 3 & 022808.60+725904.5 &    1.774\,$\pm$\,0.005 &   2.873\,$\pm$\,0.007 &   2.982\,$\pm$\,0.013 &   1.888\,$\pm$\,0.010 &  1.775\,$\pm$\,0.132 & 596.200\,$\pm$\,16.900  \\  
 4 & 022818.51+723506.2 &    1.586\,$\pm$\,0.004 &   6.086\,$\pm$\,0.023 &   8.608\,$\pm$\,0.020 &   9.509\,$\pm$\,0.010 &	  \nodata  &	      \nodata	    \\
 5 & 022820.81+723500.5 &    0.251\,$\pm$\,0.001 &   2.019\,$\pm$\,0.015 &   4.189\,$\pm$\,0.014 &   8.067\,$\pm$\,0.025 &	  \nodata  &	     \nodata	  \\
 6 & 022825.07+730945.6 &    0.150\,$\pm$\,0.002 &   0.179\,$\pm$\,0.002 &   0.197\,$\pm$\,0.007 &   0.842\,$\pm$\,0.011 &   22.590\,$\pm$\,0.135 &  43.500\,$\pm$\,11.100	    \\
 7 & 022842.57+723544.3 &   11.560\,$\pm$\,0.010 &  19.330\,$\pm$\,0.036 &  27.290\,$\pm$\,0.034 &  28.610\,$\pm$\,0.026 &     \nodata     &	     \nodata	    \\
 8 & 022844.40+723533.5 &    3.473\,$\pm$\,0.005 &   4.359\,$\pm$\,0.024 &   7.668\,$\pm$\,0.019 &   7.259\,$\pm$\,0.018 &    \nodata	   & \nodata		 \\
\enddata        				  
\tablenotetext{a}{This flux results from our measurement.} 
\tablenotetext{b}{This source is not listed in the SEIP Source List. The identifier is based on the position in the 
MIPS 70-\,\micron\ image, and the fluxes were measured as described in Sect.~2.1.}
\tablenotetext{c}{This source is not listed in the SEIP Source List. The identifier is based on the 
position in the SEIP Super Mosaic images of the region.}
\tablenotetext{*}{H$\alpha$ emission star.}
\end{deluxetable}}

%\end{document}

\clearpage

 %\documentclass[12pt,preprint]{aastex}
% \begin{document}
\begin{deluxetable}{rccccccc}
\tabletypesize{\scriptsize}
\rotate
\tablecolumns{8}
\tablewidth{0cm}
\tablecaption{SEIP SSTSL2 fluxes and uncertainties of Flat SED sources in L1340 \label{Tab1_flat}}
\tablehead{
\colhead{No.} & \colhead{SSTSL2} & \colhead{$F_{3.6}$} & \colhead{$F_{4.5}$} & \colhead{$F_{5.8}$} & \colhead{$F_{8.0}$} & \colhead{$F_{24}$} &
\colhead{$F_{70}$\tablenotemark{a}}  \\
 & & \colhead{(mJy)} & \colhead{(mJy)} & \colhead{(mJy)} & \colhead{(mJy)} & \colhead{(mJy)} & \colhead{(mJy)} }
\startdata 
 1 & 022754.00+723535.5 &    1.744\,$\pm$\,0.004  &   1.766\,$\pm$\,0.020  & 2.957\,$\pm$\,0.012  &   0.245\,$\pm$\,0.018  &\nodata	& \nodata    \\
 2 & 022759.92+723556.4\tablenotemark{b} &    0.523\,$\pm$\,0.026  &   0.859\,$\pm$\,0.060  &	0.980\,$\pm$\,0.032  & \nodata   & \nodata   \\       
 3 & 022811.32+723631.5 &    4.896\,$\pm$\,0.086  &   6.002\,$\pm$\,0.011  & 6.895\,$\pm$\,0.011  &   7.715\,$\pm$\,0.014  &		\nodata     & \nodata	 \\  
 4 & 022816.62+723732.6 & 219.545\,$\pm$\,28.270     & 268.500\,$\pm$\,0.135  & 361.400\,$\pm$\,0.124  & 374.611\,$\pm$\,11.495\tablenotemark{a} & \nodata & \nodata   \\
 5 & 022817.85+723800.9 &   92.280\,$\pm$\,0.027  & 108.000\,$\pm$\,0.056  & 140.700\,$\pm$\,0.078  &  219.498\,$\pm$\,6.839\tablenotemark{a} & 857.300\,$\pm$\,0.612 & \nodata	\\	
 6 & 022818.51+723734.6 &      \nodata       &   9.617\,$\pm$\,0.021  &      8.287\,$\pm$\,0.019  &   9.695\,$\pm$\,0.015  &   \nodata      &  \nodata \\  
 7 & 022838.02+723740.6 &    2.805\,$\pm$\,0.005  &   2.963\,$\pm$\,0.006  & 3.217\,$\pm$\,0.012  &   5.370\,$\pm$\,0.012  &  12.010\,$\pm$\,0.516 &  \nodata \\
 8 & 022850.36+723851.2 &    0.822\,$\pm$\,0.003  &   0.883\,$\pm$\,0.004  & 0.897\,$\pm$\,0.008  &   1.218\,$\pm$\,0.009  & 4.005\,$\pm$\,0.180 & \nodata   \\
\enddata   
\tablenotetext{a}{This flux results from our measurements.} 
\tablenotetext{b}{This source is not listed in the SEIP Source List. The identifier is based on the position in the 
SEIP Super Mosaic images and the fluxes were measured as describen in Sect.~2.1.}

\end{deluxetable}
%\end{document}

\clearpage

\begin{deluxetable}{ccccccc}
\tabletypesize{\scriptsize}
\rotate
\tablecolumns{7}
\tablewidth{0cm}
\tablecaption{SEIP SSTSL2 Fluxes and uncertainties  of of Class~II sources \label{Tab1_class2}}
\tablehead{
\colhead{SSTSL2} & \colhead{$F_{3.6}$} &  \colhead{$F_{4.5}$} & \colhead{$F_{5.8}$} & \colhead{$F_{8.0}$} & \colhead{$F_{24}$} & \colhead{$F_{70}$}  \\
 & \colhead{(mJy)} & \colhead{(mJy)} & \colhead{(mJy)} & \colhead{(mJy)} & \colhead{(mJy)} & \colhead{(mJy)}}
\startdata 
022638.02+730457.5\tablenotemark{*}  &     3.517\,$\pm$\,0.147 &   3.965\,$\pm$\,0.009   &   4.335\,$\pm$\,0.015   &	5.878\,$\pm$\,0.015  &    7.710\,$\pm$\,0.116  &   36.198\,$\pm$\,3.178  \\ 
022654.73+724040.8   &     1.552\,$\pm$\,0.004 &   1.012\,$\pm$\,0.004   &   0.527\,$\pm$\,0.007   &	0.461\,$\pm$\,0.010  &    4.301\,$\pm$\,0.179  &       \nodata   \\
022659.03+725716.0   &   1.477\,$\pm$\,0.005 &   1.156\,$\pm$\,0.003   &   0.921\,$\pm$\,0.006   &	0.739\,$\pm$\,0.009  &    5.040\,$\pm$\,0.115  &       \nodata   \\ 
022659.08+724016.6\tablenotemark{*}  &     0.966\,$\pm$\,0.003 &   0.815\,$\pm$\,0.004   &   0.667\,$\pm$\,0.007   &	0.893\,$\pm$\,0.009  &    1.221\,$\pm$\,0.200  &       \nodata   \\ 
022659.35+725714.2   &     1.581\,$\pm$\,0.005 &   1.204\,$\pm$\,0.003   &   0.981\,$\pm$\,0.006   &	 1.098\,$\pm$\,0.009  &    9.337\,$\pm$\,0.117  &       \nodata	 \\ 
022700.34+724743.8\tablenotemark{*}  &     2.571\,$\pm$\,0.005 &   1.920\,$\pm$\,0.006   &   1.573\,$\pm$\,0.009   &	2.136\,$\pm$\,0.010  &    8.947\,$\pm$\,0.126  &       \nodata   \\ 
022702.11+724329.0\tablenotemark{*}  &     2.061\,$\pm$\,0.004 &   1.617\,$\pm$\,0.006   &   1.043\,$\pm$\,0.008   &	0.968\,$\pm$\,0.011  &    2.460\,$\pm$\,0.123  &       \nodata   \\
022703.17+723952.9\tablenotemark{*}  &     1.172\,$\pm$\,0.003 &   1.177\,$\pm$\,0.005   &   0.794\,$\pm$\,0.007   &	1.192\,$\pm$\,0.009  &    2.206\,$\pm$\,0.179  &       \nodata   \\
\enddata        		        		   		        		        				
\tablenotetext{*}{H$\alpha$ emission star, described in detail in Paper~III.}
\tablenotetext{a}{Our measurement.}
\tablecomments{Table \ref{Tab1_class2} is published in its  entirety in the electronic edition of the {\it Astrophysical Journal}.  A  portion is shown here for guidance regarding its form and content.}   		        		\end{deluxetable}		        		   		        		        			       
\begin{deluxetable}{rrccrrrr}
\tabletypesize{\scriptsize}
\tablecolumns{8}
\tablewidth{0cm}
\tablecaption{\textit{Herschel PACS\/} point sources associated with candidate \spitzer\ YSOs in L1340\,C \label{Table_pacs}}
\tablehead{
\colhead{No.} & \colhead{SSTSL2} & \colhead{Fitted RA}  & \colhead{Fitted Dec} & \colhead{$F_{100}$}  &  \colhead{$dF_{100}$} &  \colhead{$F_{160}$}  
&  \colhead{$dF_{160}$} \\
& & \colhead{(deg)}  & \colhead{(deg)} & \colhead{(mJy)} & \colhead{(mJy)}& \colhead{(mJy)}& \colhead{(mJy)} } 
\startdata 
 1 &  022907.88+724347.2   &  37.28481   &  72.73014  &	349.372   &   \nodata &  1073.854  &	\nodata    \\	      
 2 &  022943.64+724358.6   &  37.43316   &  72.73337  &  9899.194   &  \nodata  &  11999.27  &	\nodata    \\
 3 &  023114.12+723933.3   &  37.82169   &  72.66068  &	 34.923   &  353.948  &  \nodata   &   \nodata     \\
 4 &  023127.45+723912.8   &  37.87831   &  72.65379  &	 13.773   &  101.163  &  0474.558  &	 38.295    \\
 5 &  023127.34+724013.0   &  37.86170   &  72.67144  &	918.261   &  124.522  &  1652.396  &	155.898    \\
 6 &  023146.58+723729.4   &  37.93610   &  72.62243  &	 16.760   &   57.825  &  0342.640  &	 66.379    \\
 7 &  023203.42+724131.7   &  38.01794   &  72.68809  &	  0.424   &  182.745  &  0101.090  &	 62.318    \\
 8 &  023207.96+723759.3   &  38.03459   &  72.63372  &	258.719   &  205.061  &  0500.088  &	 85.020    \\
\enddata
\end{deluxetable} 

\clearpage

\begin{deluxetable}{cccccccl}
\tabletypesize{\scriptsize}
\rotate
\tablecolumns{8}
\tablewidth{0cm}
\tablecaption{\textit{Spitzer} sources of Class~0/I SED: Associated objects \label{Tab2_class1}}
\tablehead{
\colhead{SSTSL2} & \colhead{SDSS DR9 J} & \colhead{2MASS} & \colhead{AllWISE J}  & \colhead{IRAS} & \colhead{Akari FIS}
 & \colhead{JCMTSF J} & \colhead{Other} }
\startdata
 022756.91+730354.4  &  \nodata		   &  02275695+7303542  &  022756.94+730354.6  &\nodata & 0227565+730402  &   022756.6+730404  & \nodata\\  
 022800.65+730415.2  & 022800.91+730415.4g &  02280074+7304154  &  022800.60+730415.4  &\nodata & \nodata &  \nodata & \nodata    \\	 
 022808.60+725904.5  &  \nodata	           &\nodata		&  022808.59+725904.0  & \nodata & 0228071+725858 &   022808.4+725902  & \nodata \\  
 022818.51+723506.2  &  \nodata	           &  02281842+7235061  &  022818.46+723506.3  & \nodata & \nodata &  \nodata  & \nodata  \\
 022820.81+723500.5  & \nodata	           &	\nodata		&  022820.76+723500.6  &\nodata & 0228201+723504 & \nodata & \\
 022825.07+730945.6  & \nodata	           &	\nodata		&  022825.06+730945.7  &\nodata & \nodata &  \nodata & \nodata  \\
 022842.57+723544.3  & 022842.55+723544.5g &  02284255+7235444  &  022842.54+723544.5  & \nodata & \nodata &  \nodata & \nodata \\
 022844.40+723533.5  & 022844.40+723532.9g &  02284443+7235332  &  022844.39+723533.6  & \nodata & \nodata &  \nodata & \nodata \\
\enddata 	    
\end{deluxetable}

\clearpage

\begin{deluxetable}{cccccccl}
\tabletypesize{\scriptsize}
\rotate
\tablecolumns{8}
\tablewidth{0cm}
\tablecaption{\textit{Spitzer} sources of Flat SED: Associated objects \label{Tab2_flat}}
\tablehead{
\colhead{SSTSL2} & \colhead{SDSS DR9 J} & \colhead{2MASS} & \colhead{AllWISE J}  & \colhead{IRAS} & \colhead{Akari FIS}
 & \colhead{JCMTSF J} & \colhead{Other} }
\startdata
022754.00+723535.5  & 022753.97+723535.5*   & 02275399+7235354    &  022753.96+723535.7&  \nodata & \nodata &  \nodata &\nodata \\  
022759.92+723556.4  & 022759.78+723555.6*   & 02275976+7235561    &  022759.92+723556.4 &  \nodata & \nodata &  \nodata & HH\,488S \\
022811.32+723631.5  & 022811.27+723631.6g   & 02281130+7236316    &  022811.29+723631.6 &  \nodata & \nodata &  \nodata & \nodata \\  
022816.62+723732.6  & 022816.63+723733.0*   & 02281661+7237328	  & 022816.62+723732.8   &  02236+7224 & \nodata & \nodata & [KOS94]~HA\,1,RNO7-5 \\
022817.85+723800.9  & 022817.85+723801.0*   & 02281782+7238009	  & \nodata            &  \nodata & \nodata &  \nodata & [KOS94] HA 2, RNO7-7 \\   
022818.51+723734.6  & \nodata               & 02281847+7237347    &  \nodata           &  \nodata & \nodata &  \nodata & \nodata  \\
022838.02+723740.6  & 022838.03+723740.8g   & 02283804+7237407    & 022838.01+723740.8  &  \nodata & \nodata &  \nodata & \nodata \\
022850.36+723851.2  & 022850.35+723851.0*   & 02285031+7238506    & 022850.29+723851.4 &  \nodata & \nodata &  \nodata & \nodata \\
\enddata   
\end{deluxetable}
\newpage

\begin{deluxetable}{cccccccc}
\tabletypesize{\scriptsize}
\tablecolumns{8}
\tablewidth{0cm}
\tablecaption{Extinctions, extinction-corrected spectral indices, bolometric temperatures and luminosities of
Class~0/I sources of L1340 \label{Tab3_class1}}
\tablehead{
\colhead{SSTSL2} & \colhead{A$_\mathrm{V}$} & \colhead{$\alpha(K-24)$} & \colhead{$\alpha(3.6-24)$} & \colhead{$\alpha(3.6-8.0)$} &  \colhead{$\alpha(24-70)$}
& \colhead{$T_\mathrm{bol}$}  & \colhead{$L_\mathrm{bol}$}  \\
 & (mag) &  & & & & \colhead{(K)} & \colhead{($L_{\sun}$)} }
\startdata	      
 022756.91+730354.4 &  4.6  & 	 0.79  &  \phs0.81    &  \phs1.16      & \phs1.20    & 108  &   4.40  \\
 022800.65+730415.2 &  4.7  & 	 0.16  &  \phs0.44    &  \phs0.17      & $-$0.71  &  664   &   2.72  \\ 
 022808.60+725904.5 &  2.2  & \nodata  & $-$1.06  &  $-$1.14    &  \phs4.28   & 55     &   3.54  \\
 022818.51+723506.2 &  3.6  &  1.18\tablenotemark{a} & \phs1.38\tablenotemark{b} & \phs1.52 & \nodata & \nodata &   \nodata  \\
 022820.81+723500.5 &  3.6  & \nodata  & \phs2.34\tablenotemark{b} &  \phs2.98  & \nodata   &  42  & 22.81   \\
 022825.07+730945.6 &  2.8  & \nodata  & \phs1.59     &  \phs1.93   & $-$0.41  &  149 &   0.15  \\
 022842.57+723544.3 &  1.7  & 0.71\tablenotemark{a}  & \phs0.63\tablenotemark{b} & \phs0.14 & \nodata   & \nodata & \nodata  \\
 022844.40+723533.5 &  1.7  & 0.48\tablenotemark{a}  & \phs0.64\tablenotemark{b}   &  \nodata  &   \nodata & \nodata  \\
\enddata                      	  			  				       
\tablenotetext{a}{$\alpha(K-22)$, since this source is outside of the MIPS images.}
\tablenotetext{b}{$\alpha(3.4-22)$, since this source is outside of the MIPS images.}
\end{deluxetable}

\begin{deluxetable}{ccccccc}
\tabletypesize{\scriptsize}
\tablecolumns{7}
\tablewidth{0cm}
\tablecaption{Extinctions, extinction-corrected spectral indices, bolometric temperatures and luminosities of Flat SED sources \label{Tab3_flat}}
\tablehead{
\colhead{SSTSL2} & \colhead{$A_\mathrm{V}$} &  \colhead{$\alpha(K-24)$} & \colhead{$\alpha(3.6-24)$} & \colhead{$\alpha(3.6-8.0)$}  & \colhead{$T_\mathrm{bol}$} & \colhead{$L_\mathrm{bol}$} \\
 & \colhead{(mag)} & &  &   &  \colhead{(K)} & \colhead{($L_{\sun}$)}  }
\startdata 
022754.00+723535.5  &  3.3   & $-$0.03\tablenotemark{a}   & \phs0.19\tablenotemark{b}    & \nodata    &      2393   &  0.15    \\
022759.92+723556.4  &  0.9   &  $-$0.17\tablenotemark{a}  & $-$0.19\tablenotemark{b} & \nodata        &      2674   &  0.06    \\
022811.32+723631.5  &  2.6   & $-$0.14\tablenotemark{a}	  & \phs0.03\tablenotemark{b}    & $-$0.43    &      2726   &  0.43    \\   
022816.62+723732.6  &  3.8   &  \phs0.25\tablenotemark{a} & \phs0.13\tablenotemark{b}  & 0.02         &      3472   & 28.48   \\
022817.85+723800.9  &  2.0   & \phs0.00			  &   \phs0.02	     & $-$0.18                &      1377   & 10.16    \\      
022818.51+723734.6  &  3.6   &  \nodata 		  & 	\nodata  &  $-$0.14                   &     \nodata & \nodata  \\  
022838.02+723740.6  &  3.5   & $-$0.32			  &   $-$0.24	   &  $-$0.32       	      &      2706   &  0.30    \\
022850.36+723851.2  &  3.0   &  $-$0.13 		  &   $-$0.17	   &  $-$0.51       	      &      \nodata & \nodata   \\
\enddata   
\tablenotetext{a}{$\alpha(K-22)$, since this source is outside of the MIPS images.}
\tablenotetext{b}{$\alpha(3.6-22)$, since this source is outside of the MIPS images.}  				  
\end{deluxetable}

\clearpage

\begin{deluxetable}{ccccccccccc}
\tabletypesize{\scriptsize}
\rotate
\tablecolumns{11}
\tablewidth{0cm}
\tablecaption{Properties of Class~II objects derived from the SEDs \label{Tab3_class2}}
\tablehead{
\colhead{SST SL2} & \colhead{Sp}  & \colhead{$A_\mathrm{V}$} & \colhead{$T_\mathrm{eff}$} 
& \colhead{$(L_{star}$} & \colhead{$T_\mathrm{bol}$} & \colhead{$L_\mathrm{bol}$} &  \colhead{$\alpha(K-24)$} & 
\colhead{$\alpha(3.6-8.0)$} & \colhead{$\alpha(8-24)$} & \colhead{SED subtype} \\
 & & \colhead{(mag)} & \colhead{(K)} & \colhead{($L_{\sun}$)}  & \colhead{(K)} & \colhead{($L_{\sun}$)} } 
 \startdata								
 022638.02+730457.5\tablenotemark{*} &    K4  &   0.7  & 4330  &      0.33    &  2556 &    0.56  &   $-$0.76  &   $-$0.38		    &  $-$0.77   &   II\,P  \\       
 022654.73+724040.8                  &    M1  &   1.4  & 3630  &      0.30    &  3184 &    0.34  &   $-$0.96  &   $-$2.58		    &\phs1.04	 &  II\,T  \\	     
 022659.03+725716.0                  &    M0  &   1.6  & 3770  &      0.34    &  3198 &    0.41  &   $-$0.96  &   $-$1.94		    &	\phs0.75	 &  II\,T  \\	     
 022659.08+724016.6\tablenotemark{*} &    M1  &   1.5  & 3630  &      0.14    &  3183 &    0.16  &   $-$1.15  &   $-$1.18		    &  $-$0.73   &   II\,A   \\      
 022659.35+725714.2                  &    M1  &   1.7  & 3630  &      0.33    &  3008 &    0.44  &   $-$0.71  &   $-$1.53		    &	\phs0.95	 &  II\,T \\	     
 022700.34+724743.8\tablenotemark{*} &    K3  &   1.3  & 4550  &      0.79    &  3916 &    0.83  &   $-$0.87  &   $-$1.27		    &	\phs0.30	 &  II\,T  \\	     
 022702.11+724329.0\tablenotemark{*} &    K9  &   2.2  & 3770  &      0.41    &  3447 &    0.40  &   $-$1.20  &   $-$2.06		    &  $-$0.17   &  II\,T  \\	     
 022703.17+723952.9\tablenotemark{*} &    M0  &   1.5  & 3770  &      0.13    &  3052 &    0.15  &   $-$0.82  &   $-$1.05		    &  $-$0.45   &   II\,P  \\
\enddata				        		 	      	       
\tablenotetext{*}{H$\alpha$ emission star.}
\tablenotetext{a}{$\alpha(K-22)$, since this source is outside of the MIPS images.}
\tablenotetext{b}{$\alpha(3.6-5.8)$}
\tablenotetext{c}{$\alpha(3.6-24)$}
\tablecomments{Table \ref{Tab3_class2} is published in its entirety in the
electronic edition of the {\it Astrophysical Journal}.  A portion is
shown here for guidance regarding its form and content.}
\end{deluxetable}

\begin{deluxetable}{rcccccl}
\tabletypesize{\scriptsize}
\tablecolumns{7}
\tablewidth{0cm}
\tablecaption{NIR-excess stars identified in the Omega-Cass images \label{Table_oc}}
\tablehead{
\colhead{N} & \colhead{$J \pm \Delta J$}  & \colhead{$H \pm \Delta H$} & \colhead{$K \pm \Delta K$} 
& \colhead{2MASS} & \colhead{Other Id/Position}}
\startdata 
 1  &   15.776\,$\pm$\,0.020	 &  15.258\,$\pm$\,0.016    &   14.736\,$\pm$\,0.011  &   02280836+7237183 & SSTSL2 J022808.36+723718.1	 \\   
 2  &   13.034\,$\pm$\,0.011	 &  11.828\,$\pm$\,0.030    &   10.780\,$\pm$\,0.030  &   02281182+7236447 & RNO7-3 \\
 3\tablenotemark{*}  &   15.625\,$\pm$\,0.017	 &  15.192\,$\pm$\,0.024    &   14.562\,$\pm$\,0.030  &	  \nodata  & 02 28 12.50 +72 36 44.0 \\
 4\tablenotemark{*}  &   14.631\,$\pm$\,0.020	 &  13.170\,$\pm$\,0.020    &   12.264\,$\pm$\,0.020  &   02281643+7237517   &  	\\   
 5\tablenotemark{*}  &   15.531\,$\pm$\,0.027	 &  14.921\,$\pm$\,0.015    &   13.893\,$\pm$\,0.022  &   02281673+7238090   &  \\  
 6  &   11.400\,$\pm$\,0.025	 &  10.236\,$\pm$\,0.014    &    9.100\,$\pm$\,0.011  &   02281661+7237328   & RNO7-5 \\
 7\tablenotemark{*}  &   17.097\,$\pm$\,0.018	 &  15.827\,$\pm$\,0.017    &   14.961\,$\pm$\,0.023  &	  \nodata	 & 02 28 17.35 +72 37 54.5  \\  
 8\tablenotemark{*}  &   16.375\,$\pm$\,0.010	 &  15.050\,$\pm$\,0.015    &   14.037\,$\pm$\,0.030  &   02281720+7237243   &   	\\
\enddata        			    		 
\tablenotetext{*}{New candidate YSO.} 
\end{deluxetable}

\begin{deluxetable}{ccccc}
\tabletypesize{\scriptsize}
\tablecolumns{5}
\tablewidth{0cm}
\tablecaption{\textit{AllWISE\/} point sources selected as candidate YSOs \label{Table_wise}}
\tablehead{
\colhead{AllWISE} & \colhead{[3.4]} & \colhead{[4.6]} & \colhead{[12.0]}  & \colhead{[22.0]} \\
  & \colhead{(mag)} & \colhead{(mag)} & \colhead{(mag)} & \colhead{(mag)}}
\startdata 
 022759.92+723556.4\tablenotemark{a,b}  & 14.060\,$\pm$\,0.026    & 12.460\,$\pm$\,0.023   & 10.328\,$\pm$\,0.076    & 8.475\,$\pm$\,0.322  \\
 023037.18+723037.0\tablenotemark{a}  & 11.198\,$\pm$\,0.023 & 10.652\,$\pm$\,0.021)  & 8.699\,$\pm$\,0.026)  & 6.673\,$\pm$\,0.076) \\
 023043.49+722812.2  & 10.593\,$\pm$\,0.024    & 10.179\,$\pm$\,0.020)  &	9.014\,$\pm$\,0.029)  & 6.280\,$\pm$\,0.062) \\
 023044.85+722639.8  & 11.975\,$\pm$\,0.024    & 11.503\,$\pm$\,0.021)  &  10.095\,$\pm$\,0.054)  & 8.145\,$\pm$\,0.199) \\ 
 023147.07+722722.0  & 11.360\,$\pm$\,0.022    & 10.944\,$\pm$\,0.021)  &	8.830\,$\pm$\,0.028)  & 6.953\,$\pm$\,0.079) \\
 023202.38+722820.3  & 12.646\,$\pm$\,0.024    & 12.103\,$\pm$\,0.023)  &  10.272\,$\pm$\,0.075)  & 7.656\,$\pm$\,0.145) \\
 023209.62+722858.8  & 11.206\,$\pm$\,0.022    & 10.896\,$\pm$\,0.020)  &	9.179\,$\pm$\,0.033)  & 6.312\,$\pm$\,0.060) \\ 
 023212.59+723013.7\tablenotemark{a}  & 11.086\,$\pm$\,0.023  & 10.352\,$\pm$\,0.021) & 8.275\,$\pm$\,0.022)  & 6.286\,$\pm$\,0.050) \\ 
 023227.63+723841.4\tablenotemark{b}  & 14.231\,$\pm$\,0.027    & 12.630\,$\pm$\,0.023   &  12.249\,$\pm$\,0.339   & 6.632\,$\pm$\,0.064 \\ 
\enddata        			    		 
\tablenotetext{a}{H$\alpha$ emission star (Paper~III)}
\tablenotetext{b}{This source is located within the field of view of the \spitzer\ observations, but missing from the SEIP Source List.}
\end{deluxetable}

\clearpage

\begin{deluxetable}{ccccccc}
\tabletypesize{\scriptsize}
\tablecolumns{7}
\tablewidth{0cm}
\tablecaption{Properties of Class~II \textit{AllWISE\/} sources, derived from the SEDs \label{Tab2_wise}}
\tablehead{
\colhead{AllWISE} & \colhead{Sp.}  & \colhead{$A_\mathrm{V}$} & \colhead{$T_\mathrm{eff}$} 
& \colhead{$(L_\mathrm{star}/L_{\sun})$} &  \colhead{$\alpha(K-22)$} &  \colhead{SED subtype} } 
\startdata 
 023037.18+723037.0  &  K4  &  3.3 &   4330 &   1.377  &  $-$1.101  &  II\,A \\ 
 023043.49+722812.2  &  K4  &  1.0 &   4330 &   2.070  &  $-$1.063  &  II\,A \\ 
 023044.85+722639.8  &  M0  &  1.8 &   3770 &   0.278  &  $-$1.112  &  II\,A \\ 
 023147.07+722722.0  &  K4  &  1.5 &   4330 &   1.064  &  $-$0.930  &  II\,A \\ 
 023202.38+722820.3  &  K6  &  1.8 &   4020 &   0.222  &  $-$0.713  &  II\,T \\ 
 023209.62+722858.8  &  K4  &  1.0 &   4330 &   1.089  &  $-$0.797  &  II\,T \\
 023212.59+723013.7  &  K1  &  3.8 &   4920 &   0.864  &  $-$0.811  &  II\,P \\ 
\enddata
\end{deluxetable}

\begin{deluxetable}{lccccccr} 
\tabletypesize{\scriptsize}
\tablecolumns{8} 
\tablewidth{0pc} 
\tablecaption{Average properties of candidate PMS stars with different SED slopes \label{Table_sedcomp}}
\tablehead{ 
\colhead{SED slope}& \colhead{$\langle K_s \rangle$} & \colhead{$\langle A_\mathrm{V} \rangle$ } & \colhead{$\langle T_\mathrm{eff} \rangle$} & \colhead{$\langle L/L_{\sun} \rangle$}& \colhead{$\langle T_\mathrm{bol} \rangle$} & \colhead{$\langle L_\mathrm{bol}/L_{\sun} \rangle$}  & \colhead{N} } 
\startdata  
Primordial (II\,P)  & 12.225 & 2.83 & 4160 & 1.02  & 2634 & 1.49  & 46 \\
Evolved (II\,A)     & 13.474 & 2.83 & 3620 & 0.69  & 2460 & 0.84 &  80 \\
Evolved (II\,T)     & 13.343 & 2.34 & 3660 & 0.34  & 2378 & 0.51  & 29  \\
All Class~II        & 13.124 & 2.74 & 3806 & 0.71  & 2480 & 0.96 &  155 \\
\enddata  
\end{deluxetable}

\begin{deluxetable}{lcccc}
\tabletypesize{\scriptsize}
\tablecolumns{4}
\tablewidth{0cm}
\tablecaption{Major properties of the YSO groups associated with the cloud clumps\label{Table_clump}}
\tablehead{
\colhead{Clump}  &  \colhead{L1340\,A} &  \colhead{L1340\,B}  & \colhead{L1340\,C} & \colhead{L1340 Total\tablenotemark{a}} }
\startdata
 Area (pc$^2$)  &  5.4 & 10.8  &  7.0  &  90\tablenotemark{b}  \\
 N(Class 0/I)    &  8   &  23   &  14   &  45 \\ 
 N(Flat)       &  12  &  9    &   6   & 27 \\
 N(Class II)    &  51  &  47   &  50   &  170  \\
 N(Sp $\le$ F5)  &  3  &   7   &   1   & 11 \\
 N(total)       &  74  &  87  &  71   &  253 \\
Mean surface density (stars\,pc$^{-2}$)   & 41.1 & 5.25 & 13.4 & 2.8   \\
 NII / (NI+NFlat)     &  2.55 & 1.47 & 2.5  &  2.36 \\
SFE\tablenotemark{c} & 0.17 & 0.03 & 0.05 & 0.03 \\
\enddata  
\tablenotetext{a}{Stars outside of the surface density contour of 2~stars\,pc$^{-2}$ are also included in this column.}
\tablenotetext{b}{Total area of L1340 with $A_\mathrm{V} > 1.0$~mag (Paper III).}
\tablenotetext{c}{Star formation efficiency, assuming average stellar mass of 0.5\,M$_{\sun}$, and taking the clump masses from Paper~III.}
\end{deluxetable}

\begin{deluxetable}{lcccccc}
\tabletypesize{\scriptsize}
\tablecolumns{7}
\tablewidth{0cm}
\tablecaption{Positions, sizes and stellar contents of the clusters embedded in L1340 \label{Table_cluster}}
\tablehead{
Object & \colhead{RA(J2000)} & \colhead{Dec(J2000)} & \colhead{a(pc)\tablenotemark{a}}  & \colhead{AR\tablenotemark{b}} &  \colhead{N(total)} &  \colhead{N(ClassII)/N(ClassI)}}
\startdata
RNO 7  & 02 28 31.2 & 72 37 34 & 0.8 & 1.9 &  56 & 2.5  \\
RNO 8  & 02 30 42.0 & 72 59 50 & 0.21 & 1.17 & 8 & 2   \\
RNO 9  & 02 31 21.6 & 72 40 08 & 0.55 & 1.8  & 26 & 5   \\
IRAS 02276+7225 & 02 32 32.6 & 72 39 02 & 0.34 & 1.48 &  12 & 1.0   \\
SSYSC\tablenotemark{c} median & \nodata & \nodata & 0.39 & 1.82 & 26 & 3.7 \\
\enddata  
\tablenotetext{a}{Major axis of the smallest ellipse containing the member stars.}
\tablenotetext{b}{Aspect ratio of the same ellipse.} 
\tablenotetext{c}{A Spitzer Survey of Young Stellar Clusters within one kiloparsec of the Sun, \citep{Gutermuth09}}
\end{deluxetable}

\end{document}